\documentclass[11pt,a4paper]{article}
\usepackage[utf8]{inputenc}
\usepackage{authblk}
\usepackage[scale=0.7]{geometry}
\usepackage[T1]{fontenc}\usepackage{hyperref} 
\usepackage{amssymb}
\usepackage{amsmath,amsthm} 
\usepackage[english]{babel}
\usepackage{graphicx}
\usepackage{natbib} 
\usepackage{caption}
\usepackage{breakcites}
\usepackage{hyperref}
\usepackage{todonotes}
\usepackage{bm}
\usepackage{stmaryrd} 
\usepackage{color}
\usepackage{ulem} 
\usepackage{gensymb}
\usepackage{tikz}
\usepackage{color}
\usepackage{multirow}
\usepackage{ulem} 
\newcommand{\new}[1]{#1}
\usepackage[toc,page]{appendix}
\usepackage{tocloft}
\usepackage{array} 

\usetikzlibrary{calc,shapes,backgrounds,arrows,automata,shadows,positioning}
\providecommand{\keywords}[1]{\textbf{\textit{Keywords---}} #1}

\newcommand{\prior}{\textit{prior }}
\newcommand{\posterior}{\textit{posterior }}

\newcommand{\EqRef}[1] {Equation (\ref{#1})}
  
\edef\hc{\string: }

\title{\huge Bayesian calibration of a numerical code for prediction \vspace*{0.2cm} \\ \small Theory of code calibration and application to the prediction of a photovoltaic power plant electricity production}

\author[1,2,3]{Mathieu Carmassi}
\author[1]{Pierre Barbillon}
\author[3]{Merlin Keller}
\author[1]{Eric Parent}
\author[2]{Matthieu Chiodetti}
\affil[1]{UMR MIA-Paris, AgroParisTech, INRA, Paris }
\affil[2]{EDF R\&D, TREE department, Moret-sur-Loing}
\affil[3]{EDF R\&D, PRISME department, Chatou}

\begin{document}
\maketitle

\begin{abstract}
Les difficultés de mise en \oe uvre d'expériences de terrain ou de laboratoire, ainsi que les coûts associés, conduisent les sociétés 
industrielles à se tourner vers des codes numériques de calcul. Ces codes, censés être représentatifs des phénomènes physiques en jeu, 
entraînent néanmoins tout un cortège de problèmes. Le premier de ces problèmes provient de la volonté de prédire la réalité à partir d'un 
modèle informatique. En effet, le code doit être représentatif du phénomène et, par conséquent, être capable de simuler des données proches de  
la réalité. Or, malgré le constant développement du réalisme de ces codes, des erreurs de prédiction subsistent. Elles sont de deux natures différentes. La
première provient de la différence entre le phénomène physique et les valeurs relevées expérimentalement. La deuxième concerne l'écart entre le code développé et le phénomène physique. Pour diminuer cet écart, souvent qualifié de biais ou d'erreur de modèle, les développeurs complexifient en général les codes, les rendant très chronophages dans certains cas. De plus, le code dépend de paramètres à fixer par l'utilisateur qui doivent être choisis pour correspondre au mieux aux données de terrain. L'estimation de ces paramètres propres au code s'appelle le calage.
Ce papier propose une revue des méthodes de calage bayésien et s'appuie sur un cas d'application qui permet de discuter les divers choix méthodologiques et d'illustrer leurs divergences. Cet exemple s'appuie sur un code de calcul servant à prédire la puissance d'une centrale photovoltaïque.

\end{abstract}
  

\vspace{1cm}

\begin{abstract}
Field experiments are often difficult and expensive to carry out. To bypass these issues, industrial companies have developed computational codes. These codes are intended to be representative of the physical system, but come with a certain number of problems. Despite continuous code development, the difference between the code outputs and experiments can remain significant. Two kinds of uncertainties are observed. The first one comes from the difference between the physical phenomenon and the values recorded experimentally. The second concerns the gap between the code and the physical system. To reduce this difference, often named model bias, discrepancy, or model error, computer codes are generally complexified in order to make them more realistic. These improvements increase the computational cost of the code. Moreover, a code often depends on user-defined parameters in order to match field data as closely as possible. This estimation task is called calibration. This paper proposes a review of Bayesian calibration methods and is based on an application case which makes it possible to discuss the various methodological choices and to illustrate their divergences. This example is based on a code used to predict the power of a photovoltaic plant. 

\end{abstract}


\section{Introduction}

Numerical experiments have become increasingly popular in many (if not all) industrial fields, as setting up field experiments can represent a huge investment for a company. Numerical simulations are generally considered as a substitute to bypass physical or field experiments \citep{santner2013,fang2005}. However, the complexity of the computer codes used in such simulations increases with the capacity of computer processors, and sometimes at a much higher rate. As a result, some codes have become greedy in computational time \citep{sacks1989}. Moreover, a gap between computer code outputs and field measures of the physical process that the code seeks to simulate is routinely observed. 
Checking the accuracy of the code by confronting it with field experiments is called validation \citep{bayarri2007}. This task is difficult since the requisite field data are scarce and it is based on a computational code that often has a long runtime. 
Throughout this paper, we will use the word ``code'' as a proxy for numerical code, sometimes also called numerical model, simulator or computational code and field experiment for real world experiment.
\newline


The code generally depends on two kinds of inputs\hc variables and parameters.
The variables are input variables (observable and often controllable) which are set during a field experiment and can encompass environmental variables that can be measured.
The parameters are generally interpreted as physical constants defining the mathematical model of the system of interest, but can also contain so-called tuning parameters, which have no physical interpretation. They have to be set by the user to run the code and need to be chosen carefully to make the code mimic the real physical phenomenon. 
The code can be mathematically represented by a function $f_c$. Let us note, in what follows, $\boldsymbol{\theta}\in\mathcal{Q}\subset\mathbb{R}^p$ to represent the parameter vector and $\mathbf{x}\in\mathcal{H}\subset \mathbb{R}^d$ for the variable vector. The space $\mathcal{Q}$ is called the input parameter space and $\mathcal{H}$ the input variable space.
The physical phenomenon is denoted by $\zeta$ and only depends on variables in vector $\boldsymbol{x}\in\mathcal{H}$, the parameter vector $\boldsymbol{\theta}$ having no counterpart in field experiments. \newline

A code output is then written as $f_c(\boldsymbol{x},\boldsymbol{\theta})$ whereas $\zeta(\boldsymbol{x})$ denotes the output of the physical phenomenon for the same variable $\boldsymbol{x}$. This is of course an idealized formalization, in which we assume that the code variables $\boldsymbol{x}$ are exhaustive to describe the phenomenon of interest, in that the quantity to be predicted can take a single deterministic value $\zeta(\boldsymbol{x})$ for a given $\boldsymbol{x}$. 
In this paper, the quantity of interest is assumed to be a scalar but 
extensions with high dimension outputs are
possible \citep{higdon2008}. 
Therefore, in what follows, the outputs of $\zeta$ and $f_c$ lie in $\mathbb{R}$.
\newline

We consider that a vector of field data ($\boldsymbol{y}_{exp}$), which are noisy measurements of $\zeta$, is observed as a realization of the statistical model\hc

\begin{equation*}
\mathcal{M}_0\ : \ \forall i \in \llbracket1,\dots,n\rrbracket \quad y_{exp_i}=\zeta(\boldsymbol{x}_i)+\epsilon_i
\end{equation*}

where $\forall i \in \llbracket1,\dots,n\rrbracket \quad \epsilon_i\overset{iid}{\sim} \mathcal{N}(0,\sigma_{err}^2)$. The corresponding values of the variables $\boldsymbol{x}_i$ are also observed.
\newline

Calibrating the code consists
in making the parameter vector $\boldsymbol{\theta}$ consistent in some sense with these $n$ field data. In an industrial framework, uncertainty quantifications \citep{rocquigny2009quantifying} can be decomposed into a step by step procedure and calibration is identified as a key element of the so-called \textit{step B'}. As illustrated in \citet{damblin2015}, this step concerns calibration,
verification and validation (V\&V). V\&V can be further split into 3 phases  described in \citet{roache1998}, \citet{bayarri2007} and \citet{oberkampf1998}.
Calibration then aims at finding the "best" parameter vector $\boldsymbol{\theta}=\boldsymbol{\theta}^*$ such that the error term made by the
code in a statistical model is minimal. Several statistical modeling strategies have been proposed in the literature. When only measurement errors
are considered, \citet{cox2001} use a rather simple model, considering that the code does not differ from the phenomenon under study while
\citet{higdon2004}, \citet{kennedy2001} and \citet{bayarri2007} advocate for some extensions which encompass a \textit{model bias} 
or a \textit{model error} term, also dubbed as \textit{discrepancy} in the following. All of these models are reviewed and discussed in Section
\ref{sec:calibration}. The identifiability issues between the parameter $\boldsymbol{\theta}$ and the discrepancy were already discussed in the written discussions of \citet{kennedy2001}. \citet{tuo2015efficient} consider the calibration task as a minimization of a loss function between the code and the physical reality. 
In \citet{tuo2016theoretical}, they show that this loss function leads to an estimation of $\boldsymbol{\theta}$ depending on the chosen \textit{prior} distribution of the discrepancy. Then,  \citet{plumlee2017} advises an orthogonality 
specification for the discrepancy \textit{i.e.} the discrepancy should be orthogonal to the gradient of the computer code with respect to a loss function.
\newline

As an industrial illustration, we will focus, in this paper, on predicting the energy production from a photovoltaic (PV) plant \citep{martin2001}. The industrial context is that of an electricity producer and supplier who has to consider bids from PV plants. The selling price announced by the industrial company has to be competitive enough to be successful in the bidding process. Of course, the building costs and the production over the plant lifetime have to be known to evaluate the profit margins. Although a best guess-estimate, the deterministic evaluation of PV production through a sophisticated computer code does not fulfill the needs of a financial investor in  PV projects. To evaluate the financial risks of the investment and consequently to make a decision, the investor needs to assess the uncertainty around the expected production estimation. To do so, all the sources of uncertainties have to be identified and treated in their current form. Calibration will be useful for hunting down the uncertainties related to the modeling and quantifying the range of the main potential errors made in predicting the profit ratio. \new{This real case study will allow us to emphasize the differences between the statistical models used for calibration in a context where the validity of the statistical hypotheses usually made is  not guaranteed.}
\newline


This paper first presents the illustrative case study (Section \ref{PVsection}). The issues at stake are described with the explanation of the code and the source of the experimental data. Section \ref{sec:calibration} deals with the presentation of the many different statistical models one can find in the literature which have been developed for calibration. Then, the different likelihoods and conditional densities needed for parameter estimation are highlighted. In Section \ref{sec:application}, the different statistical models are implemented and tested for the PV application case, in order to illustrate the various ways of reasoning behind calibration and point out their differences.

\section{Production estimation from a PV plant \label{PVsection}}

\subsection{Stakes}

The electricity production market has become increasingly competitive. Environmental issues have brought changes in such a way that producing electricity by exploiting solar power has become a popular and a major vector of green production. However, building a PV plant represents a considerable financial risk. Many factors must be taken into account before computing the return rate. The overall building cost is the first figure needed but can easily be estimated. Once it is evaluated, a prediction of the PV plant production will set up the return rate. To compute such a prediction, a code has been developed, implementing a mathematical model which aims at reproducing the physical system of the plant. \newline

However, there are two major sources of uncertainty linked to this method. The first one is the meteorological data which are difficult to predict, especially in an environment that is changing due to global warming. In a project framework, we will usually use the meteorological data based on the previous years with past scenarios adjusted if necessary to take into account the temperature increase. The second source of uncertainties comes from the modeling errors. The code may encounter difficulties in mimicking the physical system. As discussed above, this can be explained by the fact that the mathematical model implemented is only a simplified representation of the physical world, which might not take into account all the existing influential variables, and also depends on uncertain physical constants, which are precisely the parameters we wish to estimate. \newline

Consequently, the error made by the output of the code directly impacts the uncertainty on the estimation of production. In this article, we will focus only on the modeling errors. In practice, these errors are responsible for one half ($4\%$) of the error made on the total energy given by the plant ($8\%$).

\subsection{Source of the code}

To understand how the phenomenon has been coded, some explanation about how the PV cell works is needed. A PV cell is mainly composed of a semi-conductor material. For most technologies, this material is silicon \citep{luque2011}. The energy supply from the sun is remarkable at a quantum level since the energy from the light spectrum will modify the energy levels of the silicon atoms until one electron appears. In a single semiconductor crystal, two parts are visible. The ``p'' (positive) side which contains an excess of holes and the "n" side which contains an excess of electrons. A hole is an excess of a positive charge. The electron is attracted to the hole by diffusion, creating an electron/hole pair. The principle is to capture enough solar energy to create such an electron/hole pair. The displacement of an electron is directly  translated into electric current. The so called ``energy gap'' corresponds to the difference between the energy of the conduction band and the valence band. To create electricity, the incident solar ray on the PV cell has to have an energy spectrum higher than the energy gap. That is why cloudy days are not favorable for PV plant production. The PV cell has a plastic film and a glass cover to protect the silicon. Otherwise, the lifetime of the cell would be too short because of its degradation. These protections act as a filter for the sun's rays and some of the energy spectrum is lost. All in all, many conditions have to be met and only $20\%$ of the initial spectrum is at best transformed into electricity \citep{martin2001}. \newline

The code, hereafter considered as a ``black box'', is a solver of the main equations mimicking the electrical behavior of the PV plant. In this article the code will represent a PV test stand with 12 panels connected together. The power considered will be the one before the inverter (multiplication of the continuous current and continuous voltage). Fortunately, for an in-depth exploration of the complete range of situations encountered in the domain of Bayesian calibration of computer codes, the ``black box'' code $f_c$ appears to be fast for the case study (one launch needs only $39\mu s$ to run). To investigate what happens when the code is time consuming, we will simply slow it down by restructuring the number of runs allowed for the computer code $f_c$. The code depends on some parameter vector $\boldsymbol{\theta}$ and input variables $\boldsymbol{x}$ detailed as follows (also called general inputs in \citet{plumlee2017}): $\boldsymbol{\theta}=\begin{pmatrix}
\eta \\
\mu_t \\
n_t \\
a_l \\
a_r\\
n_{inc}
\end{pmatrix}$ and $\boldsymbol{x}=\begin{pmatrix}
t\\
L\\
l\\
I_g\\
I_d \\
T_e
\end{pmatrix}$. \newline

The physical meaning of the parameters is explained below \citep{duffie2013}\hc 
\begin{itemize}
\item $\eta$\hc module photo-conversion efficiency,
\item $\mu_t$\hc module temperature coefficient (the efficiency decreases when the temperature rises) in $\%/^{\circ}C$,
\item $n_t$\hc reference temperature for the normal operating conditions of the module in $^{\circ}C$,
\item $a_l$\hc reflection power of the ground (albedo),
\item $a_r$\hc transmission of the radiation as a function of the incidence angle of solar rays, which depends on optical properties and the cleanliness,
\item $n_{inc}$\hc transmission factor for normal incidence. \newline
\end{itemize}

The input variables contain all measurable data\hc 
\begin{itemize}
\item $t$\hc the UTC time since the beginning of the year in $s$,
\item $L$\hc the latitude in $^{\circ}$,
\item $l$\hc the longitude in $^{\circ}$,
\item $I_g$\hc global irradiation (normal incidence of the sun ray to the panel) in $W/m^2$,
\item $I_d$\hc diffuse irradiation (horizontal incidence of the sun ray to the panel) in $W/m^2$,
\item $T_{e}$\hc ambient temperature in $^{\circ}C$.\newline
\end{itemize}

Note that the temporal aspect is taken into account to a certain extent through the input variables. We do not consider any delay in the PV reaction to the forcing conditions. Time $t$ indicates here a snap shot corresponding to the instant when the power has to be computed. This code only focuses on a specific time and if the evolution of the power over a day is what we look for, a repetition over the specific duration has to be made. This operation has to consider the number of time steps available. For example, if $300$ configurations of $\boldsymbol{x}$ are accessible for one day, the code will have to be executed $300$ times to obtain the power evolution over a day. For the rest of the article, we will denote the code output referring to the i\textsuperscript{th} time step by $f_c(\boldsymbol{x}_i,\boldsymbol{\theta})$ and by $f_c(\textbf{X},\boldsymbol{\theta})$ the code outputs corresponding to the whole time frame contained in matrix $\textbf{X}$. \newline

A sensitivity analysis performed according to the screening method of Morris \citep{morris1991} showed that only the variations of $\eta$, $\mu_t$ and $a_r$ have a significant impact on the power. This sensitivity analysis consists in evaluating with elementary displacements in a normalized input space, the impacts of these displacements on the output. To demonstrate that $\eta$, $\mu_t$ and $a_r$ have a significant impact over a whole duration but not instantaneously a PCA (Principal Component Analysis) is performed on all outputs generated for the duration and for all combinations of the design of experiments (DOE) of Morris. On the new uncorrelated basis of the space, the information is summed up by a new Morris plot. This representation makes it possible to visualize all the information summarized over the duration on one plot. However, to  implement such a method, the input space parameter needs to be well defined. That is why the work of the experts is extremely important. They have to define the range of each parameter as best they can.\newline

\subsection{Available data}

\begin{figure}[htbp!]
\begin{center}
  \begin{tabular}{c @{} c c}
    \rotatebox{90}{ \hspace{7em} \small Power in $W$}
    & \includegraphics[width=.4\textwidth]{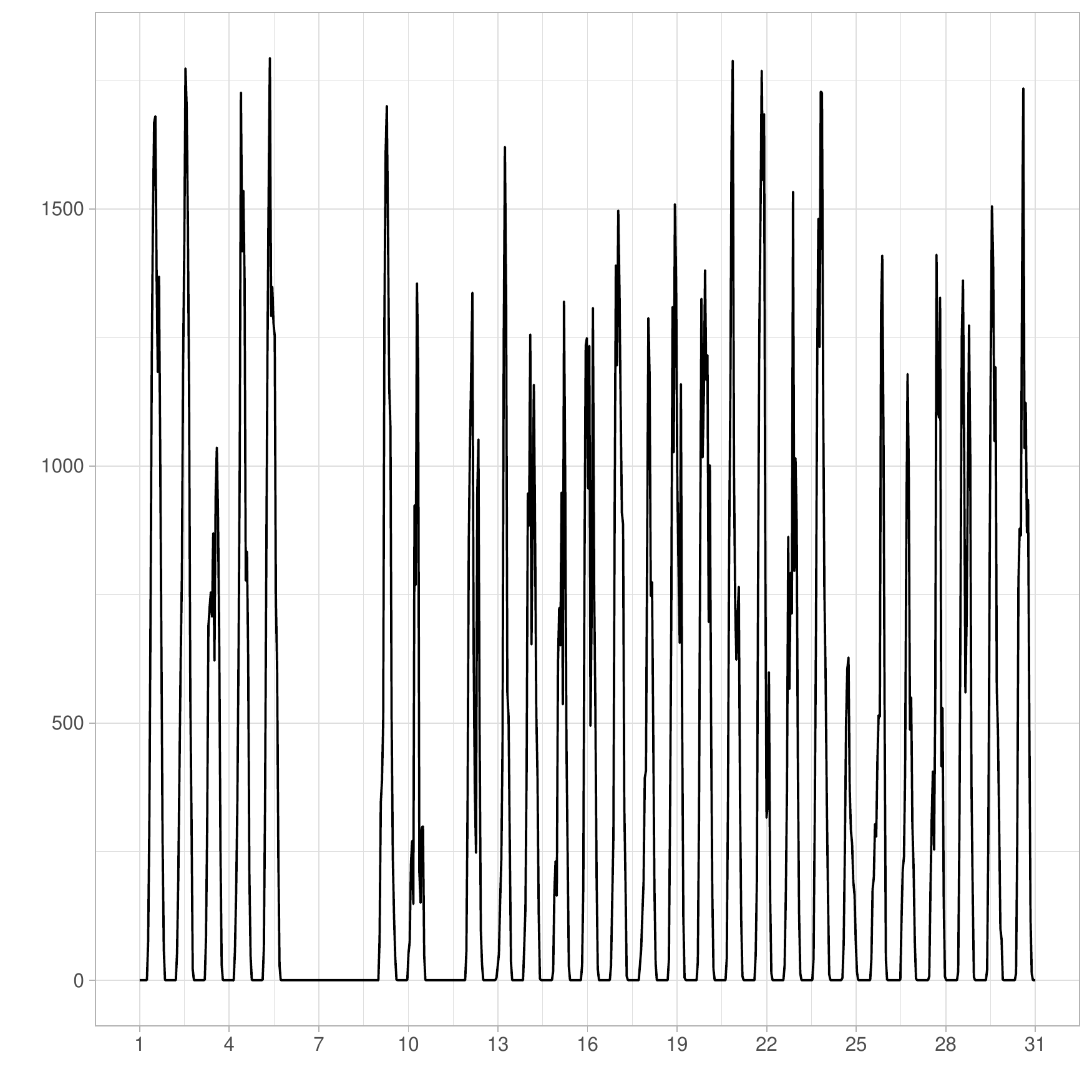} 
    &  \includegraphics[width=.4\textwidth]{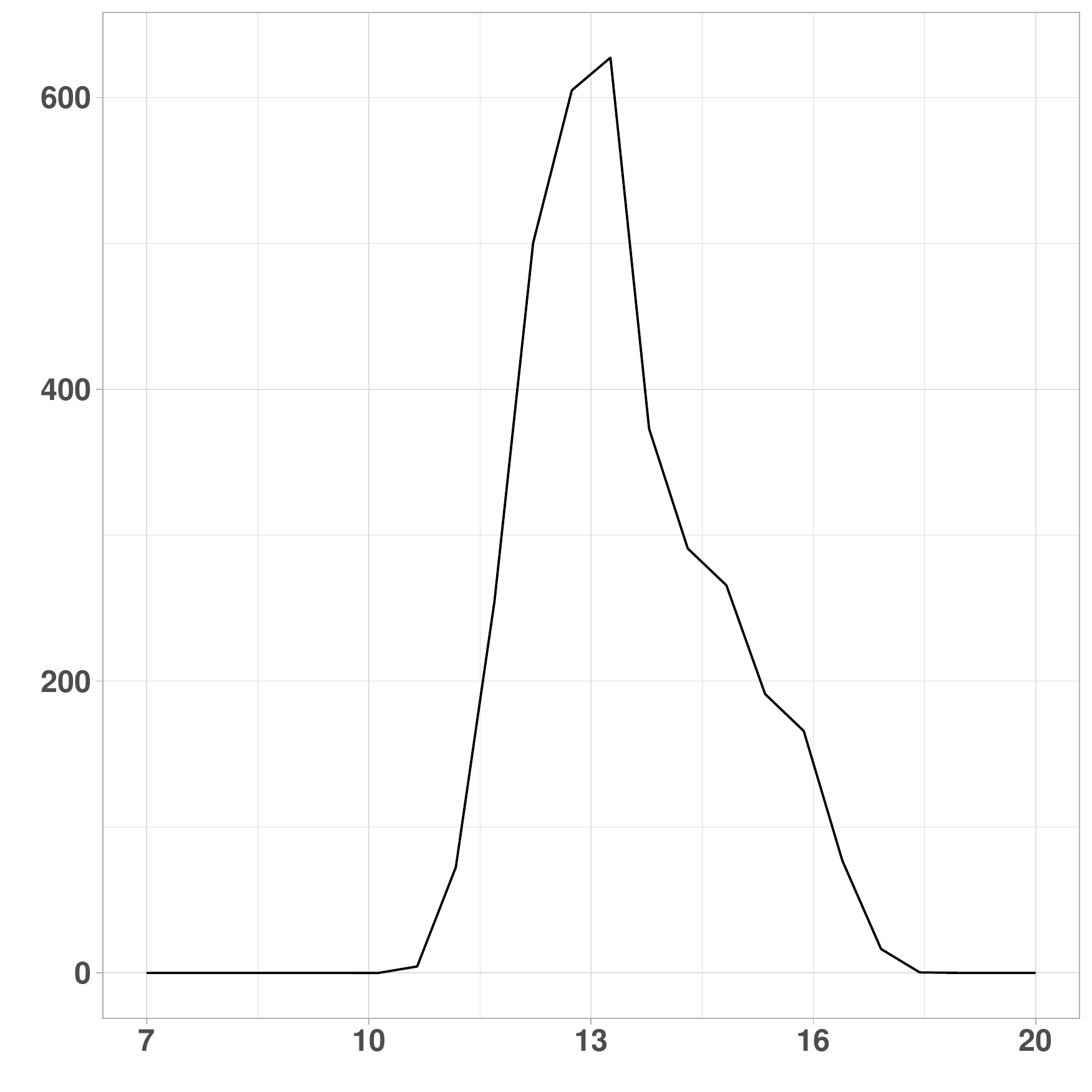}\\
    & Days & Hours \\
  \end{tabular}   
\caption{Power production by PVzen for August 2014 (left) and \\for August 25\textsuperscript{th} 2014 (right)}
\label{fig:PowerExample}
\end{center}
\end{figure}

As mentioned above, the code represents a test stand of 12 panels. Data are available over 2 months and instantaneous power is collected every $10s$, which makes around $777,600$ points to process. When the recording facilities are interrupted for some reason, a specific data processing is carried out. Figure \ref{fig:PowerExample} shows the kind of data collected from the stand over one month (on the left) and detailed for one day (on the right). On certain days the production remains stuck at $0$. This typically happens when recording errors occur. In fact, the power saved is aberrant with too much high or 
negative power. These errors can be detected and sorted by data cleaning. The panel on the right in Figure \ref{fig:PowerExample} shows the typical behavior of an assembly of solar panels. When the irradiation of the sun is high, so is the production. As expected, the maximum production happens around noon when the irradiation is high.

\subsection{Estimation of the error}

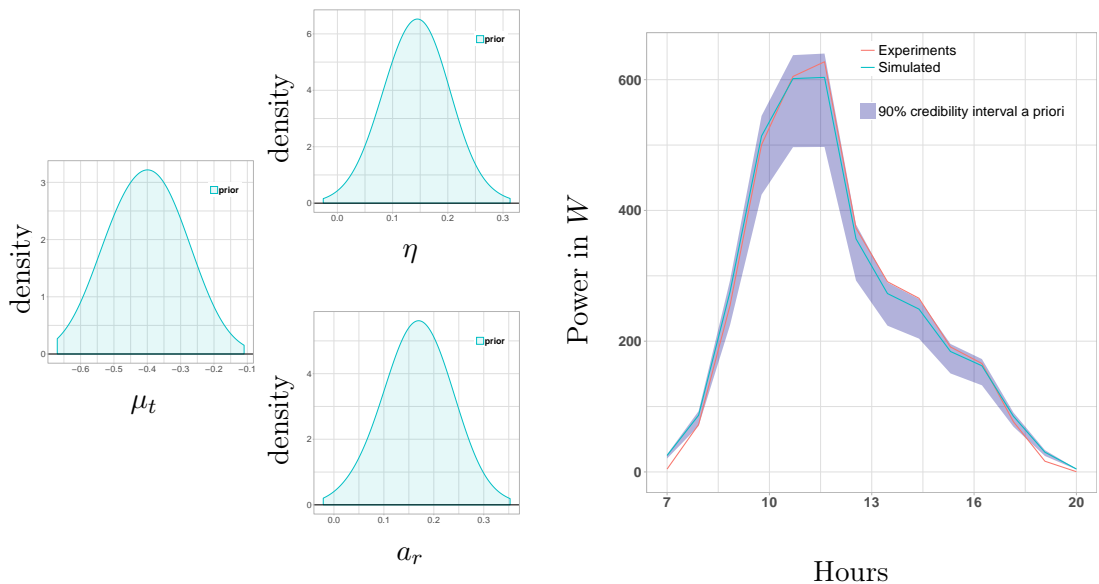
\begin{figure}[htbp!]
\centering
    \begin{tikzpicture}
		\tikzstyle{m1}=[]
		
		\node[m1] (N1) at (0,0) {\includegraphics[width=.2\textwidth]{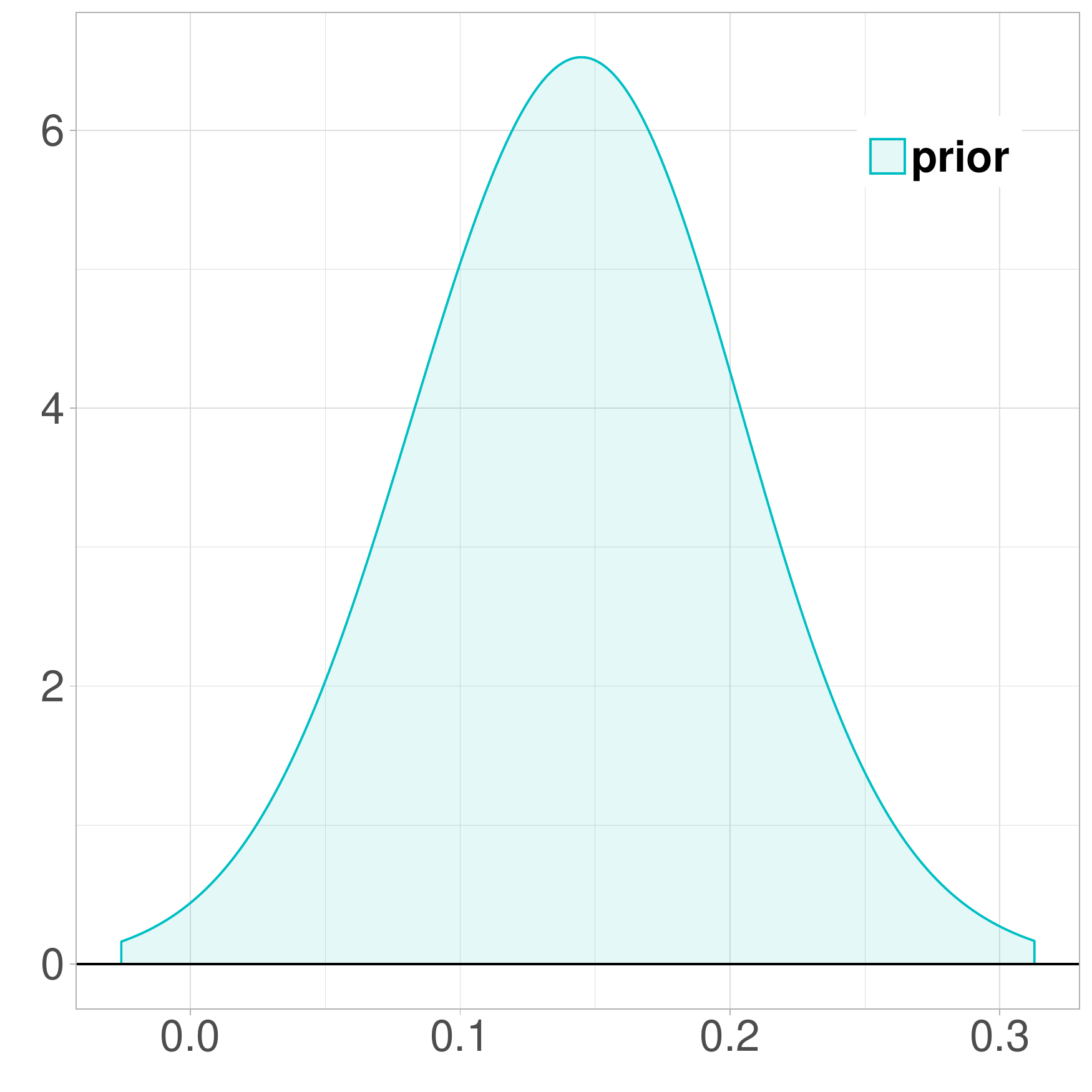}};
		\node[m1] (N2) at (-1.7,0) {\rotatebox{90}{density}};
		\node[m1] (N3) at (0,-1.8) {$\eta$};
		\node[m1] (N11) at (-5.1,-2) {\rotatebox{90}{density}};
		\node[m1] (N10) at (-3.5,-2) {\includegraphics[width=.2\textwidth]{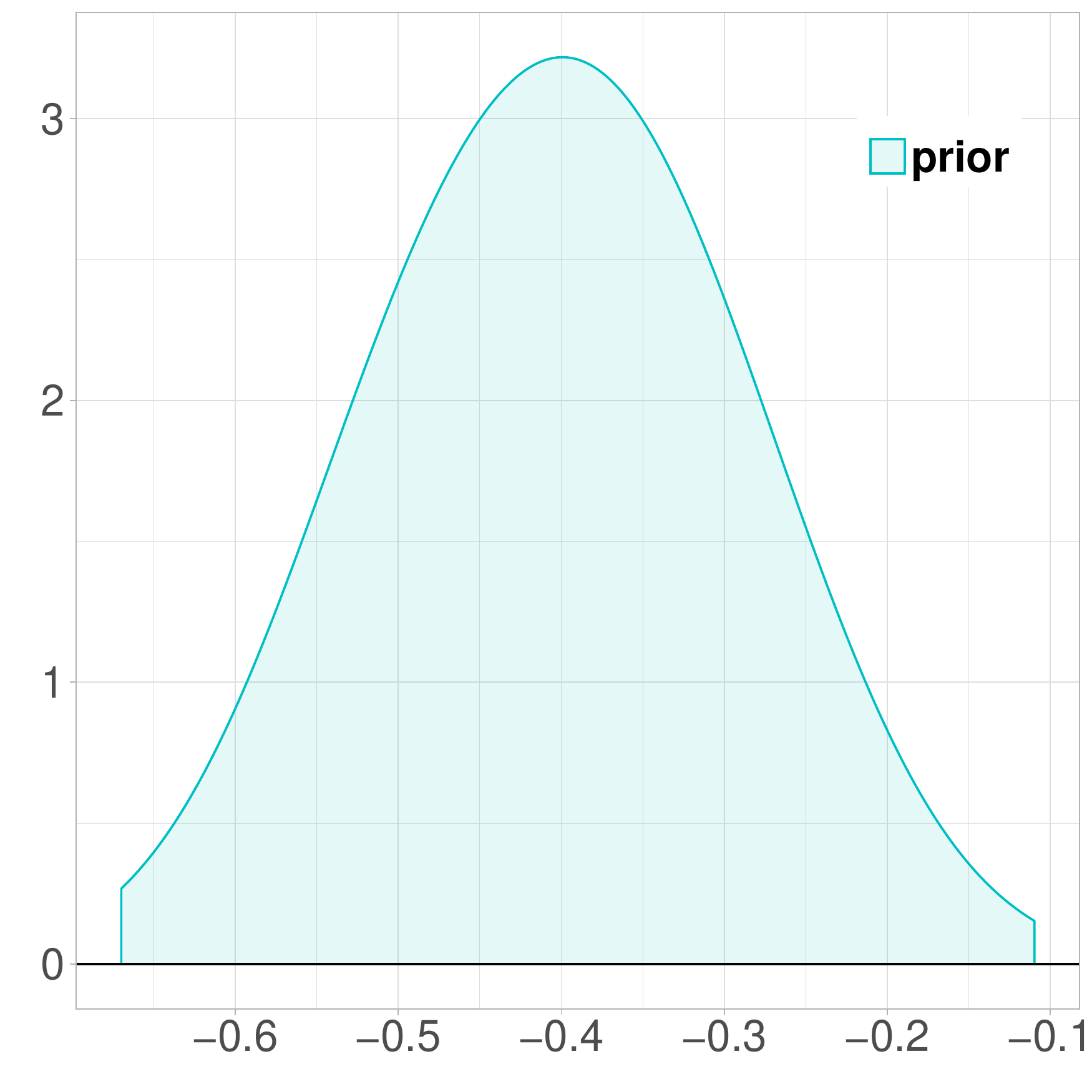}};
		\node[m1] (N12) at (-3.5,-3.8) {$\mu_t$};
		\node[m1] (N9) at (-1.7,-4) {\rotatebox{90}{density}};
		\node[m1] (N4) at (0,-4) {\includegraphics[width=.2\textwidth]{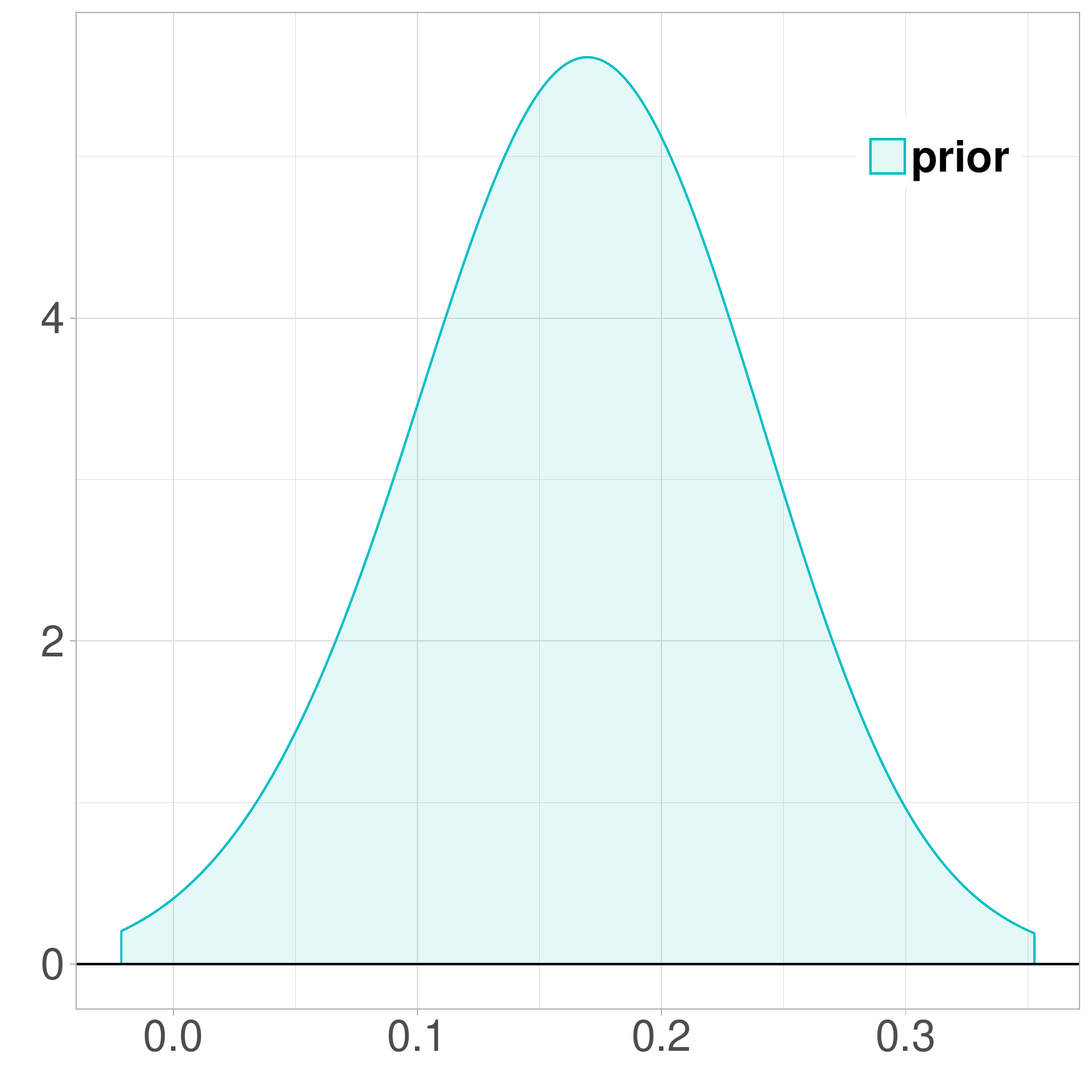}};
		\node[m1] (N5) at (0,-5.8) {$a_r$};
		\node[m1] (N8) at (2.2,-2) {\rotatebox{90}{Power in $W$}};
		\node[m1] (N6) at (5.8,-2.1) {\includegraphics[width=.45\textwidth]{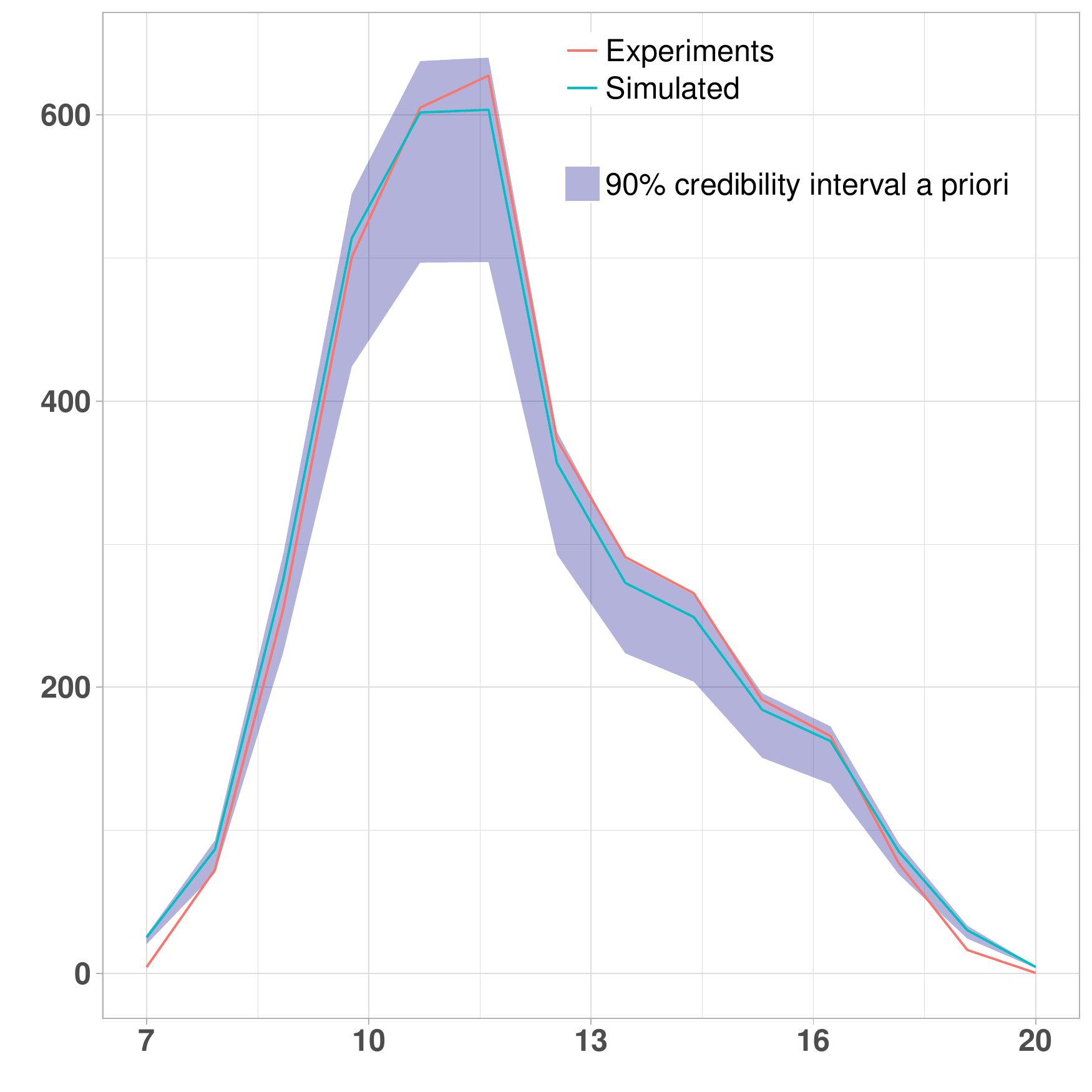}};
		\node[m1] (N7) at (5.8,-6) {Hours};
	
    \end{tikzpicture}
    
  \caption{$\pi(\eta)$, $\pi(\mu_t)$ and $\pi(a_r)$ \textit{prior} densities (represented on the left panel) and induced credibility interval of the instantaneous power (right panel)}
  \label{fig:ParamError}
\end{figure}

So far, experts have used the code with some parameter values with the knowledge that these parameters are uncertain (the so called reference values). They can also provide more expertise on the nature of the parameter. For example for $a_r$, the nominal value is $0.17$ and experts state that the parameter lies within the $95\%$ confidence interval $[0.05,0.29]$. We chose to consider $a_r$ as Gaussian with $a_r\sim\mathcal{N}(\mu=0.17,\sigma^2=3.6.10^{-3})$. The standard deviation is chosen equal to $0.06$ because we considered the upper bound and the lower bound of the given interval as respectively the quantiles $a_{r_{0.975}}$ and $a_{r_{0.025}}$. Similarly, $\eta$ and $\mu_t$ are taken as Gaussian such that $\eta\sim\mathcal{N}(\mu=0.143,\sigma^2=2.5.10^{-3})$ and $\mu_t\sim\mathcal{N}(\mu=-0.4,\sigma^2=10^{-2})$. If $100$ realizations are drawn from the joint distribution of $\eta$, $\mu_t$ and $a_r$, the production curve and the \textit{prior} credibility interval can be simultaneously plotted on the same graph to see how uncertain the predicted power is over a day. Figure \ref{fig:ParamError} illustrates on the left the distribution of $\eta$, $\mu_t$ and $a_r$ and on the right the production curve obtained for reference values and the \prior credibility interval at $90\%$. On the right side, experiments collected that same day are also displayed. Figure \ref{fig:ParamError} shows that the \textit{prior} credibility interval, built thanks to the experts, looks coherent with respect to the experimental data.\newline

If one is interested in the energy produced rather than the power (the energy in $kWh$ is the power in $kW$ multiplied by a duration), one can easily compute the maximum and the minimum energy for say $100$ realizations. The energy for collected power is $W_{exp}=3.44 kWh$, the maximum energy computed $W_{max}=3.65 kWh$ and the minimum energy $W_{min}=2.93 kWh$. $W_{min} < W_{exp} < W_{max}$ which means that the experts' interval seems correct for that day. With the considered uncertainty on $\eta$, $\mu_t$ and $a_r$, the error made is about $20\%$ over only one day. 
Considering this error over a day, the cumulative error over the lifetime of a plant could be too prejudicial. The aim of the calibration is to quantify this error and, at the same time, increase the knowledge on the parameter distribution. The calibration results for this application case are detailed in Section \ref{sec:application}.

\section{Calibration through statistical models \label{sec:calibration}}

Calibration aims to find the ``best fitting'' parameters of a computational code, in order to minimize the difference between the output and the experiments.  It can be used in two cases. In a forecasting context \citep{craig2001}, where the code calibrated on data collected on site can be used to compute the behavior of the power plant over the next time period. But also, in a prediction context, where data from an experimental stand are used to predict the behavior of a non-existing stand (assuming they have the same features). \newline

A simple way to express calibration is to write down a first simple model. The computational code is set up to entirely replace the physical system. Intuitively, we can assume that $\forall \boldsymbol{x} \in \mathcal{H}, \zeta(\boldsymbol{x})=f_c(\boldsymbol{x},\boldsymbol{\theta})$ for some well-chosen $\boldsymbol{\theta}$, which leads to the following equation\hc 

\begin{equation}
\mathcal{M}_1\ : \ \forall i \in \llbracket1,\dots,n\rrbracket \quad  y_{exp_i}=f_c(\boldsymbol{x}_i,\boldsymbol{\theta})+\epsilon_i,
\label{eq:model1}
\end{equation}

with $\forall i \in \llbracket1,\dots,n\rrbracket \quad \epsilon_i\overset{iid}{\sim}\mathcal{N}(0,\sigma_{err}^2)$.\newline

\new{Calibration consists then in estimating $\boldsymbol{\theta}$ in this statistical model. Moreover,
the variance of the measurement $\sigma_{err}^2$ error is also unknown and has to be estimated as well as the parameters but will be considered as a nuisance parameter.}
The likelihood of such a model depends on $f_c$. In methods such as Maximum Likelihood Estimation (MLE) or Bayesian estimation (which resorts to many MCMC iterations), it becomes intractable to work with a time consuming $f_c$. 
For the sake of simplicity we will consider the code as deterministic in what follows. It means that for the same inputs, the output of the code is identical, which is generally the case. Even in a deterministic context, a gap between the code and the physical system is often unavoidable. This gap is called code error or discrepancy. Some papers advocate adding this discrepancy to statistical models \citep{kennedy2001,higdon2004,bayarri2007,bachoc2014}. In the following, we present three other models which take into account a time consuming code and/or an additional discrepancy.
\newline

\subsection{Presentation of the models}

\subsubsection{A time consuming code}

\noindent Let us consider a time consuming code. As said above, in this particular case, the computational burden become too huge to perform calibration. That is why \citet{sacks1989} introduced an emulation of the, not yet computed, outputs from the code by a random function, \textit{i.e.} a stochastic process. The common choice is a Gaussian process because the conditional Gaussian process is still a Gaussian process (see Appendix \ref{ap:GaussianProcesses} for more details). It is, parsimoniously, defined by its mean and covariance functions. The first ``simple'' model was introduced by \citet{cox2001} which uses this emulation of $f_c$.

\begin{eqnarray}
\mathcal{M}_2 \ : \ \forall i \in \llbracket1,\dots,n\rrbracket \quad y_{exp_i} &=& F(\boldsymbol{x}_i,\boldsymbol{\theta}) + \epsilon_i,
\label{eq:M2}\\
F(\bullet,\bullet) & \sim &{\mathcal{GP}}{\Big(m_S(\bullet,\bullet),c_S\{(\bullet,\bullet),(\bullet,\bullet)\}\Big)},
\nonumber
\end{eqnarray}
where $\forall i \in \llbracket1,\dots,n\rrbracket \quad \epsilon_i\overset{iid}{\sim}\mathcal{N}(0,\sigma_{err}^2)$ and the random function $F(\boldsymbol{x}_i,\boldsymbol{\theta})$ stands for a Gaussian process (GP) over the joint domain of $\boldsymbol{x}_i$ and $\boldsymbol{\theta}$. \new{For the following, we consider that the measurement error is independent on the error made by the Gaussian process}. The mean function $m_{S}(\boldsymbol{x}_i,\boldsymbol{\theta})$ is generally a linear form of simple functions of $\boldsymbol{x}_i$ and $\boldsymbol{\theta}$. Its covariance function $c_S\{(\boldsymbol{x}_i^*,\boldsymbol{\theta}^*),(\boldsymbol{x}_i,\boldsymbol{\theta})\}=\sigma_S^2 r_{\boldsymbol{\psi}_S}\{(\boldsymbol{x}_i^*,\boldsymbol{\theta}^*),(\boldsymbol{x}_i,\boldsymbol{\theta})\}$ is such that the function $r_{\boldsymbol{\psi_S}}\{(\bullet,\bullet),(\bullet,\bullet)\}$ is the correlation function with a parameter vector $\boldsymbol{\psi}_S$. This parameter vector represents the scale and the regularity of the kernel and where $\sigma_S^2$ represents the variance. The mean $m_{S}(\boldsymbol{x}_i,\boldsymbol{\theta})$ can be written as\newline

\begin{equation}
m_S(\boldsymbol{x}_i,\boldsymbol{\theta})=m_{\boldsymbol{\beta}_S}(\boldsymbol{x}_i,\boldsymbol{\theta})=\mathbb{E}[F(\boldsymbol{x}_i,\boldsymbol{\theta})]=\beta_{S_0}+\sum_{j=1}^{M}\beta_{S_j}h_{S_j}(\boldsymbol{x}_i,\boldsymbol{\theta})=\boldsymbol{h}_S(\boldsymbol{x}_i,\boldsymbol{\theta})\boldsymbol{\beta}_S,
\label{eq:MeanLin}
\end{equation}

where $\boldsymbol{\beta}_S^T=(\beta_{S_0},\dots,\beta_{S_M})$ is the coefficient vector to be estimated and $\boldsymbol{h}_S(\bullet,\bullet)=(h_{S_0}(\bullet,\bullet),\dots\allowbreak,h_{S_M}(\bullet,\bullet))$ the row vector of regression functions where $h_{S_0}=1$. Similarly, we define the $n\times(M+1)$ matrix $\boldsymbol{H}_{S}(\boldsymbol{X},\boldsymbol{\theta})$ such that its $i^{th}$ row is $\boldsymbol{h}_{S}(\boldsymbol{x}_i,\boldsymbol{\theta})$. The correlation function can take multiple forms such as Gaussian or Matérn for instance \citep[see][for more examples]{santner2013}.
We will consider, for now and for all theoretical developments, the general form of $c_S\{(\bullet,\bullet),(\bullet,\bullet)\}=\sigma_S^2 r_{\boldsymbol{\psi}_S}\{(\bullet,\bullet),(\bullet,\bullet)\}$ where $\sigma_S^2$ is the variance and $r$ is the correlation function with a parameter vector $\boldsymbol{\psi}_S$. The advantage of using an emulator for $f_c(\boldsymbol{X},\boldsymbol{\theta})$ is to alleviate the computational burden, at the cost of adding an additional source of uncertainty, and of increasing the number of uncertain parameters. Specific hypotheses, for instance a known smoothness of the random field, may help to choose the size of the parametric family in which the correlation
shape is to be assessed. \newline

When the code is time consuming, a fixed number $N$ of simulations is set up. The ensuing simulated data (we will call them $\boldsymbol{y}_c$) are usually the image of a design of experiments (DOE) representative of the input space. Some interesting developments have been made on using the fewest possible points in the input space with some judicious distributions (the \textit{Latin Hilbert Space} sampling is one example; see \citet{pronzato2012} for helpful insights). \newline

Let us call $\boldsymbol{D}$ a DOE, a set of $N$ points sampled in the input space defined as the product of $\mathcal{H}$ and $\mathcal{Q}$. We can write $\boldsymbol{D}=\{ (\boldsymbol{x}_1^D,\boldsymbol{\tau}_1^D),\dots (\boldsymbol{x}_N^D,\boldsymbol{\tau}_N^D) \}$ where $\forall i \in \llbracket1,\dots,N\rrbracket \  (\boldsymbol{x}_i^D,\boldsymbol{\tau}_i^D)$ are chosen in $\mathcal{H}\times \mathcal{Q}$.
The establishment of the DOE will lead to simulated data which are defined as $\boldsymbol{y}_c=f_c(\boldsymbol{D})$. The error made by the emulator strongly depends on the numerical design of experiments used to fit the emulator. Adaptive numerical designs introduced in \citet{damblin2018} are a way to enhance the emulator when the goal is to calibrate the code. With this method, based on a Gaussian process-based optimization called Efficient Global Optimization \citep{jones1998efficient}, other points, judiciously chosen with respect to further calibration,  can be added to the original DOE. \newline

\subsubsection{With a code error}

\noindent Considering the computational code as a perfect representation of the physical system may be too strong a hypothesis and it is legitimate to wonder whether the code might differ from the phenomenon. This error (called discrepancy and introduced above) is defined as\hc
\begin{equation*}
\delta(\boldsymbol{x}_i)=\zeta(\boldsymbol{x}_i)-f_c(\boldsymbol{x}_i,\boldsymbol{\theta}).
\end{equation*}
In all the papers cited above, this unknown discrepancy is modeled as an occurrence of a Gaussian process that yields a random function over the domain  $\mathcal{H}$ of input variables only. For the sake of simplicity, we will denote by $m_s$, $c_S$ ($c_{\sigma_S^2,\boldsymbol{\psi}_S}$) and $r_S$ ($r_{\boldsymbol{\psi}_S}$) the mean, covariance (with $\sigma_S^2$ as the variance) and correlation function relative to the emulator and by $m_{\delta}$, $c_{\delta}$ ($c_{\sigma_{\delta}^2,\boldsymbol{\psi}_{\delta}}$) and $r_{\delta}$ ($r_{\boldsymbol{\psi}_{\delta}}$) the same functions relative to the discrepancy (respectively $\sigma_{\delta}^2$ for the variance in the covariance function). 
Note that $m_{\delta}$ and $c_{\delta}$ are functions of $\boldsymbol{x}$ only and not $\boldsymbol{\theta}$. The aim of adding the discrepancy lies in the fact that correlation is sometimes visible in the residuals and/or that no value of $\boldsymbol{\theta}$ brings the computer close to experiments. However, the discrepancy could lead to an identifiability issue. For example, two  different couples $(\boldsymbol{\theta},\delta(\boldsymbol{x}_i))$ and $(\boldsymbol{\theta}^*,\delta^*(\boldsymbol{x}_i))$ may verify these two equalities\hc $\delta(\boldsymbol{x}_i)=\zeta(\boldsymbol{x}_i)-f_c(\boldsymbol{x}_i,\boldsymbol{\theta})$ and $\delta^*(\boldsymbol{x}_i)=\zeta(\boldsymbol{x}_i)-f_c(\boldsymbol{x}_i,\boldsymbol{\theta}^*)$. Some papers \citep{higdon2004,bachoc2014,bayarri2007} advocate setting the mean of the discrepancy to $0$ to solve this identifiability issue. The contribution of the discrepancy is widely discussed in the literature.\newline 

When the code is not time consuming, the (real) code $f_c$ is used\hc

\begin{eqnarray}
\mathcal{M}_3 \ : \ \forall i \in \llbracket1,\dots,n\rrbracket \quad y_{exp_i}&=& f_c(\boldsymbol{x}_i,\boldsymbol{\theta})+\delta(\boldsymbol{x}_i)+\epsilon_i,
\label{eq:M3}\\
\delta(\bullet) & \sim &{\mathcal{GP}}{\Big(\boldsymbol{m}_{\delta}(\bullet),c_{\delta}(\bullet,\bullet)\Big)},
\nonumber
\end{eqnarray}

where $\forall i \in \llbracket1,\dots,n\rrbracket \quad \epsilon_i\overset{iid}{\sim}\mathcal{N}(0,\sigma_{err}^2)$, and $\delta(\bullet)$ stands for a Gaussian process which mimics the discrepancy and  only depend on the input variables $\boldsymbol{x}$. \new{For the rest of the article, we make the assumption that the discrepancy is independent on the measurement error.} Therefore, we write $\delta(\bullet)\sim\mathcal{GP}(\boldsymbol{m}_{\delta}(\bullet),c_{\delta}(\bullet,\bullet))$ with $\forall \boldsymbol{x}, \ \boldsymbol{m}_{\delta}(\boldsymbol{x})={\boldsymbol{h}_{\delta}}(\boldsymbol{x})\boldsymbol{\beta}_{\delta}$ (where $\boldsymbol{h}_{\delta}$ is a row vector and $\boldsymbol{\beta}_{\delta}$ is a column vector if we choose a parametric representation of the mean) and $c_{\delta}$ the covariance function of the discrepancy. We also denote $\boldsymbol{H}_{\delta}(\boldsymbol{X})$ the $n$ row matrix, the $i^{th}$ row of which is $\boldsymbol{h}_{\delta}(\boldsymbol{x}_i)$.\newline

When the code is time consuming, the systematic use of $f_c$ is not computationally acceptable.
Then, as for Model $\mathcal{M}_2$, the code is replaced by a Gaussian process emulator. This leads to the more generic model introduced in \citet{kennedy2001}.

\begin{equation}
\mathcal{M}_4 \ : \ \forall i \in \llbracket1,\dots,n\rrbracket \quad y_{exp_i}= F(\boldsymbol{x}_i,\boldsymbol{\theta})+\delta(\boldsymbol{x}_i)+\epsilon_i,
\label{eq:M3p}
\end{equation}
where $\forall i \in \llbracket1,\dots,n\rrbracket \quad \epsilon_i\overset{iid}{\sim}\mathcal{N}(0,\sigma_{err}^2)$, $F(\boldsymbol{x}_i,\boldsymbol{\theta})$ and $\delta(\boldsymbol{x}_i)$ are the two Gaussian
processes defined as before. \new{As before, we consider, for the following, that the measurement error, the discrepancy and the error induced by the emulator are all independent.}
In their model, \citet{kennedy2001} also used a multiplicative scale parameter $\rho$ for $F$.
This parameter is usually set to $1$ in many papers in order to achieve the best estimate on $\boldsymbol{\theta}$.
Thus, we omit this scaling parameter in the model definition.
\newline

A quantification of the bias form is the aim of both models. If we are interested in improving the computational code or its emulator, it is usually fair to set the mean of the discrepancy to zero and find the best tuning parameter vector which compensates a potential bias \citep{higdon2004,bachoc2014}. \newline

\begin{figure}[htbp!]
  \centering
	
    \begin{tikzpicture}
		\tikzstyle{m1}=[fill=gray!50!black, text=white,font=\scriptsize, circle,draw]
		\tikzstyle{m2}=[fill=red!50!black,text=white,font=\scriptsize, circle, draw]
		\tikzstyle{m3}=[fill=blue!50!black, text=white, font=\scriptsize, circle, draw]
		\tikzstyle{m4}=[fill=green!50!black, text= white, font=\scriptsize, circle, draw]
		
		\node[m1] (N1) at (0,0) {$\boldsymbol{y}_{exp}$};
		\node[m1] (N2) at (-2.5,1) {$\boldsymbol{y}_c(\boldsymbol{X},\boldsymbol{\theta})$};
		\node[m1] (N3) at (0,4) {$\sigma_{err}^2$};
		\node[m1] (N4) at (-2,2.5) {$\boldsymbol{\theta}$};
		\node[m2] (N5) at (3,4) {$\boldsymbol{\beta}_\delta$, $\sigma_\delta^2$, $\boldsymbol{\psi}_\delta$};

		\node[m4] (N7) at (-3,4) {$\boldsymbol{\beta}_S$, $\sigma_S$, $\boldsymbol{\psi}_S$};
		\node[m2] (N8) at (2.5,1) {$\delta(x)$};

		\draw [->] (N2) to (N1);
		\draw [->] (N3) to (N1);
		\draw [->] (N4) to (N2);
		\draw [->] (N7) to (N2);
		\draw [->] (N5) to (N8);
		\draw [->] (N8) to (N1);		
		
		\draw [dashed] (-5,3) to (5,3);
		\draw [dashed] (-5,2) to (5,2);

    \end{tikzpicture}
    
  \caption{Directed Acyclic Graph (DAG) representation of the different models}
  \label{fig:DAG}
\end{figure}
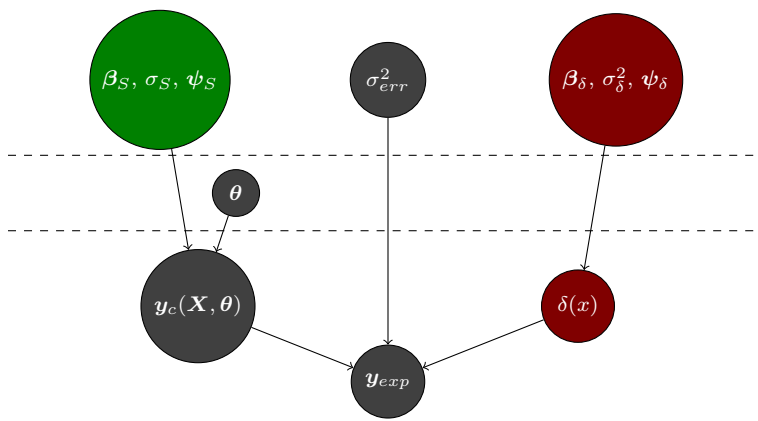

The directed acyclic graph (DAG) shown in Figure \ref{fig:DAG} summarizes and compares the structures of all the previously introduced models.
Specifically\hc if one considers only the grey nodes, the obtained DAG corresponds to Model $\mathcal{M}_1$. Adding the green node, the resulting DAG represents $\mathcal{M}_2$. Considering the grey and red nodes yields a DAG for model $\mathcal{M}_3$ and the whole DAG represents the general model $\mathcal{M}_4$. Note that two categories of parameters are considered. The tuning parameters are only related to the code and other parameters (also called nuisance parameters) concern the measurement error, the emulator or the discrepancy introduced in the models. In calibration, we only focus on the value of $\boldsymbol{\theta}$ but the other parameters introduced need to be estimated as well. We will examine these estimation issues at greater length into the next sections. 
\newline

All these models introduce new parameters and need to be estimated as well as tuning parameters. Estimation needs to delve into technical aspects such as writing the likelihood for each model. The following section provides all the elements required to go one step further and carry out the estimation.

\subsection{Likelihood}

To estimate parameters (whatever the framework used, Bayesian or Maximum Likelihood Estimation (MLE)), expressing the likelihood comes as the first requirement. Two major categories stand out. When the code is not time consuming, the main issue in code calibration (\textit{i.e.} the computational time burden) is avoided. When the code is time consuming, new parameters have to be taken into account and to be estimated. In the models $\mathcal{M}_2$ and $\mathcal{M}_4$, both numerical data ($\boldsymbol{y}_c$) and field data  ($\boldsymbol{y}_{exp}$) are available and can be collected in the whole data vector $\boldsymbol{y}^T=(\boldsymbol{y}_{exp}^T,\boldsymbol{y}_c^T)$. In the models $\mathcal{M}_1$ and $\mathcal{M}_3$, data can only represent field data ($\boldsymbol{y}_{exp}$).
In what follows, we will denote by $\boldsymbol{\theta}^*$ the true parameter vector.
Note that it is well-defined only in Models~$\mathcal{M}_1$ and $\mathcal{M}_2$, as the value of $\boldsymbol{\theta}$ which satisfies\hc $\zeta(\boldsymbol{x})=f_c(\boldsymbol{x},\boldsymbol{\theta}^*)$ for all possible $\boldsymbol{x}$, is assumed to exist and to be unique. On the other hand, the models $\mathcal{M}_3$ and $\mathcal{M}_4$ are both defined by the relation $\zeta(\boldsymbol{x})=f_c(\boldsymbol{x},\boldsymbol{\theta})+\delta(\boldsymbol{x})$, which holds for infinitely many couples $(\boldsymbol{\theta},\delta(\bullet))$, as discussed earlier. \citet{kennedy2001} avoid this issue by defining $\boldsymbol{\theta}^*$ as a ``best-fitting'' value, but it is unclear what this means exactly (see the discussion section of their paper for further details).\newline

In order to simplify the notation, for the rest of the paper we will use $\Phi=\{\sigma_S^2,\sigma_{\delta}^2,\boldsymbol{\psi}_S,\boldsymbol{\psi}_{\delta}\}$ and $\Phi_S=\{\sigma_S^2,\boldsymbol{\psi}_S\}$ and $\Phi_{\delta}=\{\sigma_{\delta}^2,\boldsymbol{\psi}_{\delta}\}$, where $\sigma_S^2$ and $\sigma_{\delta}^2$ are the variances of the two Gaussian processes respectively relative to the emulator and the discrepancy. The two parameter vectors $\boldsymbol{\psi}_S$ and $\boldsymbol{\psi}_{\delta}$ are relative to the correlation functions. Let us call $\boldsymbol{\beta}^T=(\boldsymbol{\beta}_S^T,\boldsymbol{\beta}_{\delta}^T)$ the vector of collected coefficient vectors.\newline

The likelihood equations will be written for the generic forms of $\mathcal{M}_3$ and $\mathcal{M}_4$. The likelihoods for the simpler models
$\mathcal{M}_1$ and $\mathcal{M}_2$ 
will then be derived since  
$\mathcal{M}_1\subset\mathcal{M}_3$ and $\mathcal{M}_2\subset\mathcal{M}_4$.\newline

\subsubsection{A fast code}

\noindent The generic model which deals with calibration with a code that is quick to run is given in \EqRef{eq:M3}. Only experimental data are used to compute the likelihood.
Experimental data follow a Gaussian distribution, the expectation of which
is\hc \newline

\begin{equation*}
\mathbb{E}[\boldsymbol{y}_{exp}|\boldsymbol{\theta},\boldsymbol{\beta}_{\delta};\boldsymbol{X}] = \boldsymbol{m}_{exp}^{\boldsymbol{\beta}_{\delta}}(\boldsymbol{X},\boldsymbol{\theta}) = \boldsymbol{m}_{exp}(\boldsymbol{X},\boldsymbol{\theta}) = f_c(\boldsymbol{X},\boldsymbol{\theta}) + {\boldsymbol{H}_{\delta}}(\boldsymbol{X})\boldsymbol{\beta}_{\delta}.
\end{equation*}

Then, the expression of the variance is given by\hc

\begin{equation*}
\mathbb{V}ar[\boldsymbol{y}_{exp}|\Phi_{\delta};\boldsymbol{X}] = \boldsymbol{V}_{exp}^{\Phi_{\delta},\sigma_{err}^2}(\boldsymbol{X})= \boldsymbol{V}_{exp}(\boldsymbol{X}) = \boldsymbol{\Sigma}_{\delta}(\boldsymbol{X}) + \sigma_{err}^2\boldsymbol{I}_n,
\end{equation*}

with $\forall (i,j) \in \llbracket1,\dots,n\rrbracket^2: (\boldsymbol{\Sigma}_{\delta}(\boldsymbol{X}))_{i,j} =(\boldsymbol{\Sigma}^{\Phi_{\delta}}_{\delta}(\boldsymbol{X}))_{i,j}= c_{\delta}(\{\boldsymbol{x}_i,\boldsymbol{x}_j\})$. 
The likelihood in this particular case can be written as

\begin{equation}
\begin{split}
\mathcal{L}^{F}(\boldsymbol{\theta},\boldsymbol{\beta}_{\delta},\Phi_{\delta};\boldsymbol{y}_{exp},\boldsymbol{X})=\frac{1}{(2\pi)^{n/2}|\boldsymbol{V}_{exp}(\boldsymbol{X})|^{1/2}}\exp\Bigg\{-\frac{1}{2}\Big(\boldsymbol{y}_{exp}-\boldsymbol{m}_{exp}(\boldsymbol{X},\boldsymbol{\theta})\Big)^T\boldsymbol{V}_{exp}(\boldsymbol{X})^{-1}\\
\Big(\boldsymbol{y}_{exp}-\boldsymbol{m}_{exp}(\boldsymbol{X},\boldsymbol{\theta})\Big)\Bigg\}.
\end{split}
\label{eq:LikelihoodQuick2}
\end{equation}

This likelihood is relative to Model $\mathcal{M}_3$ (\EqRef{eq:M3}). For the specific case, where no discrepancy is considered (corresponding to $\mathcal{M}_1$ \EqRef{eq:model1}) the likelihood can be written in a similar way but with $\boldsymbol{m}_{exp}(\boldsymbol{X},\boldsymbol{\theta})= f_c(\boldsymbol{X},\boldsymbol{\theta})$ and $\boldsymbol{V}_{exp}(\boldsymbol{X})=\sigma_{err}^2\boldsymbol{I}_n$. Note that the covariance matrix depends only on $\sigma_{err}^2$. It implies that if we seek to estimate the \posterior density on $\boldsymbol{\theta}$ (in a Bayesian framework), this covariance term is superfluous.\newline

Then the likelihood can be rewritten more simply\hc

\begin{equation}
\mathcal{L}^F(\boldsymbol{\theta},\sigma_{err}^2;\boldsymbol{y}_{exp},\boldsymbol{X})=\frac{1}{(2\pi)^{n/2}\sigma_{err}^{n}}\exp\Bigg\{-\frac{1}{2\sigma_{err}^2}||\boldsymbol{y}_{exp}-f_c(\boldsymbol{X},\boldsymbol{\theta})||_2^2\Bigg\}.
\label{eq:LikelihoodQuick}
\end{equation}

The models using the code with or without the discrepancy do look quite similar. For theoretical development, it might be easier to work with the one without discrepancy. From an experimental point of view, it could be interesting to study the role of the code error. \newline

\subsubsection{A time consuming code}

\noindent When a code is time consuming and is replaced by an emulator, additional parameters have to be estimated. As introduced above, a DOE is set up and aims to be a representative sample of the input space (variable and parameter input space). Simulated data from this DOE (called $\boldsymbol{y}_c$) will constitute additional data for the estimation of the nuisance parameters. Depending on how we consider that two sources of data are linked, 
multiple likelihoods can be set up. For the theoretical development, we will consider the general model $\mathcal{M}_4$ and we will detail the particular case $\mathcal{M}_2$ hereafter.\newline


The first likelihood useful in estimation is the full likelihood. This concerns the distribution of all collected data ($\boldsymbol{y}^T=(\boldsymbol{y}_{exp}^T,\boldsymbol{y}_c^T)$). That means that we are interested in estimating the parameters of the distribution $\pi(\boldsymbol{y}|\boldsymbol{\theta},\boldsymbol{\beta},\Phi,\sigma_{err}^2;\boldsymbol{X},\boldsymbol{D})$ which is Gaussian.
The expectations can be written from both expectancies of $\pi(\boldsymbol{y}_{exp}|\boldsymbol{\theta},\boldsymbol{\beta},\Phi,\sigma_{err}^2;\boldsymbol{X})$ and $\pi(\boldsymbol{y}_c|\boldsymbol{\theta},\boldsymbol{\beta}_S,\Phi_S;\boldsymbol{D})$.

\begin{equation}
\begin{cases}
    \mathbb{E}[\boldsymbol{y}_c|\boldsymbol{\beta}_S;\boldsymbol{D}]= \boldsymbol{m}_S^{\boldsymbol{\beta}_S}(\boldsymbol{D})= \boldsymbol{m}_S(\boldsymbol{D}) = {\boldsymbol{H}_S}(\boldsymbol{D})\boldsymbol{\beta}_S\\
\mathbb{E}[\boldsymbol{y}_{exp}|\boldsymbol{\theta},\boldsymbol{\beta};\boldsymbol{X}]= \boldsymbol{m}_{exp}^{\boldsymbol{\beta}}(\boldsymbol{X},\boldsymbol{\theta}) = \boldsymbol{m}_{exp}(\boldsymbol{X},\boldsymbol{\theta}) = {\boldsymbol{H}_S}(\boldsymbol{X},\boldsymbol{\theta}) \boldsymbol{\beta}_S + {\boldsymbol{H}_{\delta}}(\boldsymbol{X})\boldsymbol{\beta}_{\delta}
\end{cases}
\label{eq:ExpPartialLikelihood}
\end{equation}

This can be summed up for two component vectors $\boldsymbol{y}^T=(\boldsymbol{y}_{exp}^T,\boldsymbol{y}_c^T)$\hc
 
\begin{equation}
\begin{split}
\mathbb{E}[\boldsymbol{y}|\boldsymbol{\theta},\boldsymbol{\beta};\boldsymbol{X},\boldsymbol{D}]=\boldsymbol{m}_{\boldsymbol{y}}^{\boldsymbol{\beta}}((\boldsymbol{X},\boldsymbol{\theta}),\boldsymbol{D}) = \boldsymbol{m}_{\boldsymbol{y}}((\boldsymbol{X},\boldsymbol{\theta}),\boldsymbol{D})& =\boldsymbol{H}((\boldsymbol{X},\boldsymbol{\theta}),\boldsymbol{D})\boldsymbol{\beta}\\&=\begin{pmatrix}
\boldsymbol{H}_S(\boldsymbol{X},\boldsymbol{\theta}) & \boldsymbol{H}_{\delta}(\boldsymbol{X})\\
\boldsymbol{H}_S(\boldsymbol{D}) & 0
\end{pmatrix}\boldsymbol{\beta}.
\end{split}
\label{eq:MeanFullLikelihood}
\end{equation}
The variance matrix now includes the covariance functions of the discrepancy and the emulator.

\begin{equation}
\begin{split}
\mathbb{V}ar[\boldsymbol{y}|\boldsymbol{\theta},\Phi,\sigma_{err}^2;\boldsymbol{X},\boldsymbol{D}]&=\boldsymbol{V}^{\Phi,\sigma_{err}^2}((\boldsymbol{X},\boldsymbol{\theta}),\boldsymbol{D})=\boldsymbol{V}((\boldsymbol{X},\boldsymbol{\theta}),\boldsymbol{D})\\&=\begin{pmatrix}
 \boldsymbol{\Sigma}_{exp,exp}(\boldsymbol{X},\boldsymbol{\theta}) +\boldsymbol{\Sigma}_{\delta}(\boldsymbol{X}) +\sigma_{err}^2\boldsymbol{I}_n &  \boldsymbol{\Sigma}_{exp,c}((\boldsymbol{X},\boldsymbol{\theta}),\boldsymbol{D})\\
 \boldsymbol{\Sigma}_{exp,c}((\boldsymbol{X},\boldsymbol{\theta}),\boldsymbol{D})^T & \boldsymbol{\Sigma}_{c,c}(\boldsymbol{D})
\end{pmatrix}
\end{split}
\label{eq:VarianceFullLikelihood}
\end{equation}

\begin{sloppypar}
where \begin{itemize}
\item $\forall (i,j) \in \llbracket1,\dots,n\rrbracket^2: (\boldsymbol{\Sigma}_{exp,exp}(\boldsymbol{X},\boldsymbol{\theta}))_{i,j}=c_S\{(\boldsymbol{x}_i,\boldsymbol{\theta}),(\boldsymbol{x}_j,\boldsymbol{\theta})\}$,
\item $\forall (i,j) \in \llbracket1,\dots,n\rrbracket\times\llbracket1,\dots,N\rrbracket: (\boldsymbol{\Sigma}_{exp,c}((\boldsymbol{X},\boldsymbol{\theta}),\boldsymbol{D}))_{i,j}=c_S\{(\boldsymbol{x}_i,\boldsymbol{\theta}_i),(\boldsymbol{x}_j^D,\boldsymbol{\tau}_j^D)\}$,
\item $\forall (i,j)\in\llbracket1,\dots,n\rrbracket^2: (\boldsymbol{\Sigma}_{\delta}(\boldsymbol{X}))_{i,j}=c_{\delta}\{(\boldsymbol{x}_i,\boldsymbol{x}_j)\}$,
\item $\forall (i,j)\in \llbracket1,\dots,N\rrbracket^2: (\boldsymbol{\Sigma}_{c,c}(\boldsymbol{D}))_{i,j}=c_S\{(\boldsymbol{x}_i^D,\boldsymbol{\tau}_i^D),(\boldsymbol{x}_j^D,\boldsymbol{\tau}_j^D)\}$. \newline
\end{itemize}
\end{sloppypar}

As a reminder $\boldsymbol{D}$ is the DOE set up to build the emulator and is defined as $\boldsymbol{D}=\{ (\boldsymbol{x}_1^D,\boldsymbol{\tau}_1^D),\dots (\boldsymbol{x}_N^D,\boldsymbol{\tau}_N^D) \}$. The general expression of the full likelihood can then be expressed\hc

\begin{equation}
\begin{split}
\mathcal{L}^F(\boldsymbol{\theta},\boldsymbol{\beta},\Phi,\sigma_{err}^2;\boldsymbol{y},\boldsymbol{X},\boldsymbol{D})=\frac{1}{(2\pi)^{(n+N)/2}|\boldsymbol{V}((\boldsymbol{X},\boldsymbol{\theta}),\boldsymbol{D})|^{1/2}}&\exp\Bigg\{-\frac{1}{2}\Big(\boldsymbol{y}-\boldsymbol{m}_{\boldsymbol{y}}((\boldsymbol{X},\boldsymbol{\theta}),\boldsymbol{D})\Big)^T\\\boldsymbol{V}((\boldsymbol{X},\boldsymbol{\theta}),\boldsymbol{D})^{-1}
&\Big(\boldsymbol{y}-\boldsymbol{m}_{\boldsymbol{y}}((\boldsymbol{X},\boldsymbol{\theta}),\boldsymbol{D})\Big)\Bigg\}.
\end{split}
\label{eq:FullLikelihood}
\end{equation}

\citet{bayarri2007,higdon2004} advocate, in this particular case, considering a zero Gaussian process mean for the discrepancy. Under this condition, we have $\boldsymbol{m}_{y}((\boldsymbol{X},\boldsymbol{\theta}),\boldsymbol{D})=\begin{pmatrix}
\boldsymbol{H}_S(\boldsymbol{X},\boldsymbol{\theta})\\
\boldsymbol{H}_S(\boldsymbol{D})
\end{pmatrix}\boldsymbol{\beta}_S $ and the other terms remain the same. For the model $\mathcal{M}_2$ where an emulator is used without any discrepancy \citep{cox2001}, the expectation becomes\hc

\begin{equation}
\mathbb{E}[\boldsymbol{y}|\boldsymbol{\theta},\boldsymbol{\beta}_S;\boldsymbol{X},\boldsymbol{D}]=\boldsymbol{m}_{\boldsymbol{y}}((\boldsymbol{X},\boldsymbol{\theta}),\boldsymbol{D})=\boldsymbol{H}((\boldsymbol{X},\boldsymbol{\theta}),\boldsymbol{D})\boldsymbol{\beta}_S=\begin{pmatrix}
\boldsymbol{H}_S(\boldsymbol{X},\boldsymbol{\theta})\\
\boldsymbol{H}_S(\boldsymbol{D})
\end{pmatrix}\boldsymbol{\beta}_S
\label{eq:MeanFullLikelihood2}
\end{equation}

and the covariance\hc

\begin{equation}
\mathbb{V}ar[\boldsymbol{y}|\boldsymbol{\theta},\Phi,\sigma_{err}^2;\boldsymbol{X},\boldsymbol{D}]=\boldsymbol{V}((\boldsymbol{X},\boldsymbol{\theta}),\boldsymbol{D})=\begin{pmatrix}
\boldsymbol{\Sigma}_{exp,exp}(\boldsymbol{X},\boldsymbol{\theta}) +\sigma_{err}^2\boldsymbol{I}_n & \boldsymbol{\Sigma}_{exp,c}((\boldsymbol{X},\boldsymbol{\theta}),\boldsymbol{D})\\
\boldsymbol{\Sigma}_{exp,c}((\boldsymbol{X},\boldsymbol{\theta}),\boldsymbol{D})^T & \boldsymbol{\Sigma}_{c,c}(\boldsymbol{D})
\end{pmatrix}
\label{eq:VarianceFullLikelihood1}
\end{equation}

where the covariance matrices are the same as defined above.\newline


The estimation can be separated into different steps where the partial likelihood (Equation (\ref{eq:PartialLikelihood})) could be useful.
This only concerns simulated data and the corresponding emulator.
The partial likelihoods of the models $\mathcal{M}_2$ and $\mathcal{M}_4$ are then the same.
This means that we only need to estimate the distribution $\pi(\boldsymbol{\beta}_S,\Phi_S|\boldsymbol{y}_c)$ where $\Phi_S=\{\sigma_S^2,\boldsymbol{\psi}_S\}$. The expectation can be obtained by considering only the mean function of the emulator
(\EqRef{eq:ExpPartialLikelihood}) and the variance is straightforwardly linked to the variance of the emulator.

\begin{equation*}
\mathbb{V}ar[\boldsymbol{y}_c|\Phi_S;\boldsymbol{D}] = \boldsymbol{V}_c^{\Phi_S}(\boldsymbol{D})=\boldsymbol{V}_c(\boldsymbol{D})= \boldsymbol{\Sigma}_{c,c}(\boldsymbol{D}),
\end{equation*}

where $\forall (i,j)\in [1,\dots,N]^2: (\boldsymbol{\Sigma}_{c,c}(\boldsymbol{D}))_{i,j}=c_S\{(\boldsymbol{x}_i^D,\boldsymbol{\theta}_i^D),(\boldsymbol{x}_j^D,\boldsymbol{\theta}_j^D)\}$. Let us recall that Equation~(\ref{eq:ExpPartialLikelihood}) established that $\boldsymbol{m}_c(\boldsymbol{D}) ={\boldsymbol{H}_S}(\boldsymbol{D})\boldsymbol{\beta}_S$. It implies that the partial likelihood relative to $\mathcal{M}_4$ and $\mathcal{M}_2$ is\hc

\begin{equation}
\mathcal{L}^M(\boldsymbol{\beta}_S,\Phi_S;\boldsymbol{y}_c,\boldsymbol{D})=\frac{1}{(2\pi)^{N/2}|\boldsymbol{V}_c(\boldsymbol{D})|^{1/2}}\exp\Bigg\{-\frac{1}{2}\Big(\boldsymbol{y}_c-\boldsymbol{m}_c(\boldsymbol{D})\Big)^T\boldsymbol{V}_c(\boldsymbol{D})^{-1}\Big(\boldsymbol{y}_c-\boldsymbol{m}_c(\boldsymbol{D})\Big)\Bigg\}.
\label{eq:PartialLikelihood}
\end{equation}

From what has been introduced before, one can write the conditional distribution $\pi(\boldsymbol{y}_{exp}|\boldsymbol{y}_c)$ (see Appendix \ref{ap:GaussianProcesses} for more details) from the joint distribution $\pi(\boldsymbol{y}_{exp},\boldsymbol{y}_c)$\hc

\begin{equation*}
\begin{pmatrix}
\boldsymbol{y}_{exp}\\
\boldsymbol{y}_c
\end{pmatrix} \sim \mathcal{N}\Bigg(\begin{pmatrix}
\boldsymbol{m}_{exp}(\boldsymbol{X},\boldsymbol{\theta})\\
\boldsymbol{m}_c(\boldsymbol{D})
\end{pmatrix}, \begin{pmatrix}
\boldsymbol{\Sigma}_{exp,exp}(\boldsymbol{X},\boldsymbol{\theta}) & \boldsymbol{\Sigma}_{exp,c}((\boldsymbol{X},\boldsymbol{\theta}),\boldsymbol{D})\\
\boldsymbol{\Sigma}_{exp,c}((\boldsymbol{X},\boldsymbol{\theta}),\boldsymbol{D})^T & \boldsymbol{\Sigma}_{c,c}(\boldsymbol{D})
\end{pmatrix}\Bigg)\\
\end{equation*}
where $\boldsymbol{m}_c$ and $\boldsymbol{m}_{exp}$ are defined in Equation (\ref{eq:ExpPartialLikelihood}) and the covariance matrices defined above before equation (\ref{eq:FullLikelihood}). Then,
\begin{equation*}
\boldsymbol{y}_{exp}|\boldsymbol{y}_c\sim\mathcal{N}(\boldsymbol{\mu}_{exp|c}((\boldsymbol{X},\boldsymbol{\theta}),\boldsymbol{D}),\boldsymbol{\Sigma}_{exp|c}((\boldsymbol{X},\boldsymbol{\theta}),\boldsymbol{D}))
\end{equation*}
with\hc

\begin{equation}
\boldsymbol{\mu}_{exp|c}((\boldsymbol{X},\boldsymbol{\theta}),\boldsymbol{D})=\boldsymbol{m}_{exp}(\boldsymbol{X},\boldsymbol{\theta})+\boldsymbol{\Sigma}_{exp,c}((\boldsymbol{X},\boldsymbol{\theta}),\boldsymbol{D})\boldsymbol{\Sigma}_{c,c}(\boldsymbol{D})^{-1}(\boldsymbol{y}_c-\boldsymbol{m}_c(\boldsymbol{D})),
\label{eq:conditionnalMean}
\end{equation}
\begin{equation}
\boldsymbol{\Sigma}_{exp|c}((\boldsymbol{X},\boldsymbol{\theta}),\boldsymbol{D})=\boldsymbol{\Sigma}_{exp,exp}(\boldsymbol{X},\boldsymbol{\theta})-\boldsymbol{\Sigma}_{exp,c}((\boldsymbol{X},\boldsymbol{\theta}),\boldsymbol{D})\boldsymbol{\Sigma}_{c,c}(\boldsymbol{D})^{-1}\boldsymbol{\Sigma}_{exp,c}((\boldsymbol{X},\boldsymbol{\theta}),\boldsymbol{D})^T.
\label{eq:conditionnalVariance}
\end{equation}

The conditional likelihood can then be written as\hc

\begin{equation}
\begin{split}
\mathcal{L}^C(\boldsymbol{\theta},\boldsymbol{\beta}_{\delta},\Phi_{\delta};\boldsymbol{\beta}_S,\Phi_S,\boldsymbol{y}_{exp}|\boldsymbol{y}_c,\boldsymbol{X},\boldsymbol{D}) \propto &|\Sigma_{exp|c}((\boldsymbol{X},\boldsymbol{\theta}),\boldsymbol{D})|^{-1/2}\\ &\exp\Big\{-\frac{1}{2}(\boldsymbol{y}_{exp}-\mu_{exp|c}((\boldsymbol{X},\boldsymbol{\theta}),\boldsymbol{D}))^T \Sigma_{exp|c}((\boldsymbol{X},\boldsymbol{\theta}),\boldsymbol{D})^{-1} \\ &(\boldsymbol{y}_{exp}-\mu_{exp|c}((\boldsymbol{X},\boldsymbol{\theta}),\boldsymbol{D}))\Big\}.
\label{eq:conditionalLikelihood}
\end{split}
\end{equation}


Usually in a Bayesian framework, $\boldsymbol{\beta}$ is distributed according to a Jeffreys \textit{prior}. In this case, $\pi(\boldsymbol{\beta})=\pi(\boldsymbol{\beta}_S,\boldsymbol{\beta}_{\delta})\propto 1$ and we can integrate out $\boldsymbol{\beta}$ from the full likelihood expressed by \EqRef{eq:FullLikelihood}.
\subsection{Estimation \label{sec:estimation}}

\subsubsection{Maximum likelihood estimator}

In this section, we comment on remarkable insights developed in \citet{cox2001}. To estimate the parameters $\boldsymbol{\theta}$, $\boldsymbol{\beta}$ and $\Phi$,
a first approach (for $\mathcal{M}_1$ and $\mathcal{M}_2$) would be to maximize the full likelihood introduced in the previous section. This method is called Full Maximum Likelihood Estimator.
The major drawback of this method is dealing with a high number of parameters and in certain cases this leads to a very heavy computational operation. \newline

A second method to overcome this issue, introduced in \citet{cox2001} for $\mathcal{M}_2$ only, is called the Separated Maximum Likelihood Estimation (SMLE). 
The estimation is made in two steps. The first step is to maximize the partial likelihood (Equation (\ref{eq:PartialLikelihood})) to get estimators of the parameters of
 the Gaussian Process. Then these estimators ($\hat{\Phi}$ and $\hat{\boldsymbol{\beta}}$) are plugged  into $\boldsymbol{\mu}_{exp|c}((\boldsymbol{X},\boldsymbol{\theta}),\boldsymbol{D})$ and $\boldsymbol{\Sigma}_{exp|c}((\boldsymbol{X},\boldsymbol{\theta}),\boldsymbol{D})$ which are the mean and the variance of the conditional distribution. A likelihood is set up from those quantities and maximized to get $\hat{\boldsymbol{\theta}}$. The SMLE method can also be seen as an approximation of the  generalized non linear least squares technique. \newline

These methods are applied in \citet{cox2001} for $\mathcal{M}_2$. For models $\mathcal{M}_3$ and $\mathcal{M}_4$, \citep{wong2017} developed a new approach which deals with the identifiability problem when the discrepancy is added in this framework. Then, the estimation part is conducted in two steps. The first step consists in estimating $\hat{\boldsymbol{\theta}}$ in 

\begin{equation}
\hat{\boldsymbol{\theta}}= \underset{\boldsymbol{\theta}\in\mathcal{Q}}{argmin} \ M_n(\boldsymbol{\theta}) \quad with \quad M_n(\boldsymbol{\theta})=\frac{1}{n}\sum_{i=1}^n \{\boldsymbol{y_{exp_i}}-F(\boldsymbol{x_i},\boldsymbol{\theta})\}^2,
\end{equation}

where \citet{cox2001} propose to get this minimum numerically. Then the estimation of the discrepancy is done by applying a nonparametric regression to the data $\{\boldsymbol{x_i},\boldsymbol{y_{exp_i}}-F(x_i,\hat{\boldsymbol{\theta}})\}_{i=1,\dots,n}$. Any nonparametric regressions can offer working alternatives with this method, showing the interesting flexibility of the approach.\newline

\subsubsection{Bayesian estimation}

\noindent Under the Bayesian framework, there are several \textit{ad hoc} short cuts to find estimators without evaluating and sampling from the entire joint 
\posterior distribution of the unknowns. The idea is to consider a \prior distribution on all the parameters which we will separate into two different categories. The first category represents the nuisance parameters
which are typically $\{\sigma_S^2,\sigma_{\delta}^2,\boldsymbol{\psi}_S,\boldsymbol{\psi}_{\delta}\}$, $\sigma_{err}^2$ and $\boldsymbol{\beta}$. These parameters are added because of the modeling. The second category groups the other parameters to be estimated such as $\boldsymbol{\theta}$. We will work on the two generic models
$\mathcal{M}_3$ and $\mathcal{M}_{4}$ with the corresponding sets of parameters to be estimated.\newline

The difference between the two models lies in the fact that for $\mathcal{M}_3$ the code can be used as such and for $\mathcal{M}_4$ an emulator is used to avoid running the code. In the further developments, the parameters to be estimated will be relative to $\mathcal{M}_4$ and to return to $\mathcal{M}_3$ it will just be necessary to omit the nuisance parameters relative to the emulator. \newline

As introduced above, it is common to take a weakly informative \prior on $\boldsymbol{\beta}$ such that $\pi(\boldsymbol{\beta}_S,\boldsymbol{\beta}_{\delta})\propto 1$. It is also reasonable to assume that \prior information about $\boldsymbol{\theta}$ is independent from the \prior information about $\Phi$ and $\boldsymbol{\beta}$. The \prior density can then be expressed as

\begin{equation}
\pi(\boldsymbol{\theta},\boldsymbol{\beta},\Phi) = \pi(\boldsymbol{\theta}) \times 1 \times \pi(\Phi).
\label{eq:PriorDistribution}
\end{equation}

Once the full likelihood integrated $\mathcal{L}^F$ on the \prior distribution of $\boldsymbol{\beta}$, the \textit{posterior} distribution can be expressed (full details are pursued in \citet{kennedy2001b}).\newline

For a full Bayesian analysis, integrating $\Phi$ out is needed to finally get $\pi(\boldsymbol{\theta}|\boldsymbol{y})$. However this integration can be quite difficult because of the high number of nuisance parameters. It would also demand a full and careful consideration of the \prior $\pi(\Phi)$. Two methods are mainly used to estimate $\boldsymbol{\theta}$ and $\Phi$. In \citet{higdon2004}, the choice made is to jointly estimate all parameters from
\EqRef{eq:FullLikelihood}. The strength of this method is that it is fully Bayesian\hc
 all the collected data are used (the data simulated with the DOE and experimental data) to estimate all the parameters and nuisance parameters at the same time. \newline

However, \citet{kennedy2001} and \citet{bayarri2007} chose an estimation in separate steps. This method called modularization by \citet{liu2009} makes inference simpler but gives only a rough approximation approximation of the exact \textit{posterior} (that separates the components of parameter $\Phi$ for each Gaussian Process involved). The first step consists in maximizing the likelihood $\mathcal{L}^M(\boldsymbol{\beta}_S,\Phi_S|\boldsymbol{y}_c;\boldsymbol{D})$ (Equation (\ref{eq:PartialLikelihood})) to get the maximum likelihood estimates (MLE) $\hat{\boldsymbol{\beta}}_S$ and $\hat{\Phi}_S$ of $\boldsymbol{\beta}_S$ and $\Phi_S$.
In the second stage, these estimators are plugged into the conditional likelihood $\mathcal{L}^C(\boldsymbol{\theta},\boldsymbol{\beta}_{\delta},\Phi_{\delta};\boldsymbol{\beta}_S,\Phi_S,\boldsymbol{y}_{exp}|\boldsymbol{y}_c,\boldsymbol{X},\boldsymbol{D})$ (Equation \ref{eq:conditionalLikelihood}) from which the \textit{posterior} density is sampled with MCMC methods. Note that this last step is the only one that differs from the SMLE method in \citet{cox2001}.
\newline
  
An alternative method was developed in \citet{bayarri2007} where ``virtual'' residuals are studied ($\boldsymbol{y}_{exp}-f_c(\boldsymbol{X},\boldsymbol{\theta}_{prior})$ where $\boldsymbol{\theta}_{prior}$ is a \textit{prior} value on $\boldsymbol{\theta}$). Then the \posterior densities of $\sigma_{\delta}^2$ and $\sigma_{err}^2$ are sampled with a Gibbs algorithm based on conditional complete distribution. In practice, this estimation is very time consuming as the Gibbs sampler will compute at each iteration the full likelihood which contains a $(n+N)\times(n+N)$ matrix to invert.\newline

It seems intuitively more natural to estimate the parameters with the modularization technique. Indeed, simulated data only influence the
value of the nuisance parameters relative to the emulator. Experimental data influence the nuisance parameters contained in the whole model.

\section{Application to the prediction of power from a photovoltaic (PV) plant \label{sec:application}}

In this section, the PV plant code was used as a toy example to try out all the models. First, we test the model $\mathcal{M}_1$ (Equation (\ref{eq:model1})), in which only the initial code and the measurement error are considered. The code is assumed, in this case, to be quick to run although, in most industrial case studies, numerical codes are time consuming. This is the first issue of feasibility encountered by engineers. In a second part, we apply Model $\mathcal{M}_2$ on our example to mimic the case when the code cannot be run at will. This model introduces an emulator of the code and its characteristics will be detailed below. $\mathcal{M}_3$ is motivated by the gap between reality and the code observed, most of the time, by engineers. In this case, we will add to $\mathcal{M}_1$ an error term for the discrepancy between the code and the phenomenon. This code error will be represented by a Gaussian process also detailed below. The final case is when the two issues co-occur. This leads to considering $\mathcal{M}_4$ for the application case.\newline

The Bayesian framework starts with the elicitation of \textit{priors}' densities (that will not be discussed here \citep{Albert2012}). According to the experts we choose\hc
\begin{itemize}
\item $\eta\sim \mathcal{N}(0.143,2.5.10^{-3})$,
\item $\mu_t\sim\mathcal{N}(-0.4,10^{-2})$,
\item $a_r\sim \mathcal{N}(0.17,3.6.10^{-3})$,
\item $\sigma_{err}^2 \sim \Gamma(2,169)$,
\item $\sigma_{\delta}^2\sim \Gamma(3,1)$,
\item $\psi_{\delta} \sim \mathcal{U}(0,1)$. \newline
\end{itemize}

This section comprises two subsections. The first subsection details the practical implementation procedures of the inference for each model. In the second subsection, we discuss all the results obtained for the models that we tried out.\newline

\subsection{Inference \label{Inference}}
As mentioned in Section \ref{PVsection}, a sensitivity analysis was run on the parameter vector $\boldsymbol{\theta}$ and it turns out that only $\eta$, $\mu_t$ and $a_r$ are relevant considering the power output. The inference only concerns these three parameters and the additional nuisance parameters depending on the model. For the sake of simplicity, data, recorded every $10s$, were averaged per hour and only data corresponding to a strictly positive power were kept. The Bayesian framework was chosen for the following study. It is motivated by the availability of strongly informative \textit{priors}, elicited from experts, on the parameters we want to estimate. To perform the inference, a Markov Chain Monte Carlo algorithm was used \citep{robert1996}. The algorithm used was first introduced by
\citet{metropolis1953} for a specific case and was then extended by \citet{Hastings1970}. In this application, we were able to simulate samples
from conditional distributions (algorithm called Metropolis within Gibbs). In other words, we can sample well only one component of the parameter vector 
at a time, which makes the process rather slow. That is why a Metropolis within Gibbs was launched for $3000$ iterations.
The values of this first sampling phase were kept to improve the covariance structure of the auxiliary distribution used to make proposals by the algorithm.
This will lead to better mixing properties for the following Metropolis Hastings ($10000$ iterations including a burn in phase of $3000$).\newline

Two months of data were studied. The PV production over August and September 2014 were available. We used those two months of data averaged per hour which makes 1019 points. For the cross validation, three days of instantaneous power (51 points) were taken from the learning set and used to evaluate the predictive power of the model considering the rest of the available data. \newline

\subsubsection{\new{Emulator}}
As said in Section \ref{PVsection}, $6$ input variables are needed to run the code. These are $t$ the UTC time, $L$ the latitude, $l$ the longitude, $I_g$ the global irradiation, $I_d$ the diffuse irradiation and $T_e$ the ambient temperature. As the test stand is precisely located, the latitude and longitude are not taken into account since they remain unchanged.\newline

The major issue in emulating the behavior of the code is to deal with correlated variables. The global irradiation, diffuse irradiation and ambient temperature depend on the time which defines the sun's position. If a space filling DOE is sampled in $[0,1]^4$ and then unnormalized between the upper and lower bounds of the $4$ input variables, many configurations tested would not make any sense. For example, we could obtain a time which indicates the morning and a global irradiation value which corresponds to noon. 
\new{The projection and space-filling properties of a DOE such as a maximin Latin-Hypercube-Sample (LHS) DOE \citep{morris1995exploratory} are essentially relevant for uncorrelated inputs.
A PCA on the matrix containing the input variables $\boldsymbol{x}_i$'s (over the duration used for calibration) provides a basis of the input space with uncorrelated axes. The bounds of the hypercubic domain in which the maximin LHS DOE is sampled are then chosen with respect to these axes. Therefore, the DOE will be more concentrated in the domain corresponding to configurations for input variables that are physically relevant.
Moreover, the 
PCA could help to reduce the dimension of the input space if this dimension is large.}
\new{
The main steps of this method are\hc
\begin{enumerate}
\item a PCA is performed on the matrix $\boldsymbol{X}$, the $i^{th}$ row of which is $X_i=(t_i,I_{g,i},I_{d,i},T_{e,i})$ for ${1\le i\le n}$. This matrix corresponds to the whole set of observed input variables. The matrix is scaled (each column is centered and scaled to unit variance).
\item From the eigenvectors given by the PCA, the transition matrix between the uncorrelated basis and the original basis of the input variable space: $\boldsymbol{T} \in \mathcal{M}_{4,4}$ is computed.
\item A maximin LHS DOE $\boldsymbol{D'}$ of $N$ points in dimension $d+p=4+3=7$ is sampled 
with respect to the $4$ uncorrelated axes given by the PCA and the 3 dimensions of the parameters. The DOE is sampled in $[0,1]^7$ and then unnormalized with respect to minimal and maximal values of the coordinates of the input variables on the $4$ axes and
with respect to the considered ranges of the $3$ parameters.
\item The first $4$ coordinates of  the DOE $\boldsymbol{D'}: \ \boldsymbol{D'}_{1:4} $ corresponding to the input variables are 
transformed by the transition matrix $T$: $ \boldsymbol{D}^T_{Sc,1:4}=\boldsymbol{T} \boldsymbol{D'}^T_{1:4} $ and then
$\boldsymbol{D}_{Sc,1:4}$ is unscaled which gives $\boldsymbol{D}_{1:4}$.
It results in the DOE:
 $\boldsymbol{D}=\left( \boldsymbol{D}_{1:4} \ \boldsymbol{D'}_{5:7} \right)$ where the coordinates corresponding to the parameters are concatenated to the coordinates of the input variables expressed in the original basis. 
\item Then, $\boldsymbol{y}_c=f_c(\boldsymbol{D})$ is computed to build the emulator.\newline
\end{enumerate}
}

The Gaussian processes emulated from this method prove to work much better.
\new{
To mimic a time consuming code context, we consider that only a limited number of numerical experiments is allowed for the DOE. 
First, the number of code calls will be limited to $N=50$ to investigate the high time consuming situation and compare it to an intensive use of the code.
Since calibrations with such an emulator may be not accurate enough, we propose to improve the emulator by adding $10$ code calls with an adaptive design devoted to calibration  \citep{damblin2018}. The results with or without this adaptive procedure will also be compared.}

\subsubsection{The first model $\mathcal{M}_1$}

Model $\mathcal{M}_1$ described by Equation (\ref{eq:model1}) only deals with the measurement error. The code used in its simplest form uses only the parameters $\eta$, $\mu_t$ and $a_r$. In this case the parameters to infer on are $\eta$, $\mu_t$, $a_r$ and $\sigma_{err}^2$ (where $\epsilon_i\overset{iid}{\sim}\mathcal{N}(0,\sigma_{err}^2)$).

\subsubsection{The second model $\mathcal{M}_2$}

As defined in Section \ref{sec:calibration}, when the code is time consuming, the solution is to emulate it with a Gaussian process (GP). For the GP emulator, we chose to consider the mean function $m_S(\bullet,\bullet)$ as a linear combination of linear functions. That means $\textbf{H}_S$ is a matrix of linear functions. The correlation function $r_S$ ($c_S=\sigma_S^2r_S$) chosen is defined by the following equation that corresponds to a Matérn $5/2$ kernel:

\begin{equation}
r_{S}(\boldsymbol{x},\boldsymbol{x}^*)= \Big(1+\frac{\sqrt{5}||\boldsymbol{x}-\boldsymbol{x}^*||_2}{{\psi}_{S}}+\frac{5||\boldsymbol{x}-\boldsymbol{x}^*||_2^2}{3{\psi}_{S}^2}\Big)\exp\Big\{-\frac{\sqrt{5}||\boldsymbol{x}-\boldsymbol{x}^*||_2}{{\psi}_{S}}\Big\}.
\label{eq:Matern52Ker}
\end{equation}

where $||\bullet||_2$ stands for the Euclidean norm.
\new{We used an isotropic kernel in order to have only one range parameter ${\psi}_S$ to estimate.  In order to be consistent with this simplification, we normalized the input variables and the parameters by mapping each of them in the unit interval $[0,1]$ before computing the emulator.} 
In this case, six parameters have to be estimated\hc $\eta$, $\mu_t$, $a_r$, $\sigma_{err}^2$, $\sigma_S^2$ and ${\psi}_S$. \newline

\subsubsection{The third model $\mathcal{M}_3$}

The third model introduces another GP for the discrepancy. We chose a different covariance kernel which is Gaussian (Equation (\ref{eq:GaussianKer})).
Note that compared to Equation (\ref{eq:M3}), the discrepancy mean has been set to $0$ (\textit{i.e.} $m_{\delta}(.)=0$). These choices are motivated by the fact that the purpose of calibration is to estimate the "best-fitting" 
vector parameter $\boldsymbol{\theta}$. We do not want any compensation that might lead to an additional bias. This decision is consistent with \citet{bachoc2014} where the same hypothesis was made. \new{We chose an isotropic Gaussian kernel for the correlation:}

\begin{equation}
r_\delta\{(\boldsymbol{x},\boldsymbol{\theta}),(\boldsymbol{x}^*,\boldsymbol{\theta}^*)\}= \exp\Big\{-\frac{1}{2}\frac{||(\boldsymbol{x},\boldsymbol{\theta})-(\boldsymbol{x}^*,\boldsymbol{\theta}^*)||_2^2}{{\psi}_\delta^2}\Big\}
\label{eq:GaussianKer}
\end{equation}

In this case, there are also six parameters to be estimated, \textit{i.e.} $\eta$, $\mu_t$, $a_r$ $\sigma_{\delta}^2$, ${\psi}_{\delta}$ and $\sigma_{err}^2$.

\subsubsection{The fourth model $\mathcal{M}_4$}

This part focuses on a time consuming code with discrepancy. This model uses the same emulator and discrepancy as those defined above. The two correlation functions for the emulator and the discrepancy are chosen with different regularities in order to distinguish the two Gaussian processes. It seems relevant to assume that the discrepancy is smoother than the code. That is why a Matérn correlation function is chosen for the code and a Gaussian correlation function for the discrepancy. In this case eight parameters need to be estimated\hc  $\eta$, $\mu_t$, $a_r$, $\sigma_{err}^2$, $\sigma_S^2$, ${\psi}_S$, $\sigma_{\delta}^2$ and ${\psi}_{\delta}$.

\subsubsection{Estimation of the nuisance parameters}

In the Bayesian framework, an estimation by modularization is chosen. It concerns only the second and the fourth model. As is the case in \citet{kennedy2001}, a maximization of the probability $\pi(\Phi_S|\boldsymbol{y}_c)$ is performed to estimate $\boldsymbol{\beta}_S$, $\sigma_S^2$ and $\psi_s$ where $\boldsymbol{y}_c$ are the outputs of the code for all the points given by the DOE. This maximization is included in the R function \textit{km} from the package \textit{DiceKriging} \citep{roustant2012}.
\new{The emulation fitting procedure is run several times since the estimation of the nuisance parameters relative to the code is very sensitive to the starting point of the optimization algorithm. The $Q^2$ criterion is used to choose which estimates are kept for the emulator used in calibration \citep{da2012gaussian}.}

\subsection{Results \label{sec:results}}

\begin{figure}[htbp!]
\begin{center}
  \begin{tabular}{ccccc}
	\rotatebox{90}{ \hspace{3em} \footnotesize $\mathcal{M}_1$}
    & \includegraphics[width=.2\textwidth]{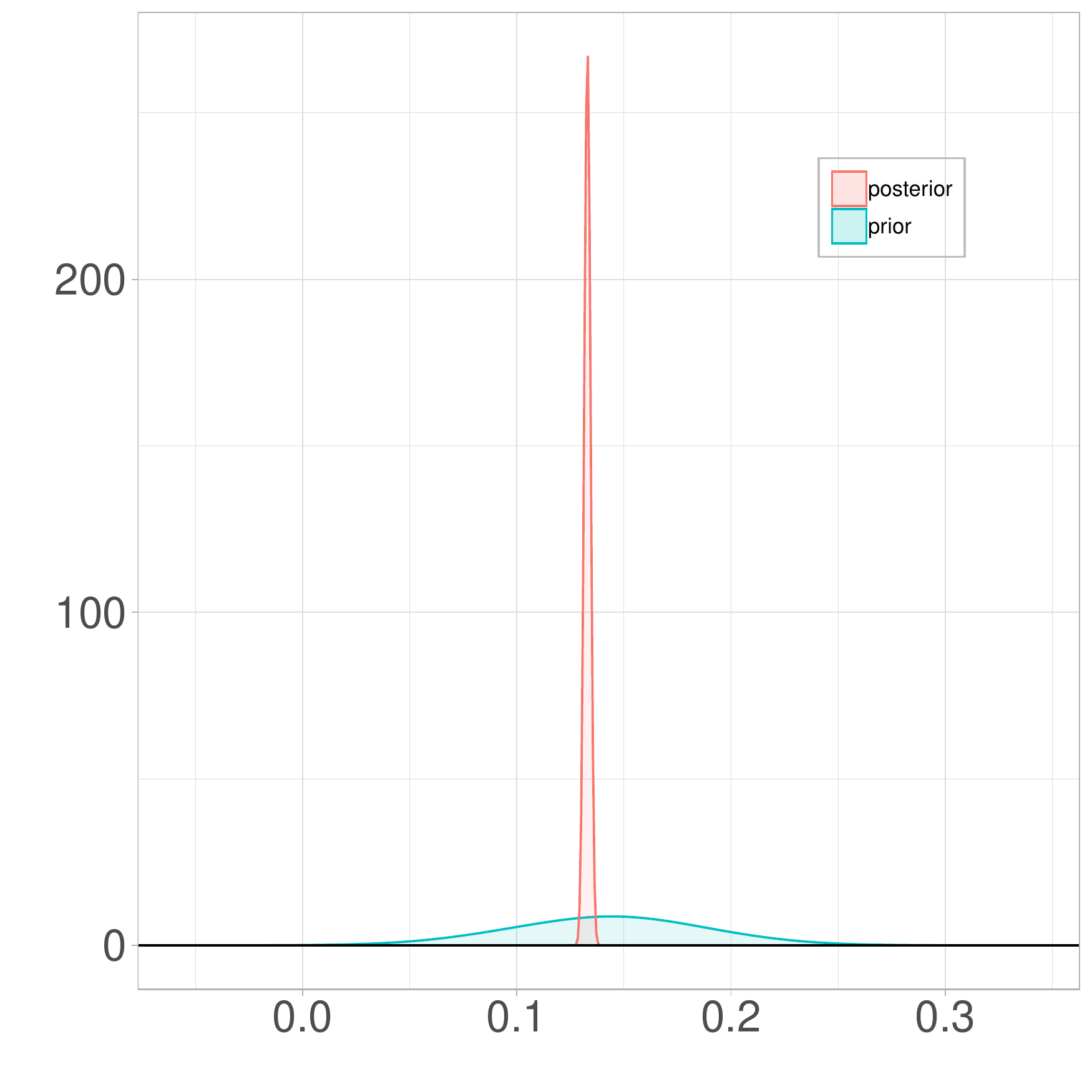}
    &  \includegraphics[width=.2\textwidth]{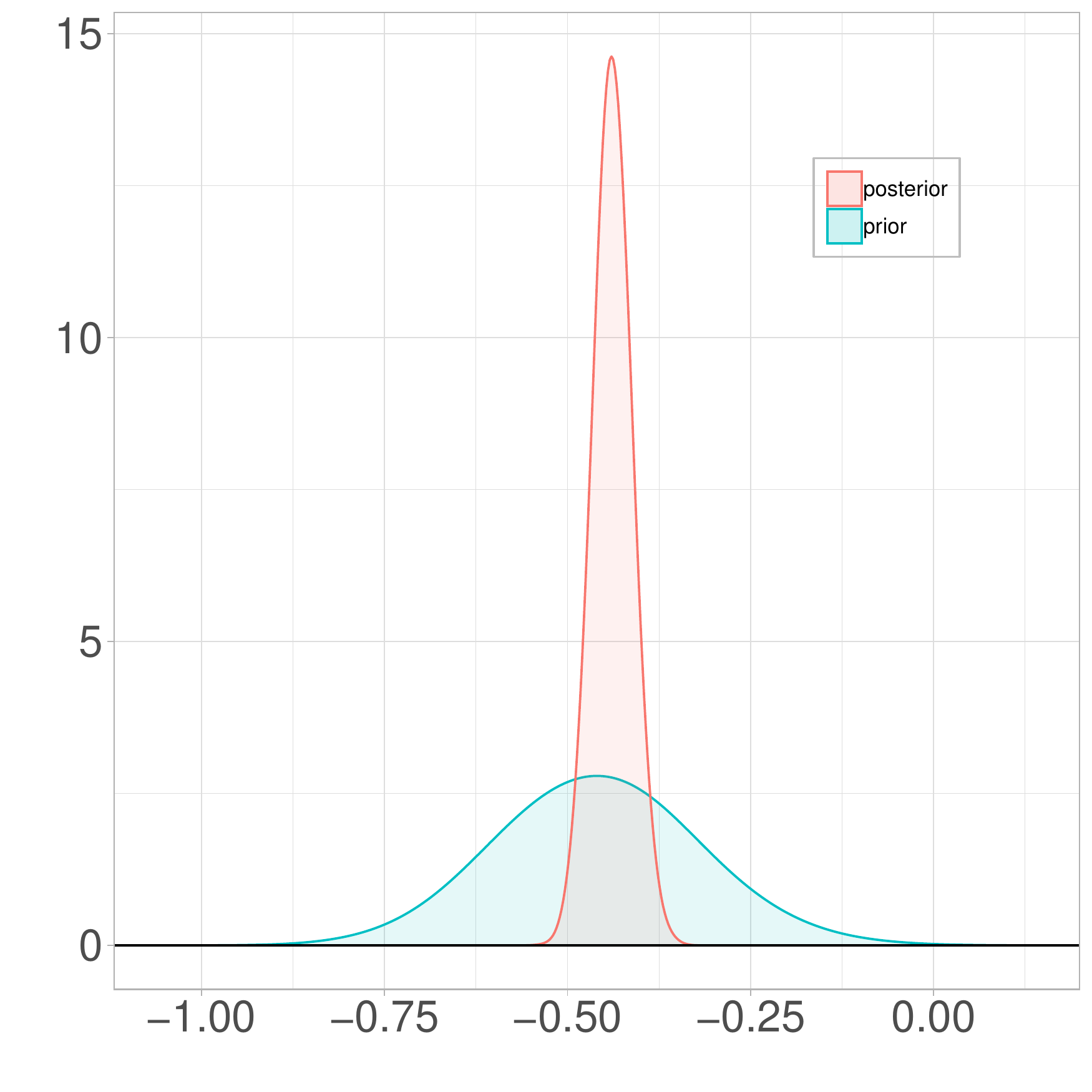}
	&  \includegraphics[width=.2\textwidth]{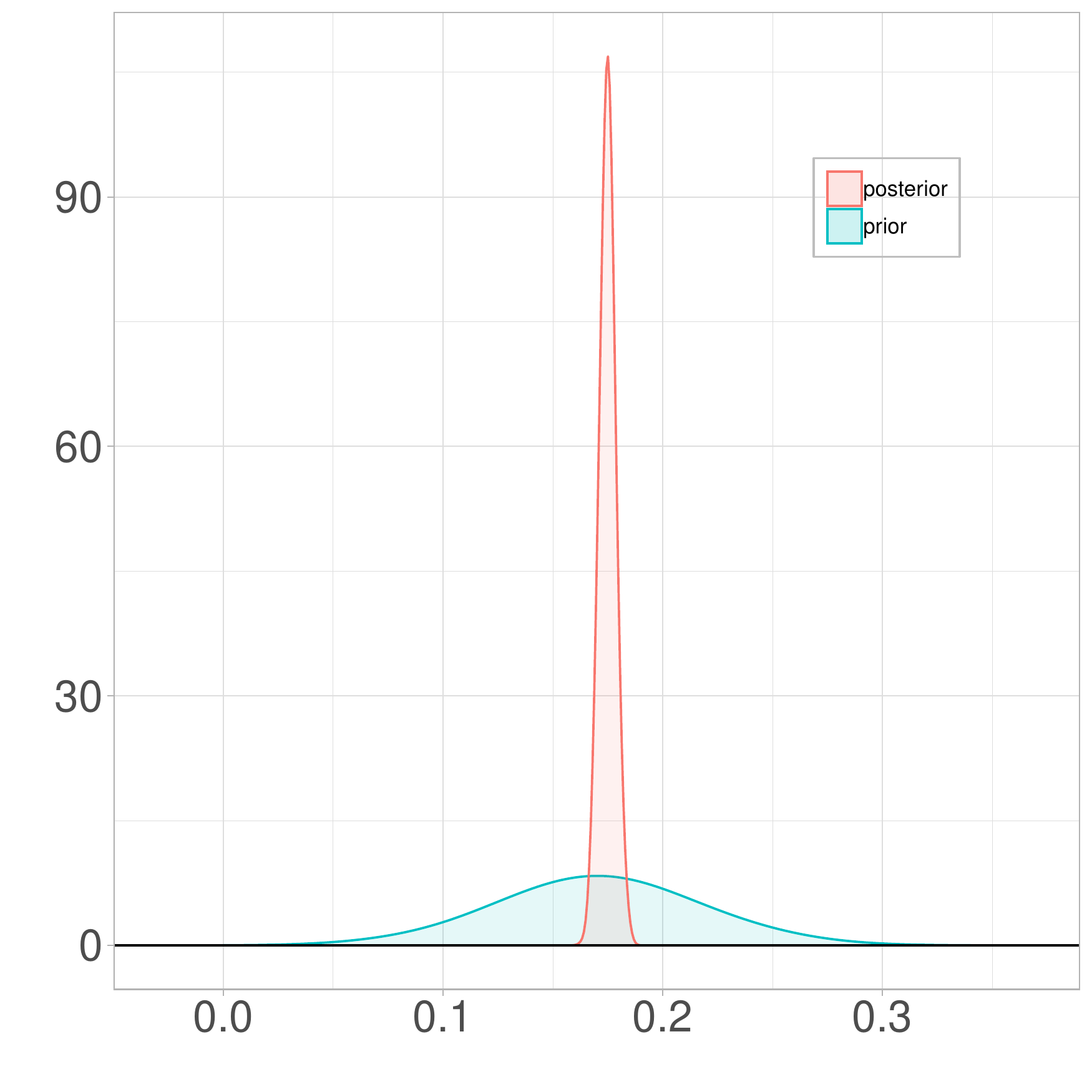}
	&  \includegraphics[width=.2\textwidth]{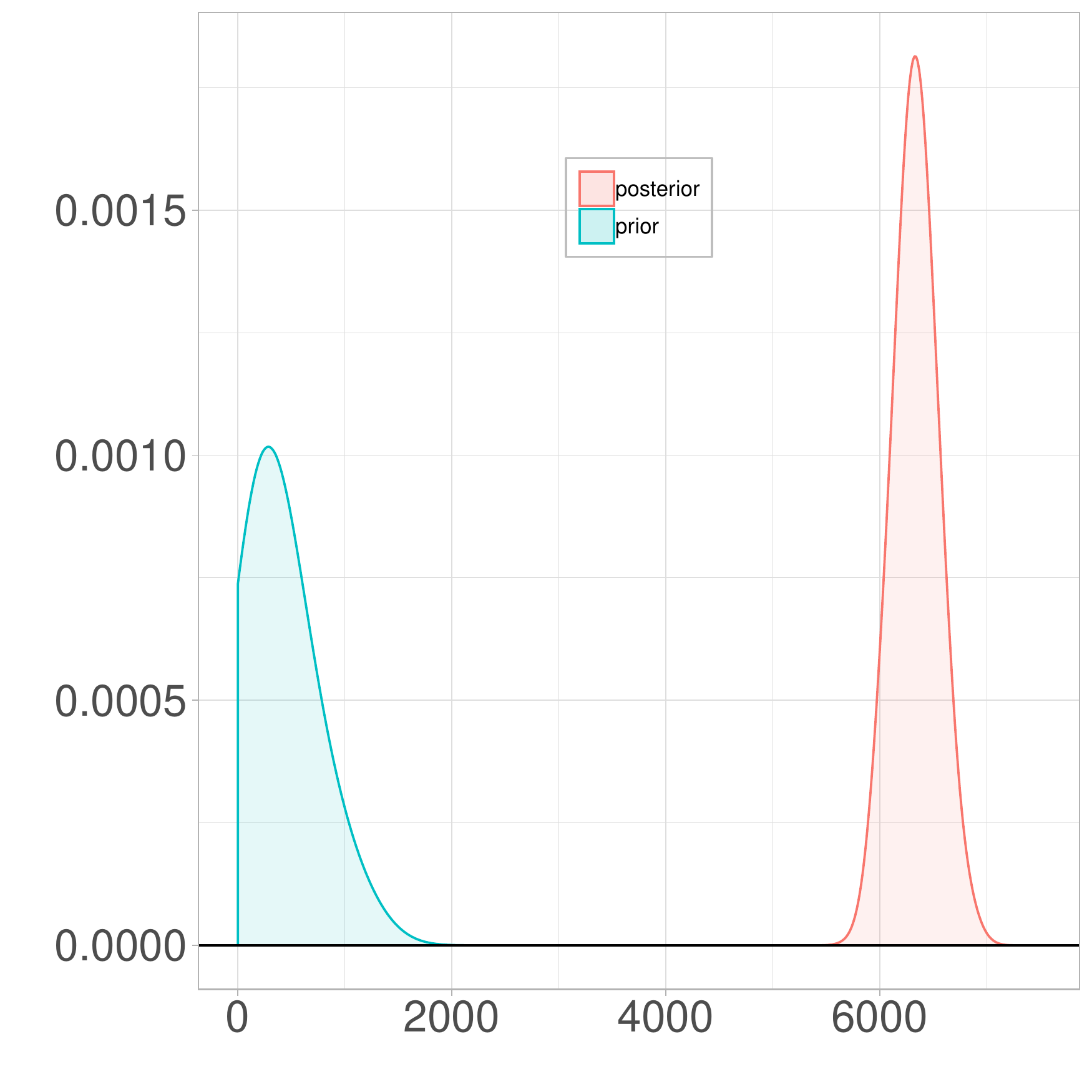}\\
		 & $\eta$ & $\mu_t$ & $a_r$ & $\sigma_{err}^2$\\
	&&&&\\
    \rotatebox{90}{ \hspace{3em} \footnotesize $\mathcal{M}_2$}
    & \includegraphics[width=.2\textwidth]{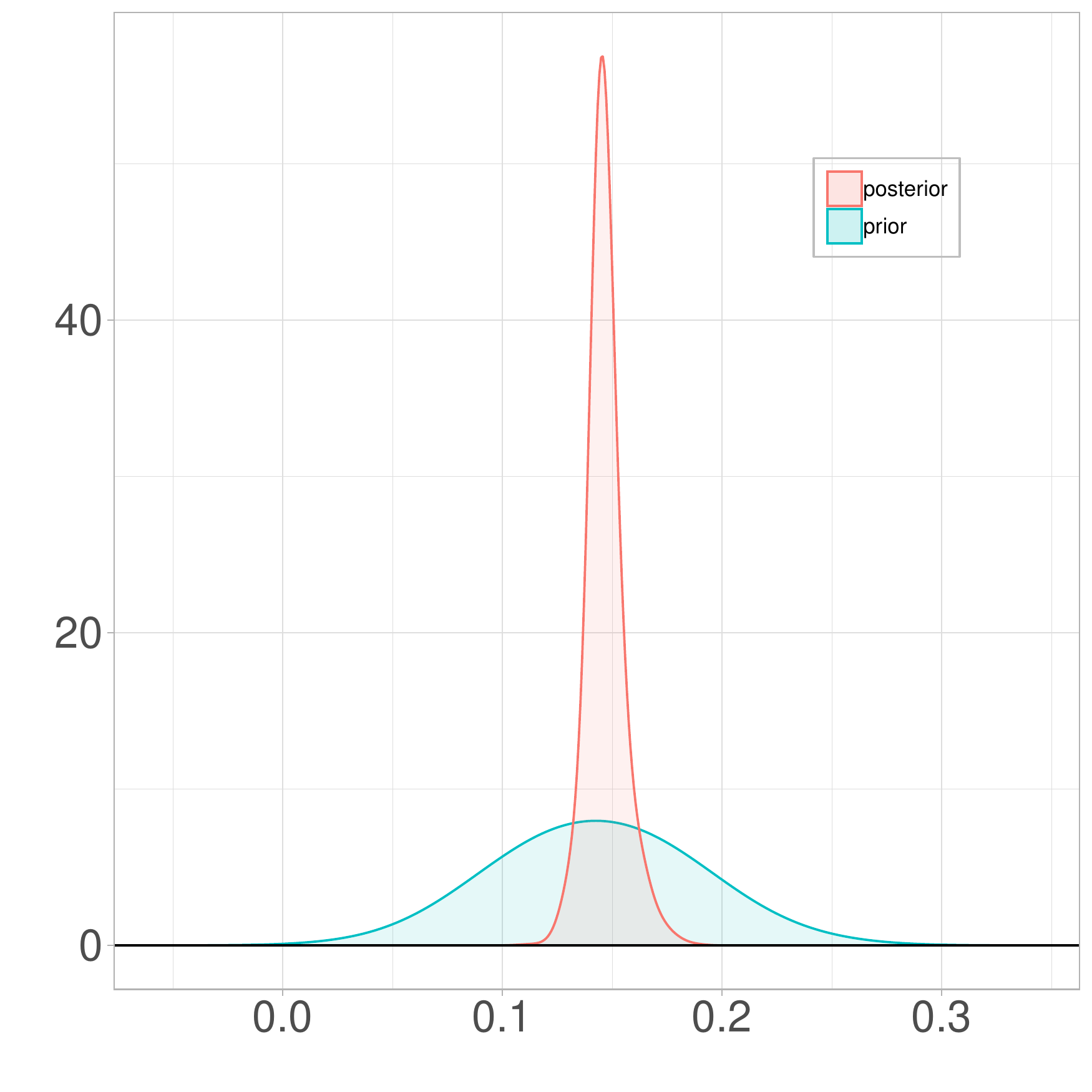}
    &  \includegraphics[width=.2\textwidth]{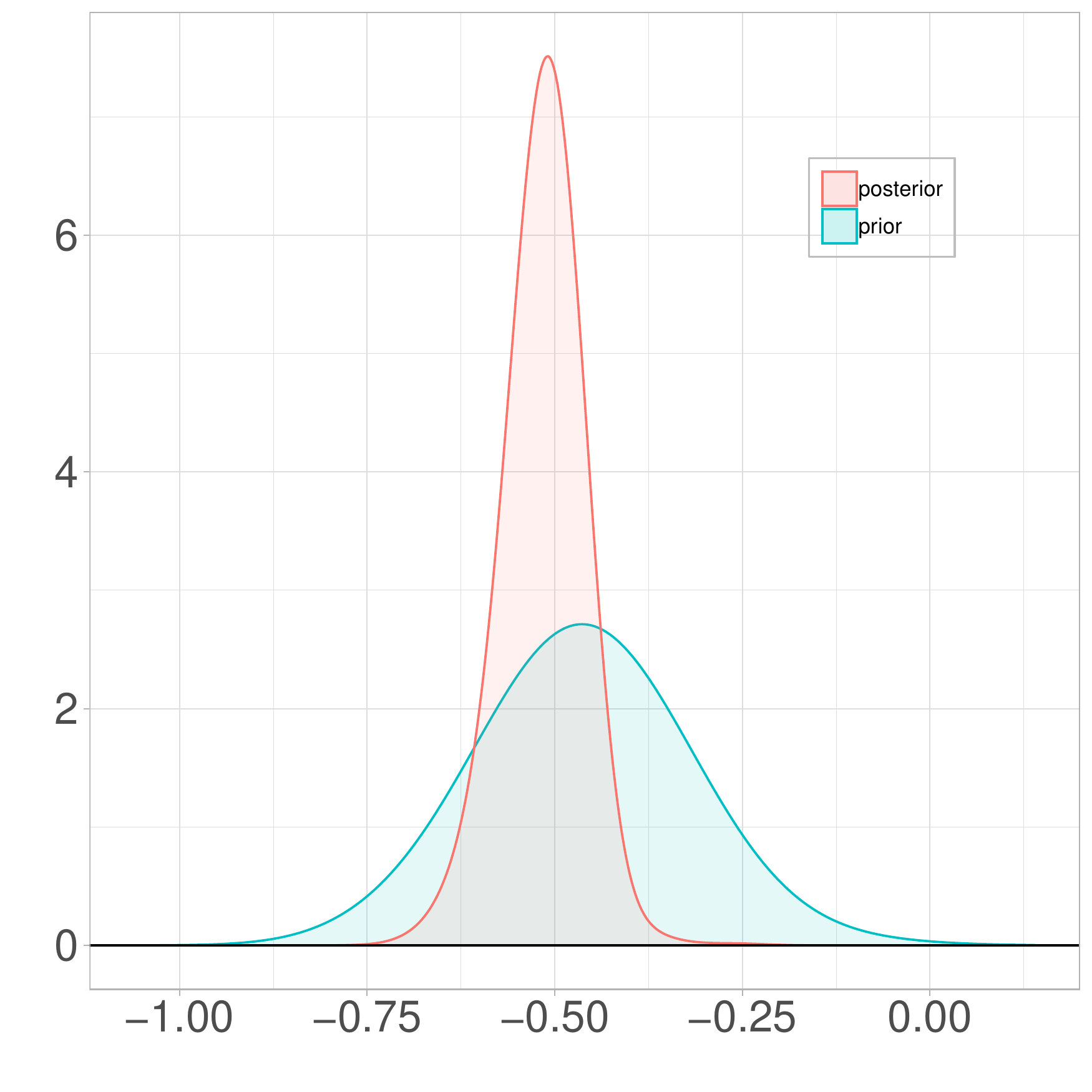}
	&  \includegraphics[width=.2\textwidth]{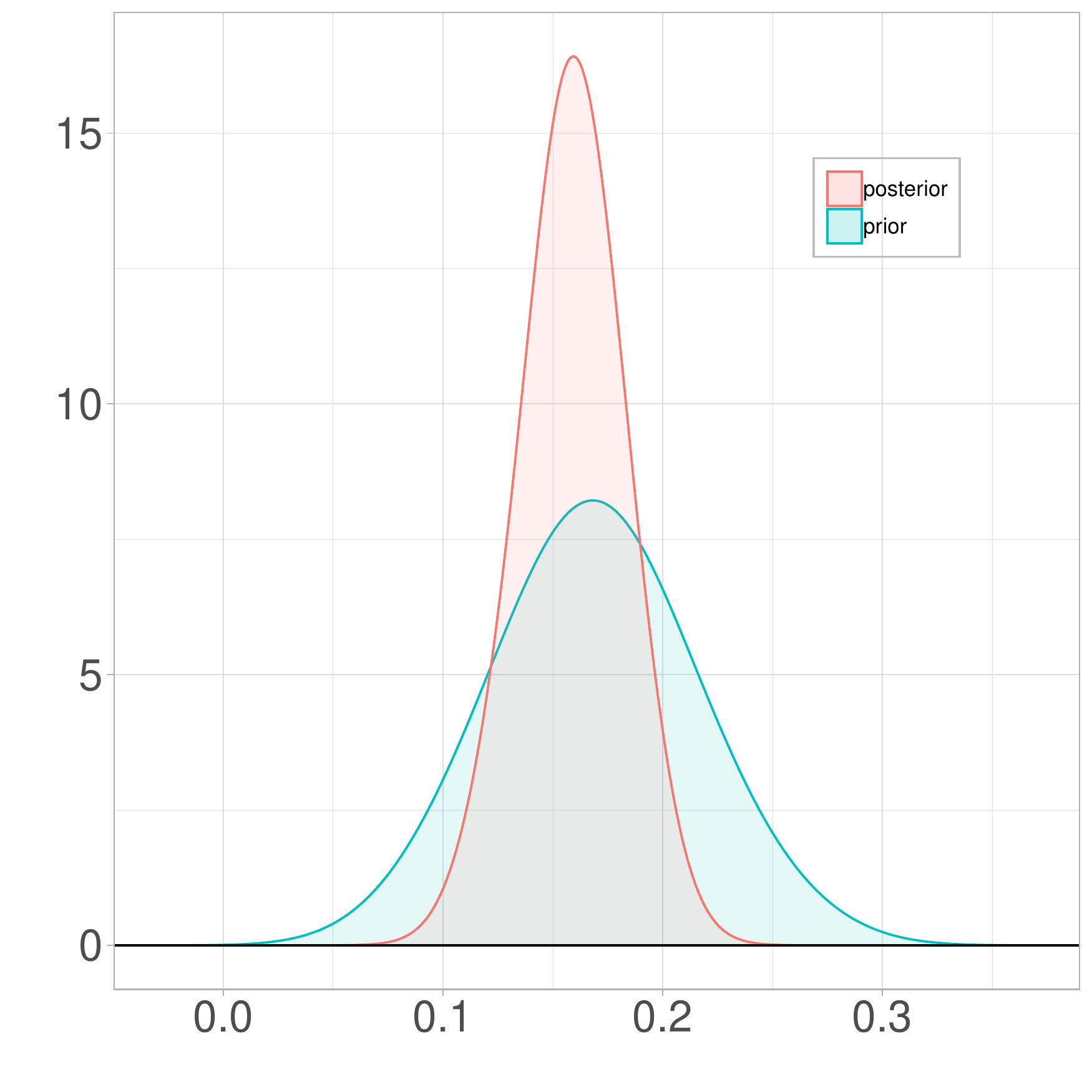}
	&  \includegraphics[width=.2\textwidth]{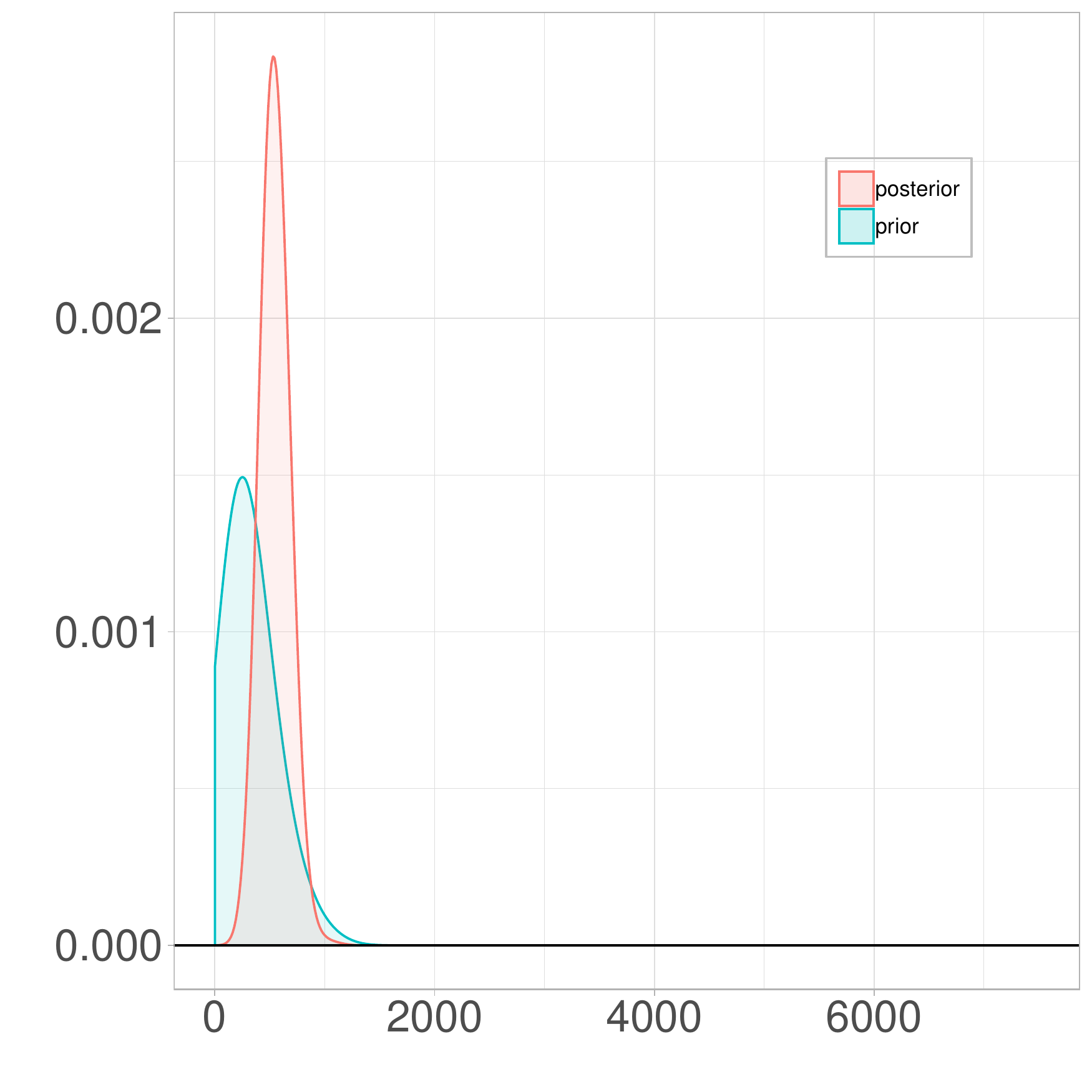}\\
		 & $\eta$ & $\mu_t$ & $a_r$ & $\sigma_{err}^2$\\
	&&&&\\
    \rotatebox{90}{ \hspace{3em} \footnotesize $\mathcal{M}_3$}
    & \includegraphics[width=.2\textwidth]{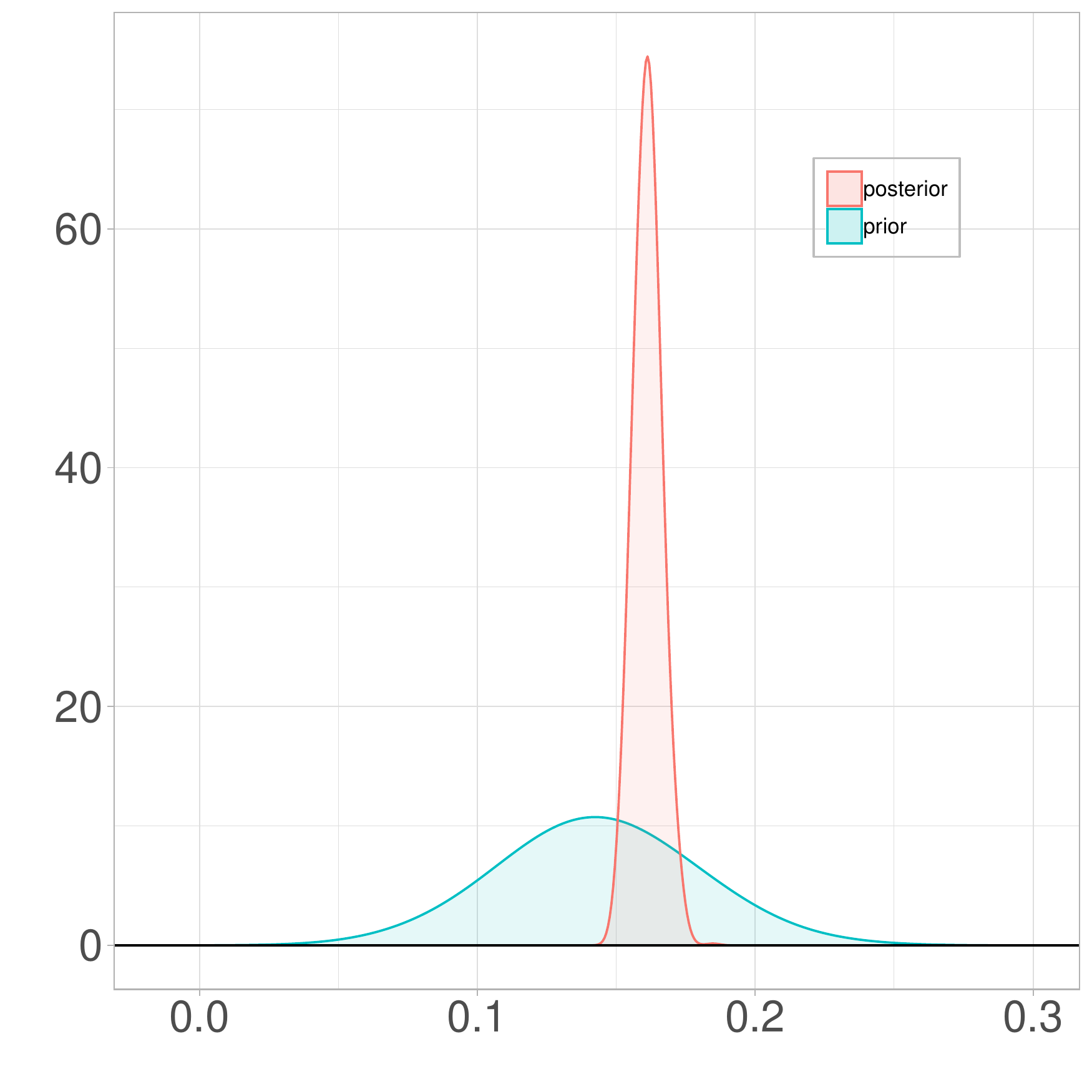} 
    &  \includegraphics[width=.2\textwidth]{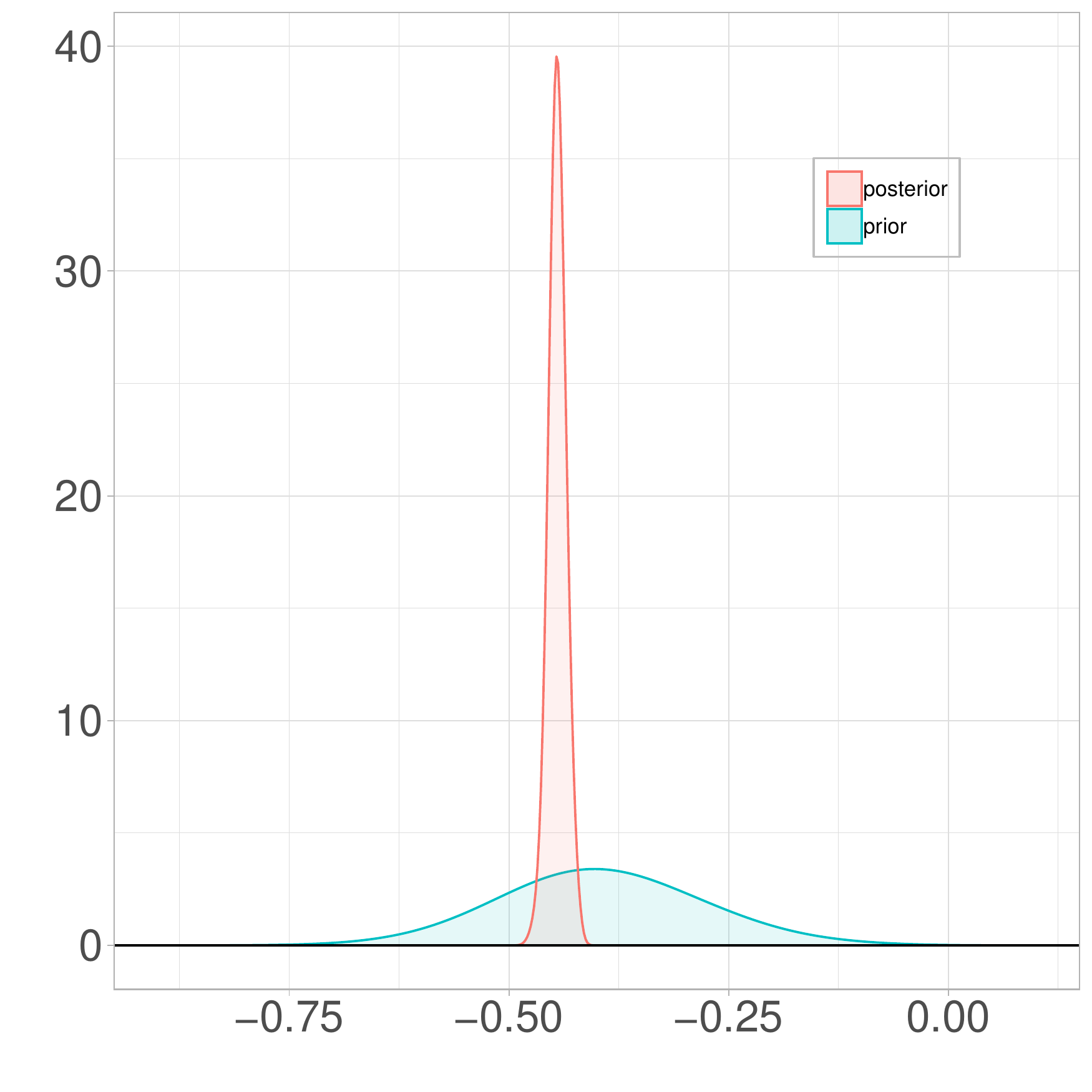}
	&  \includegraphics[width=.2\textwidth]{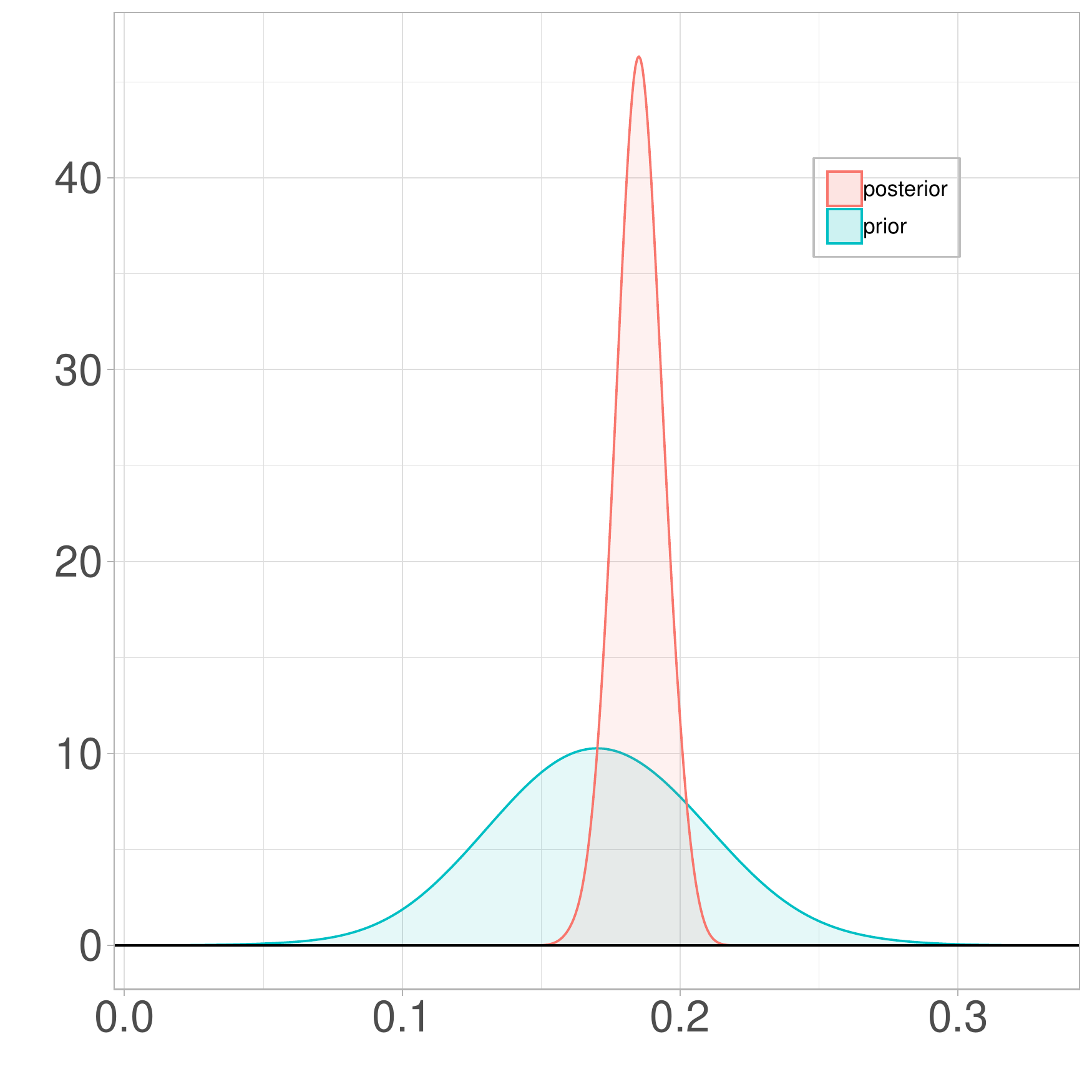}
	&  \includegraphics[width=.2\textwidth]{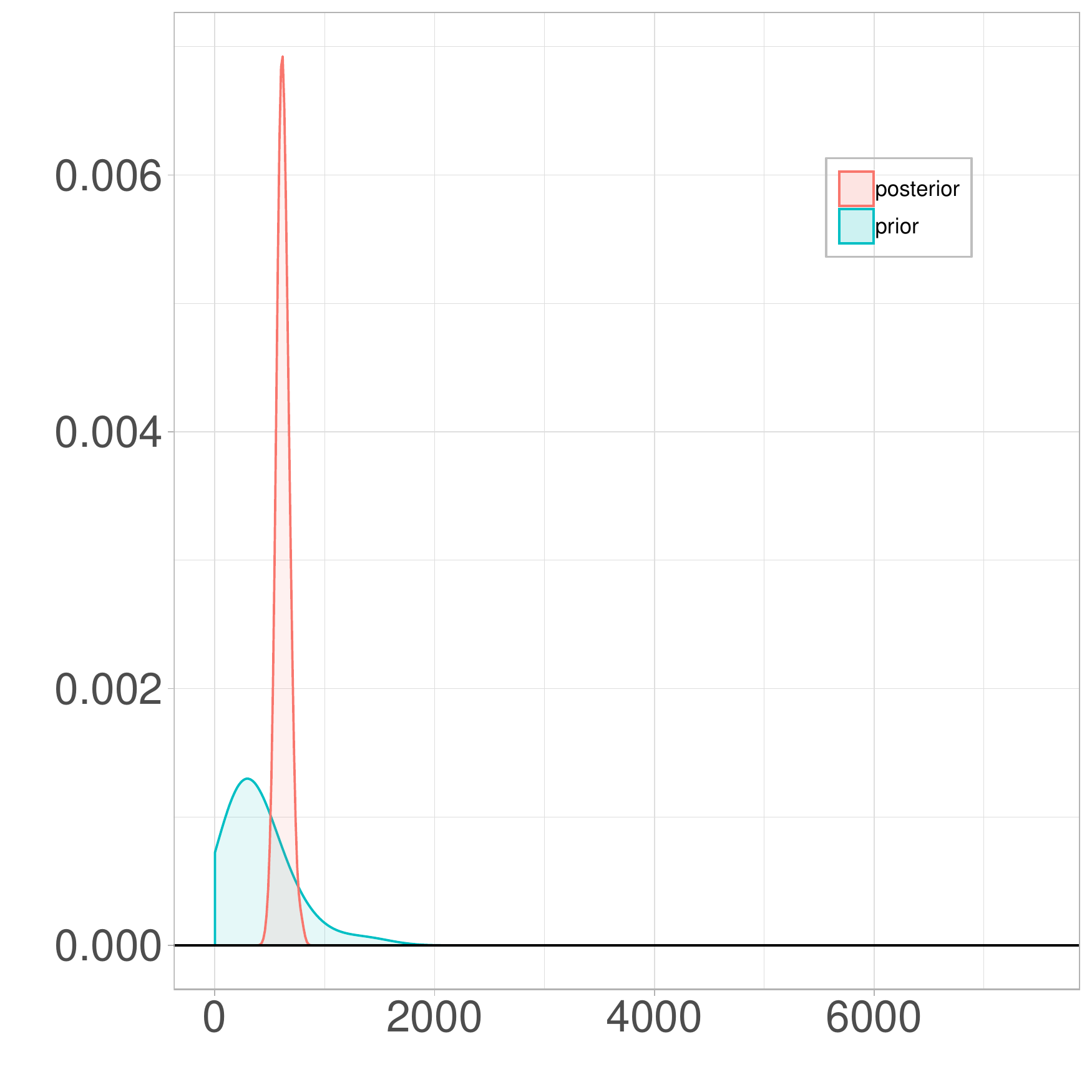}\\
		 & $\eta$ & $\mu_t$ & $a_r$ & $\sigma_{err}^2$\\
	&&&&\\
    \rotatebox{90}{ \hspace{3em} \footnotesize $\mathcal{M}_4$}
    & \includegraphics[width=.2\textwidth]{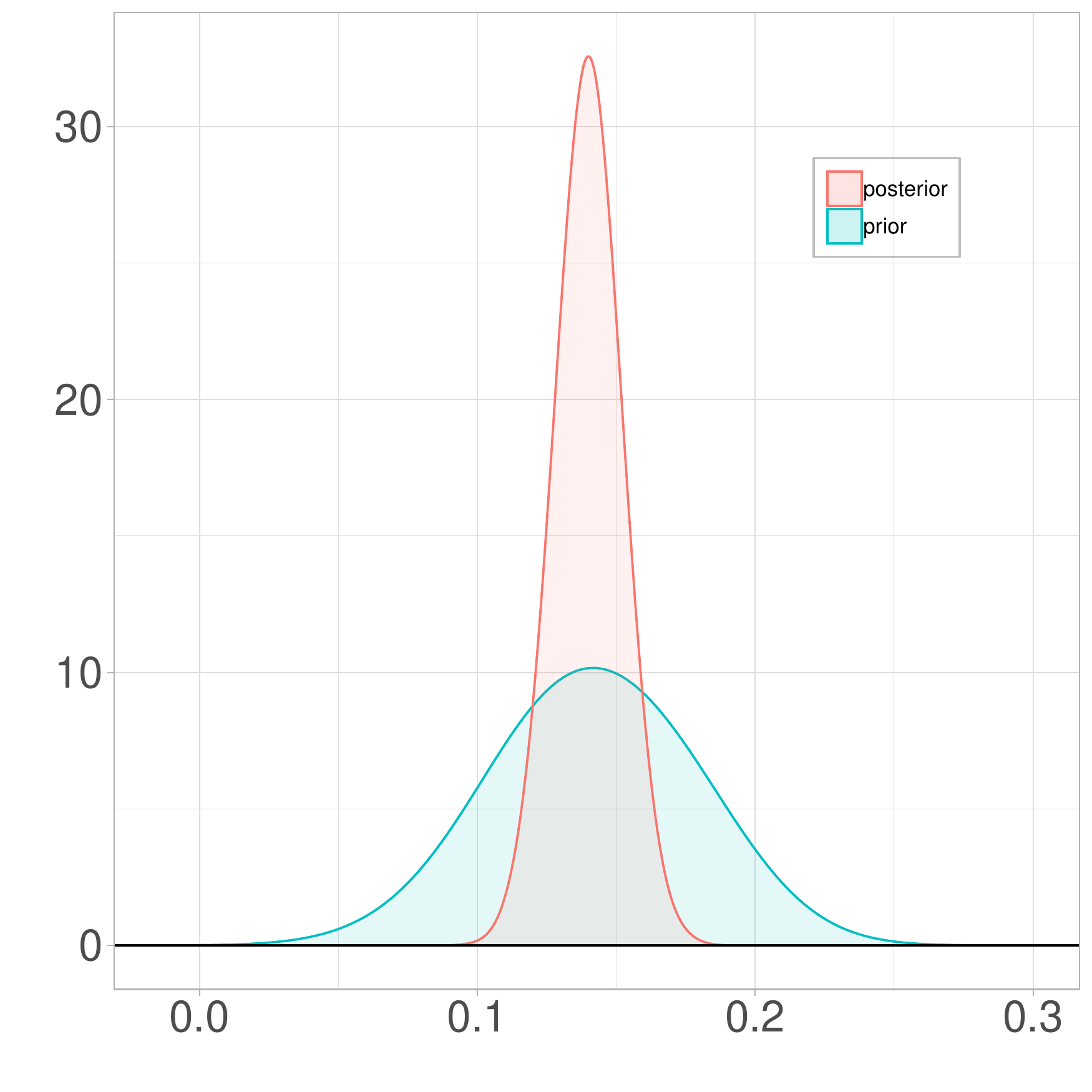} 
    &  \includegraphics[width=.2\textwidth]{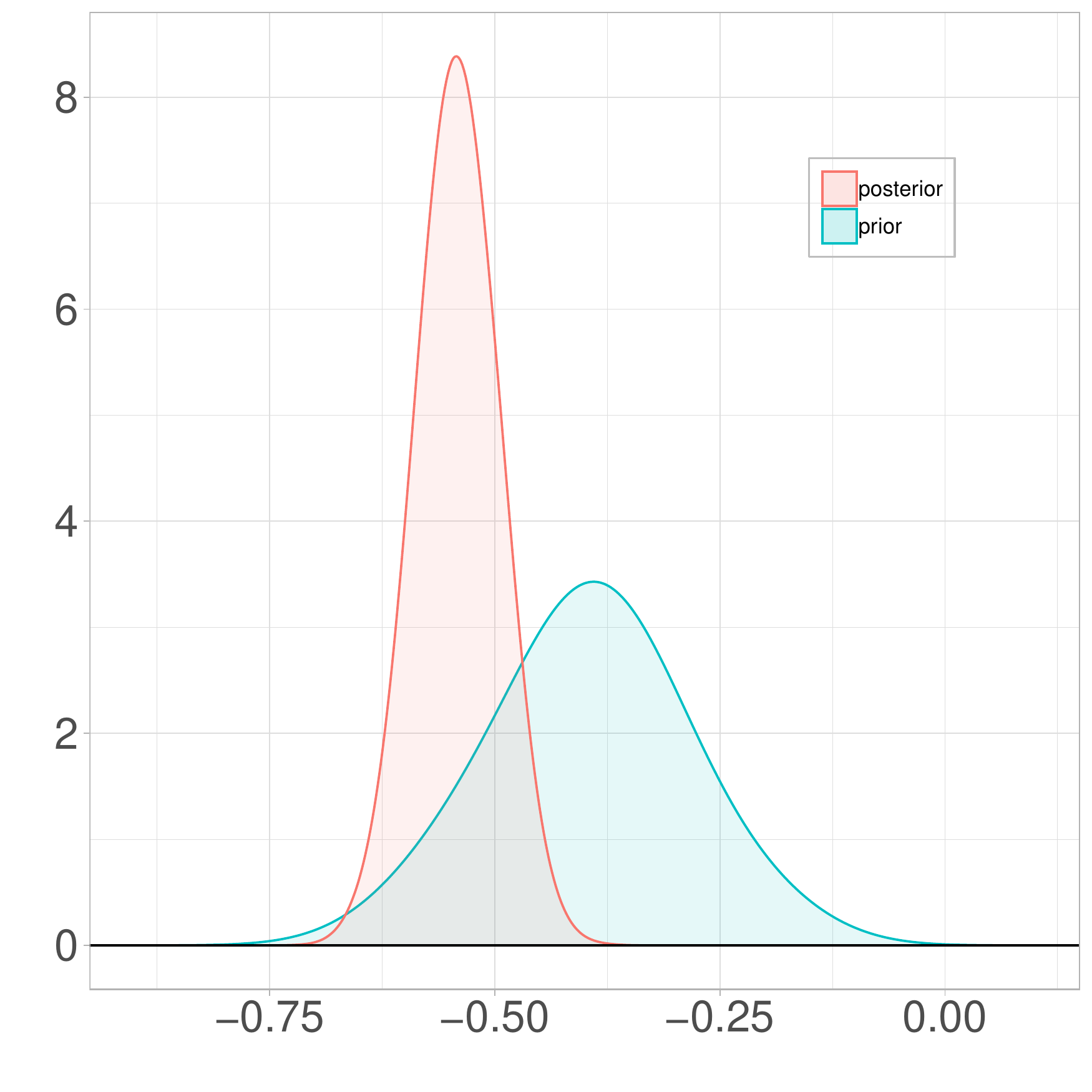}
	&  \includegraphics[width=.2\textwidth]{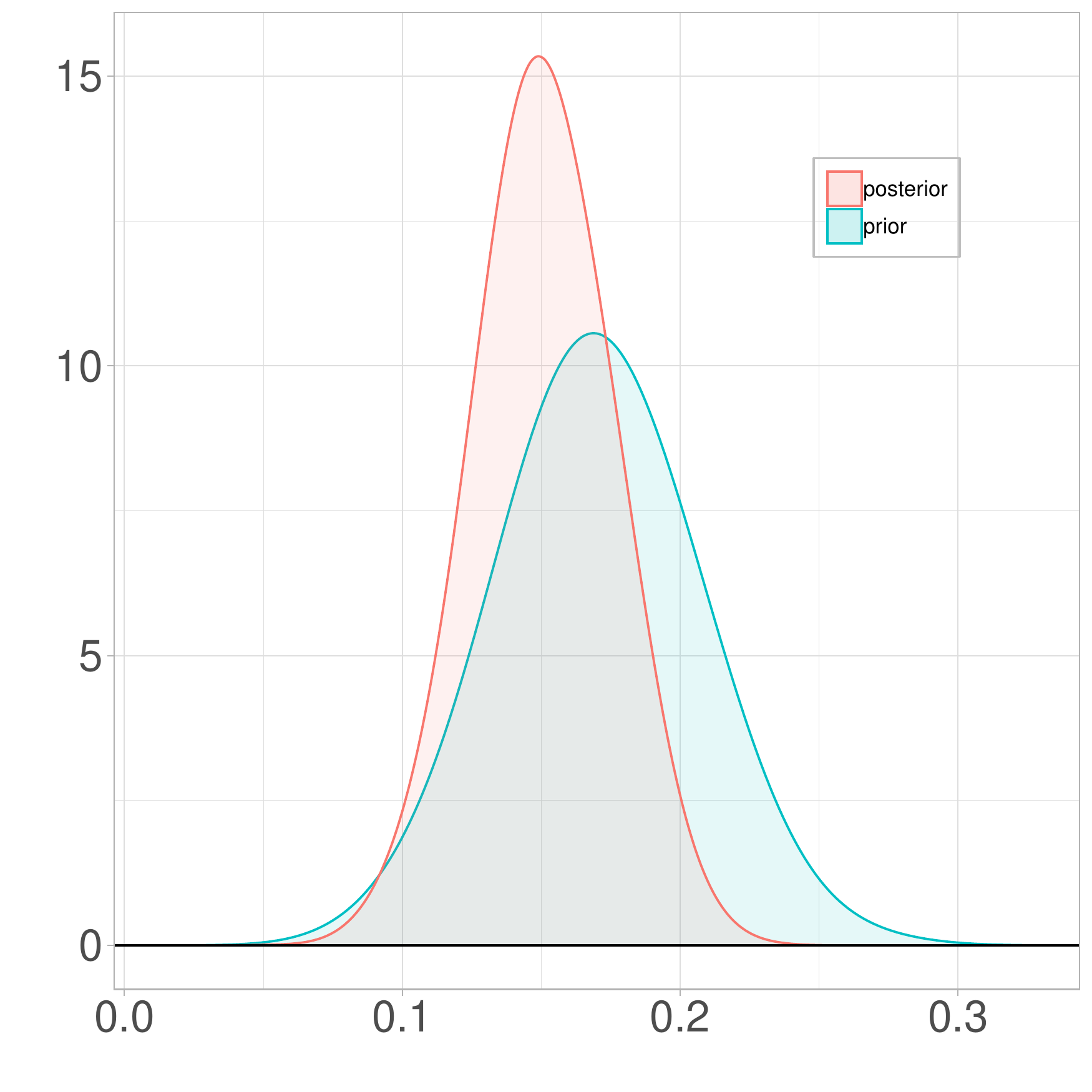}
	&  \includegraphics[width=.2\textwidth]{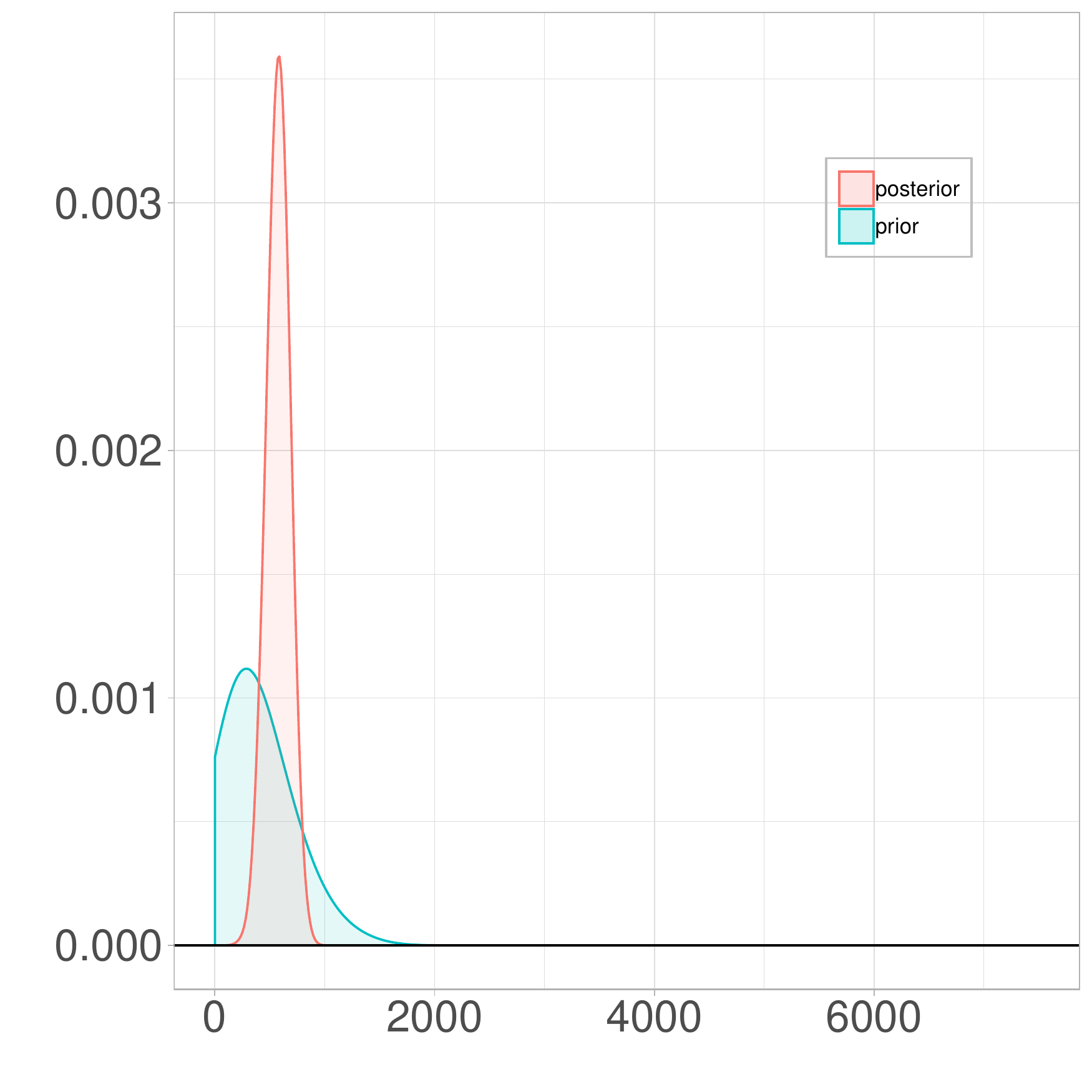}\\
	 & $\eta$ & $\mu_t$ & $a_r$ & $\sigma_{err}^2$\\
  \end{tabular}
\caption{\textit{Prior} (in blue) and \textit{posterior} (in red) densities of $\eta$, $\mu_t$, $a_r$ and $\sigma_{err}^2$ for each model.
The top two rows show the first two models (without and with emulator) which have only these four parameters to estimate.
The  bottom two rows represent the third and the fourth models which have two more parameters to estimate (see Figure \ref{fig:comparisionDensities2}).}
\label{fig:comparisionDensities1}
\end{center}
\end{figure}

\new{Figure \ref{fig:comparisionDensities1} compares the results obtained with the help of the \textbf{R} package CaliCo \citep{CaliCo}. For each parameter $\eta$, $\mu$, $a_r$ and $\sigma_{err}^2$, the MCMC chains converge. Good mixing properties are confirmed by a visual check. Figure \ref{fig:comparisionDensities1} confronts the \textit{prior} densities with the \textit{posterior} densities. In almost every model, a decrease of the variance is quantifiable, which illustrates an improvement in the knowledge of the parameter density. However, the decrease of variance is not the same for every model. Calibration performed with the model $\mathcal{M}_2$ produces \textit{posterior} densities with larger variances than for $\mathcal{M}_1$. The replacement of the numerical code by a Gaussian process has added a variance term in the full likelihood which increases the variance \textit{a posteriori}. The same phenomenon is visible from the model $\mathcal{M}_3$ to the model $\mathcal{M}_4$. Calibration with the model $\mathcal{M}_1$ also highlights a strong disagreement in the estimation of the \textit{posterior} density of $\sigma_{err}^2$. The Maximum A Posteriori (MAP) of $\sigma_{err}^2$ is $6500\ W^2$ which makes a standard deviation of $80.6\ W$. This value is too high and has no physical validity. From calibration with the model $\mathcal{M}_1$ to the model $\mathcal{M}_3$, the \textit{posterior} density of $\sigma_{err}^2$ has been corrected in accordance with the \textit{prior} distribution and the physical sense. It does not mean that calibration with the model $\mathcal{M}_1$ is incorrect, it only means that this model misses a substantial variance term due to a structural error in the experimental data. This term is the discrepancy and if one wants to be consistent with physics, one should consider Model $\mathcal{M}_3$ in this case. Note that with Model $\mathcal{M}_2$ the variance of the measurement error decreases without the addition of the discrepancy. 
This may be due to the fact that the emulator is smoother than the code and taking into account the variance of the Gaussian process could regularize the estimation problem.}
\newline

\newpage

\new{The use of a Gaussian process emulator of the code (in Models $\mathcal{M}_2$ and $\mathcal{M}_4$) has to be handled with caution. Most industrial codes are time consuming and the use of a limited number of points in the DOE is required. However, if the emulator does not represent the numerical code well enough, calibration will fail to retrieve a physical meaning for the parameter. 
The performance of the emulator is assessed through the $Q^2$ criterion \citep{da2012gaussian}. However even a $Q^2$ larger than $0.8$ does not ensure a calibration performance similar to that with the actual code. In Figure \ref{fig:comparisionDensities1}, although $Q^2=0.83$ for $\mathcal{M}_2$ and $Q^2=0.90$ for $\mathcal{M}_4$, the \textit{posterior} distributions are quite wide. Moreover, the larger \textit{posterior} variances are not the only issue\hc a shift in the \textit{posterior} mode is also observed for some parameters which could lead to quite different point estimates for these parameters. From an industrial point of view, these issues in the estimation might be unsatisfactory.
The Gaussian process can be improved by adding a small number of well-chosen points in the original DOE. A strategy, called sequential design and developed in \citet{damblin2018},  is based on the EGO algorithm \citep{jones1998efficient} to find new points regarding further calibration. From the original DOE of $50$ points used for previous calibrations (for $\mathcal{M}_2$ and $\mathcal{M}_4$) in Figure \ref{fig:comparisionDensities1}, $10$ points are appended to the original DOE by using the sequential design. Then calibration is performed on the new Gaussian process emulated with the new DOE.
}

\begin{figure}[htbp!]
\begin{center}
  \begin{tabular}{ccccc}
    \rotatebox{90}{ \hspace{3em} \footnotesize $\mathcal{M}_2'$}
    & \includegraphics[width=.2\textwidth]{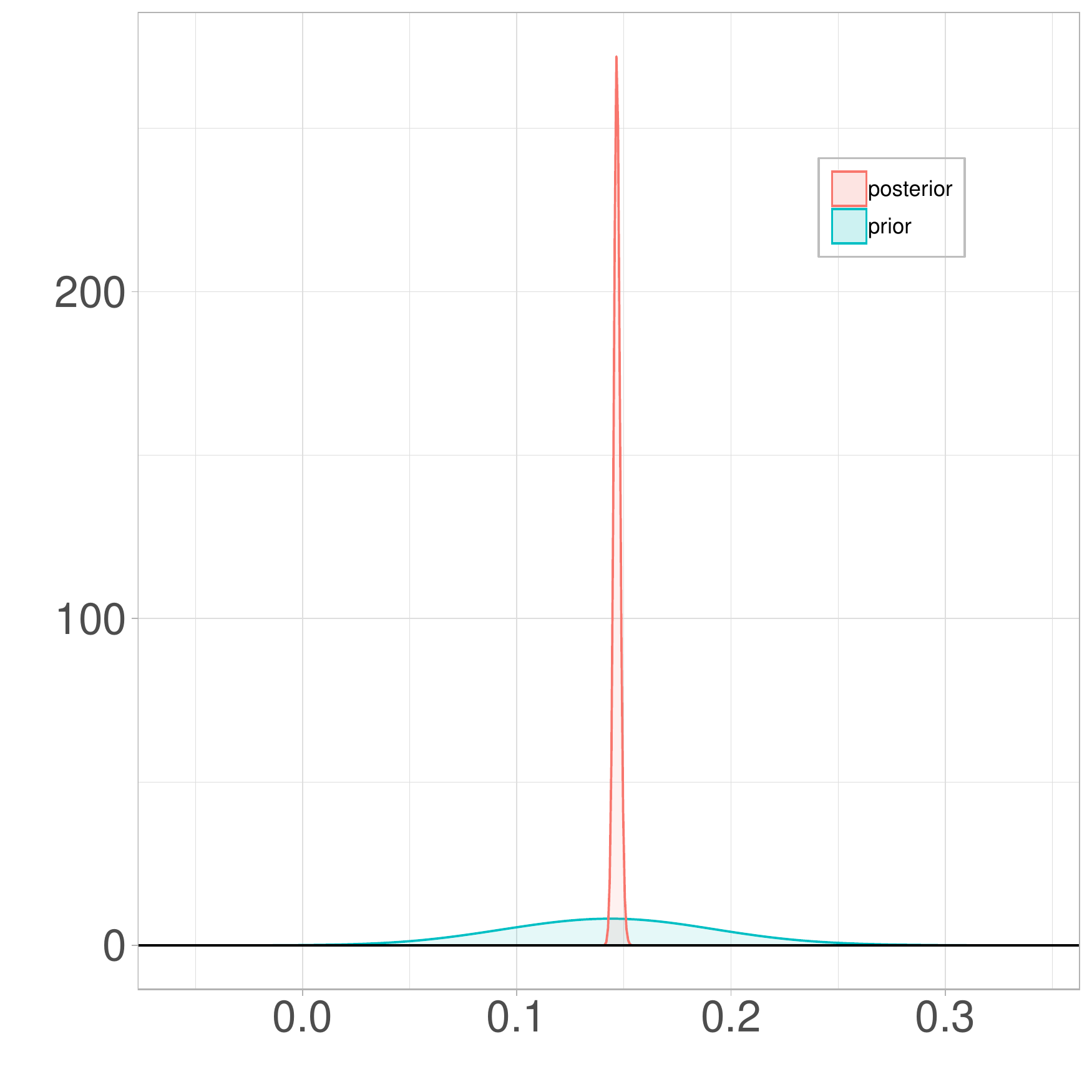}
    & \includegraphics[width=.2\textwidth]{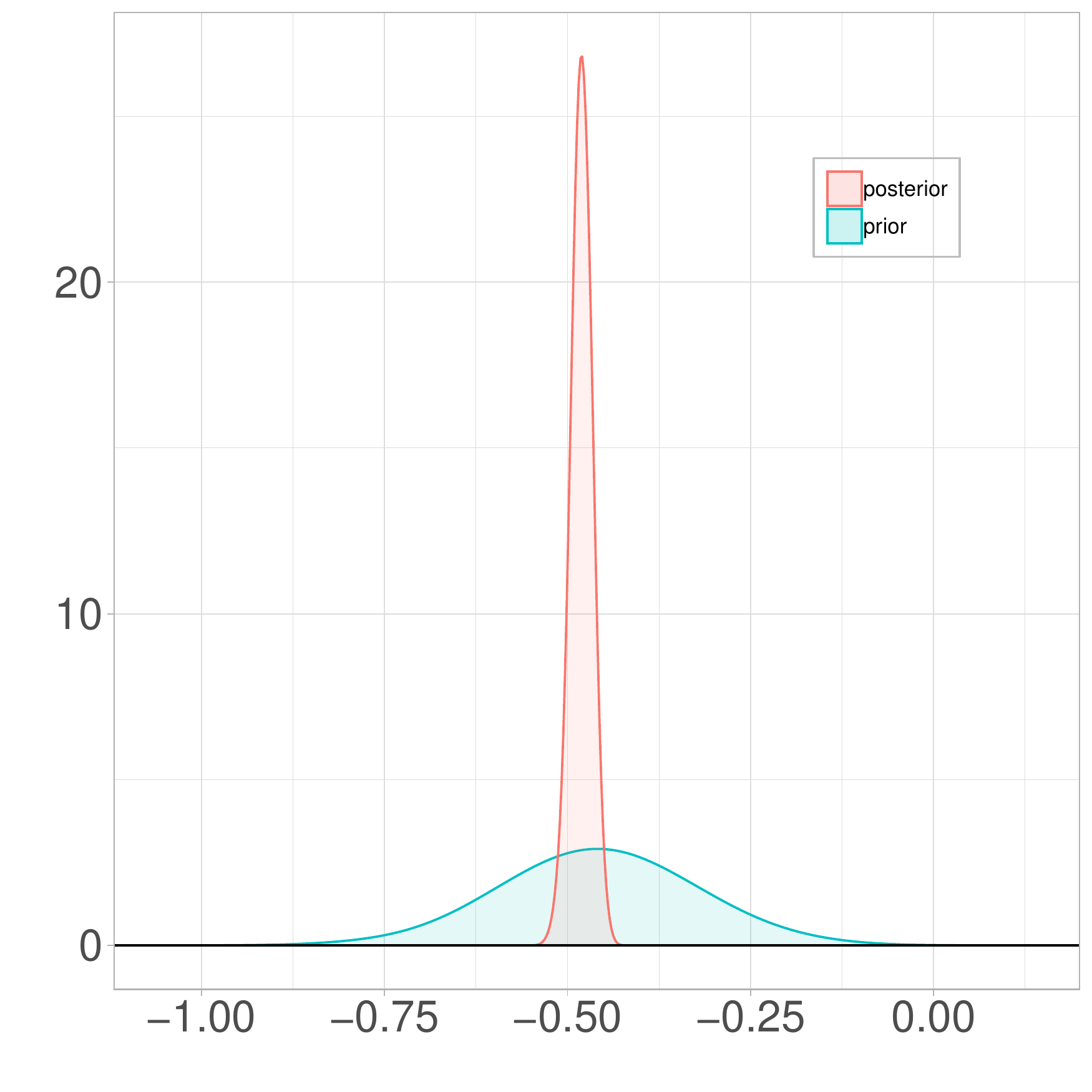}
    & \includegraphics[width=.2\textwidth]{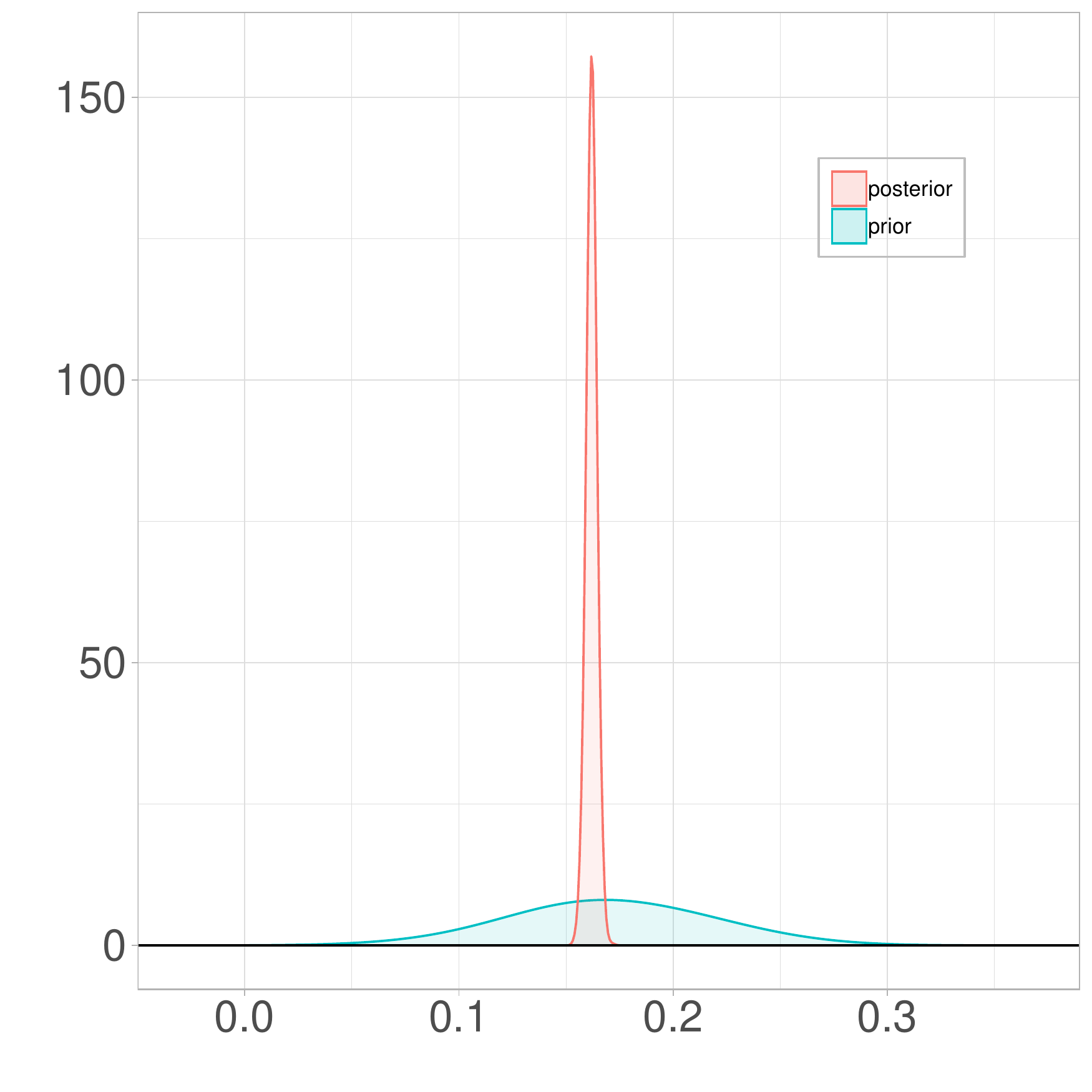}
    & \includegraphics[width=.2\textwidth]{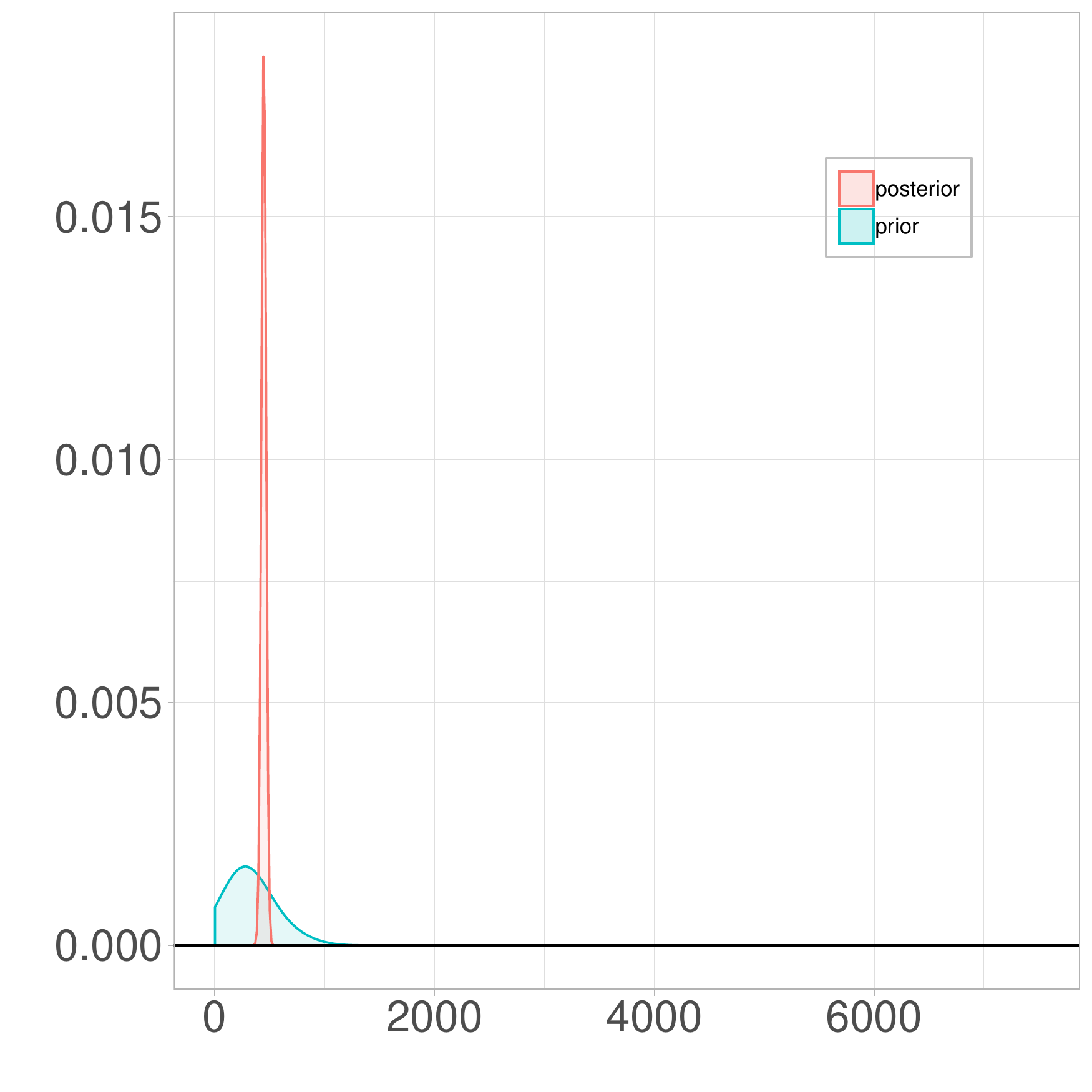}\\
	& $\eta$ & $\mu_t$ & $a_r$ & $\sigma_{err}^2$\\
	&&&&\\
    \rotatebox{90}{ \hspace{3em} \footnotesize $\mathcal{M}_4'$}
    &  \includegraphics[width=.2\textwidth]{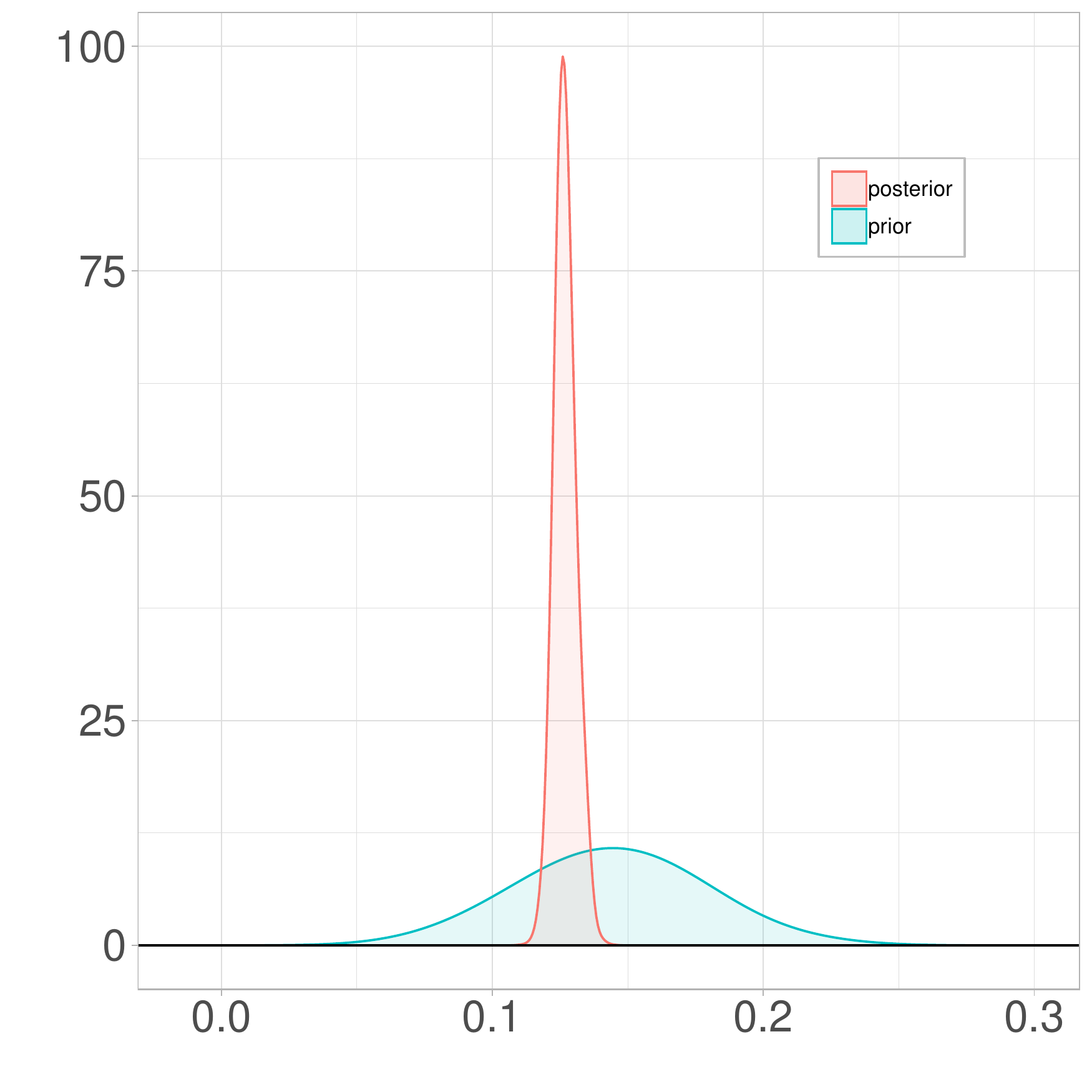}
    &  \includegraphics[width=.2\textwidth]{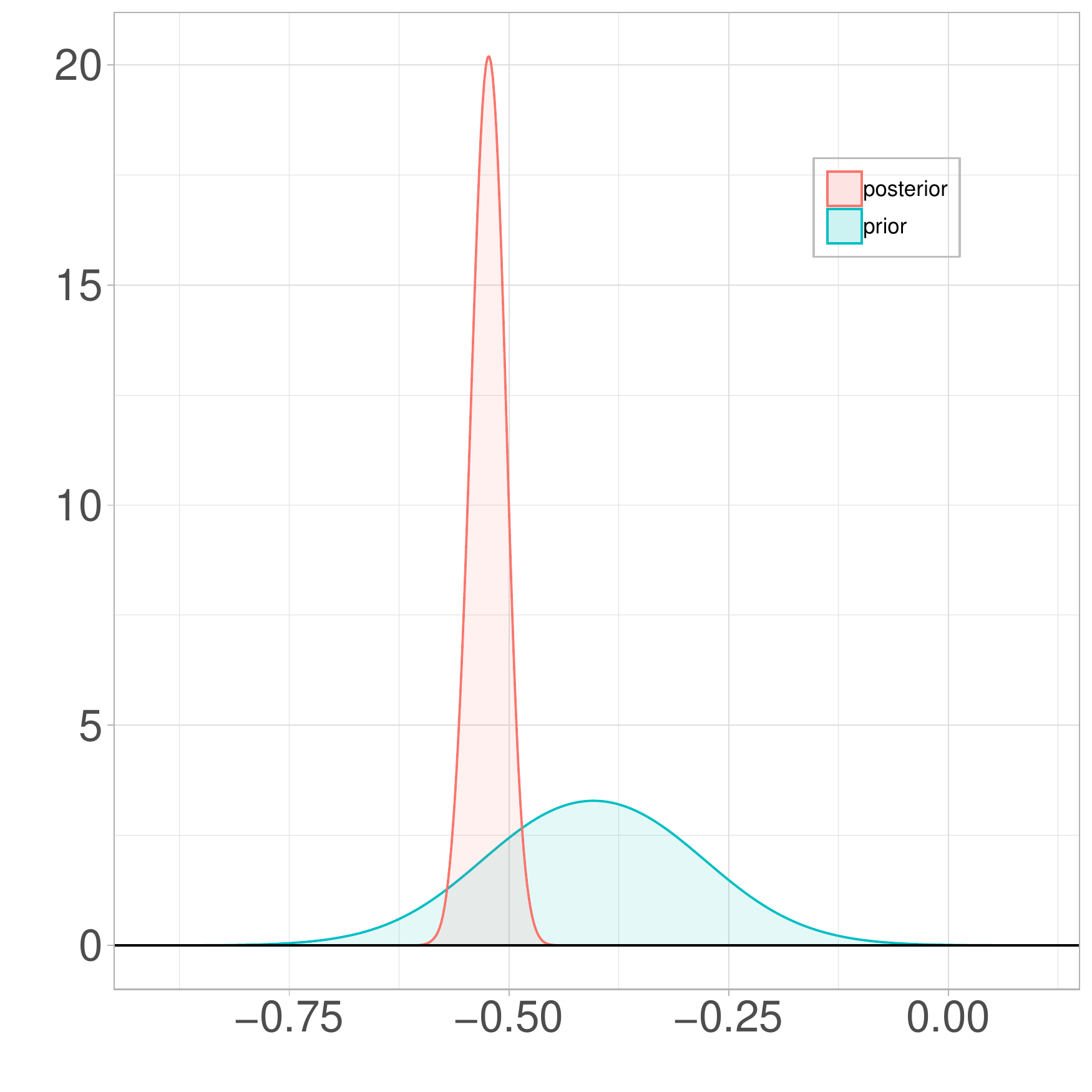}
    &  \includegraphics[width=.2\textwidth]{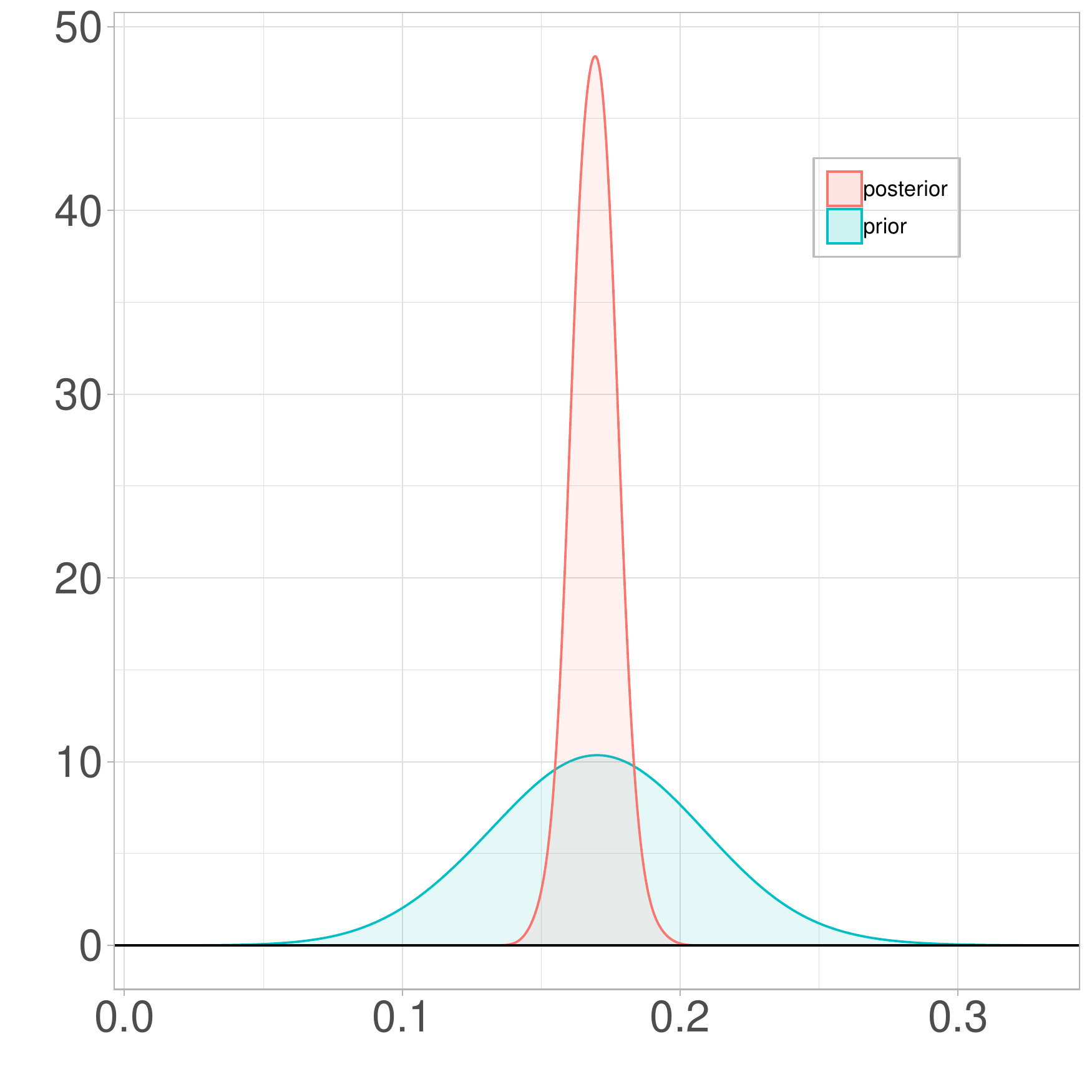}
	&  \includegraphics[width=.2\textwidth]{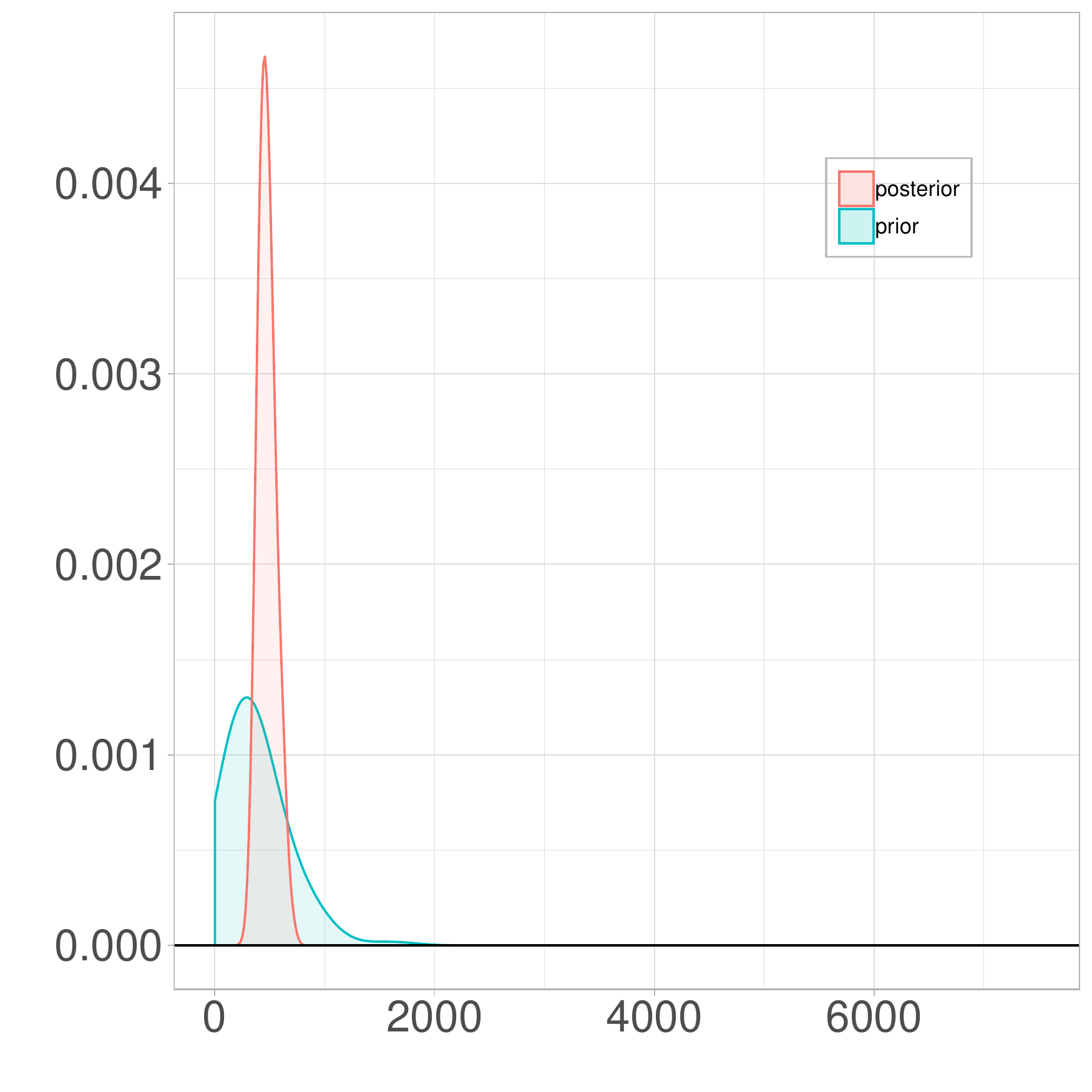}\\
		 & $\eta$ & $\mu_t$ & $a_r$ & $\sigma_{err}^2$\\
	&&&&\\
  \end{tabular}
\caption{Calibration results for $\mathcal{M}_2'$ and $\mathcal{M}_4'$ using the emulator based on the sequential design.}
\label{fig:calibrationSeq}
\end{center}
\end{figure}

\new{Figure \ref{fig:calibrationSeq} illustrates the improvement in the new results based on the Gaussian process built after the sequential design and proves that, with a better emulator, calibration appears to be consistent with the \textit{prior} densities
and thus with Models $\mathcal{M}_1$ and $\mathcal{M}_3$.\newline
}

\begin{figure}[htbp!]
\begin{center}
  \begin{tabular}{cccc}
&  $\mathcal{M}_3$ & $\mathcal{M}_4$ & $\mathcal{M}_4'$ \\
    \rotatebox{90}{ \hspace{3em} \small density}
	&  \includegraphics[width=.2\textwidth]{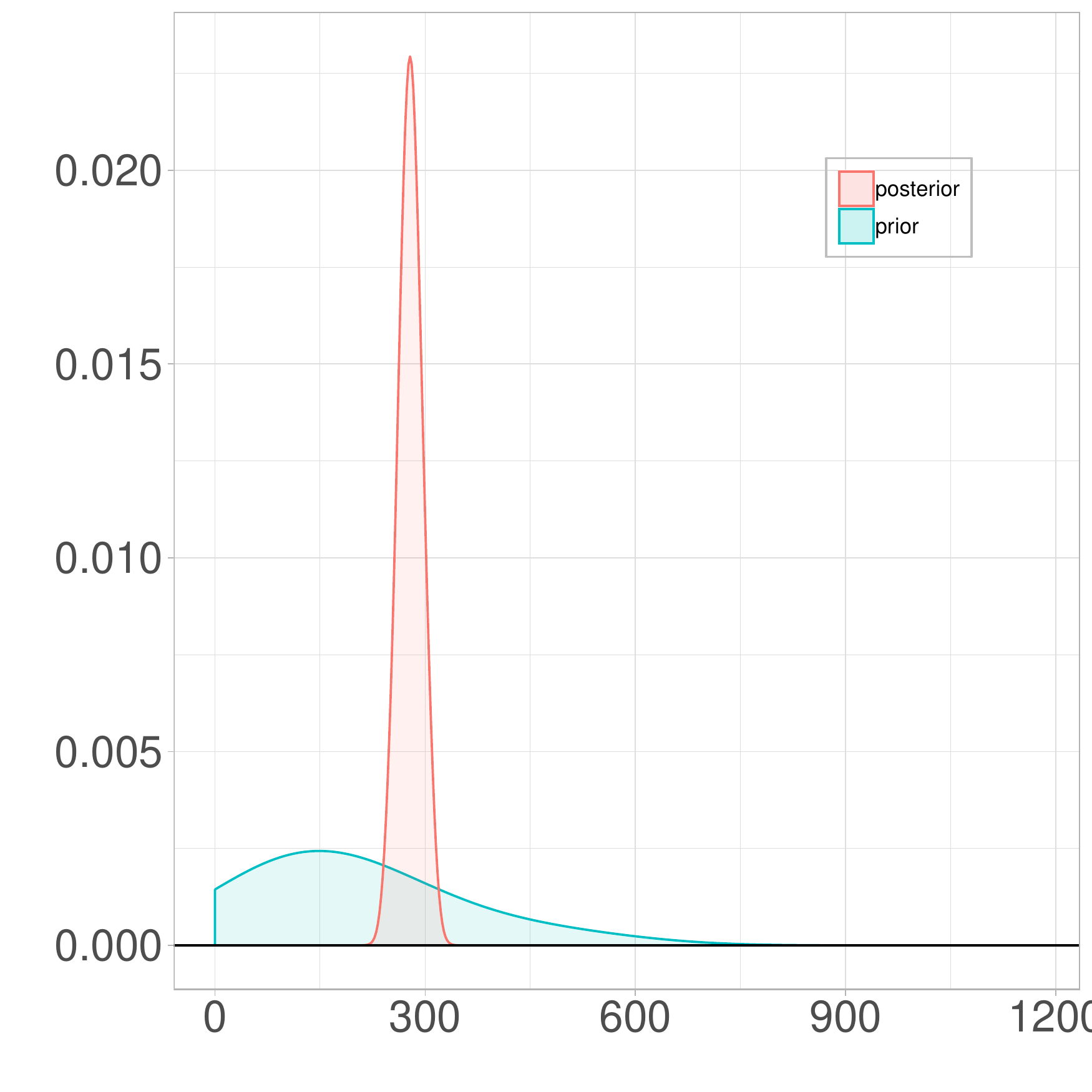}
	&  \includegraphics[width=.2\textwidth]{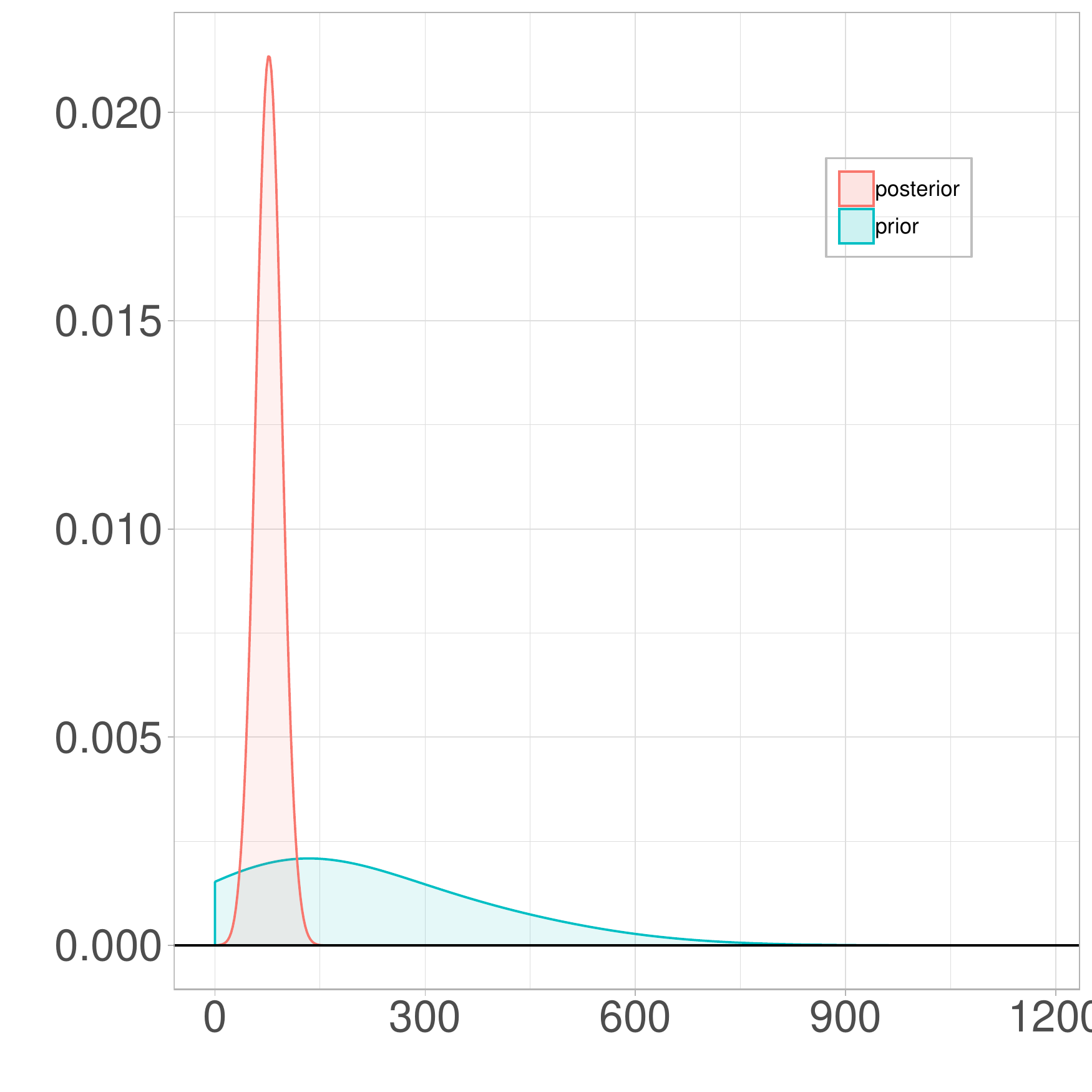}
	&  \includegraphics[width=.2\textwidth]{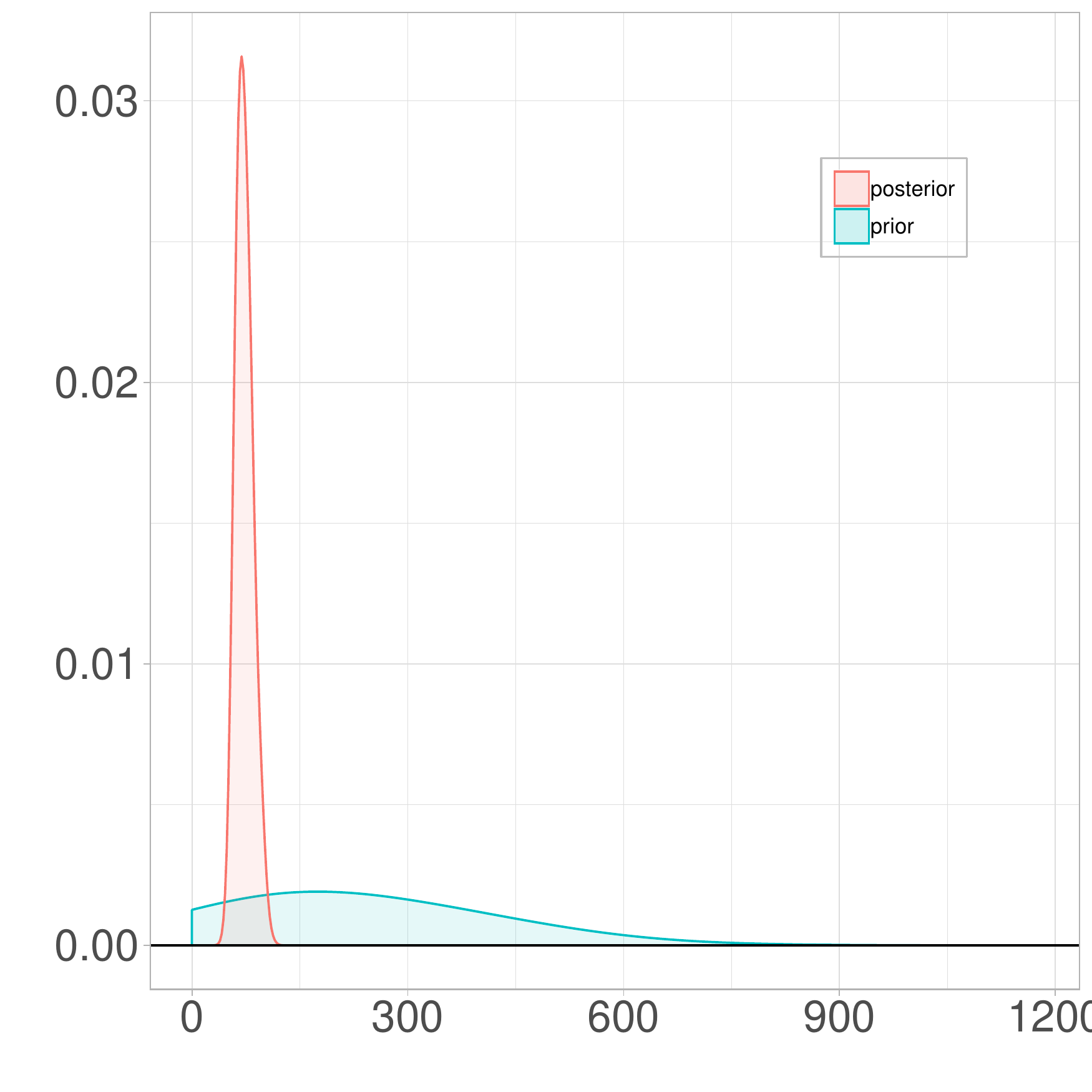}\\
	&\multicolumn{3}{c}{$\sigma_{\delta}^2$}\\
	&&&\\
    \rotatebox{90}{ \hspace{3em} \small density}
	&  \includegraphics[width=.2\textwidth]{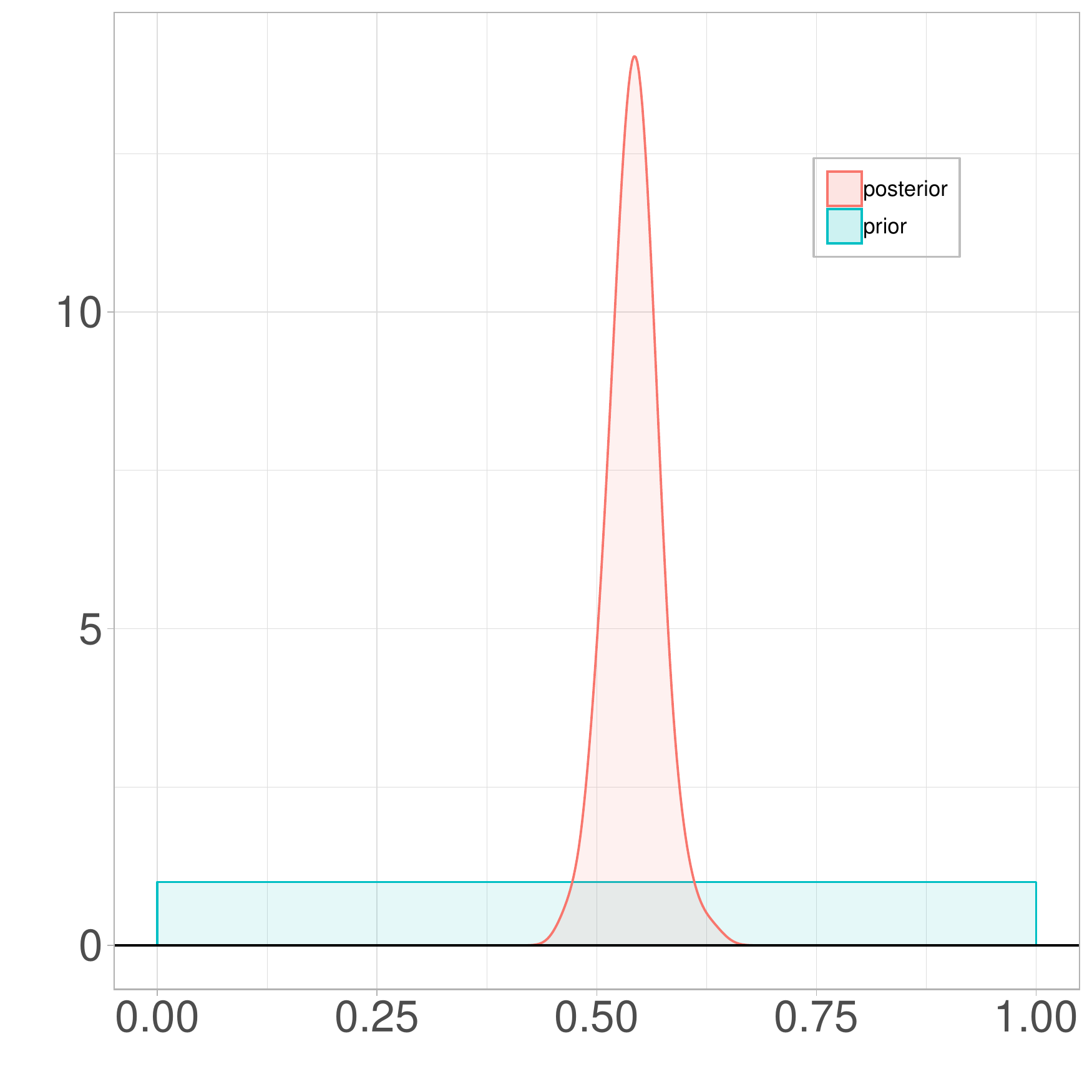}
	&  \includegraphics[width=.2\textwidth]{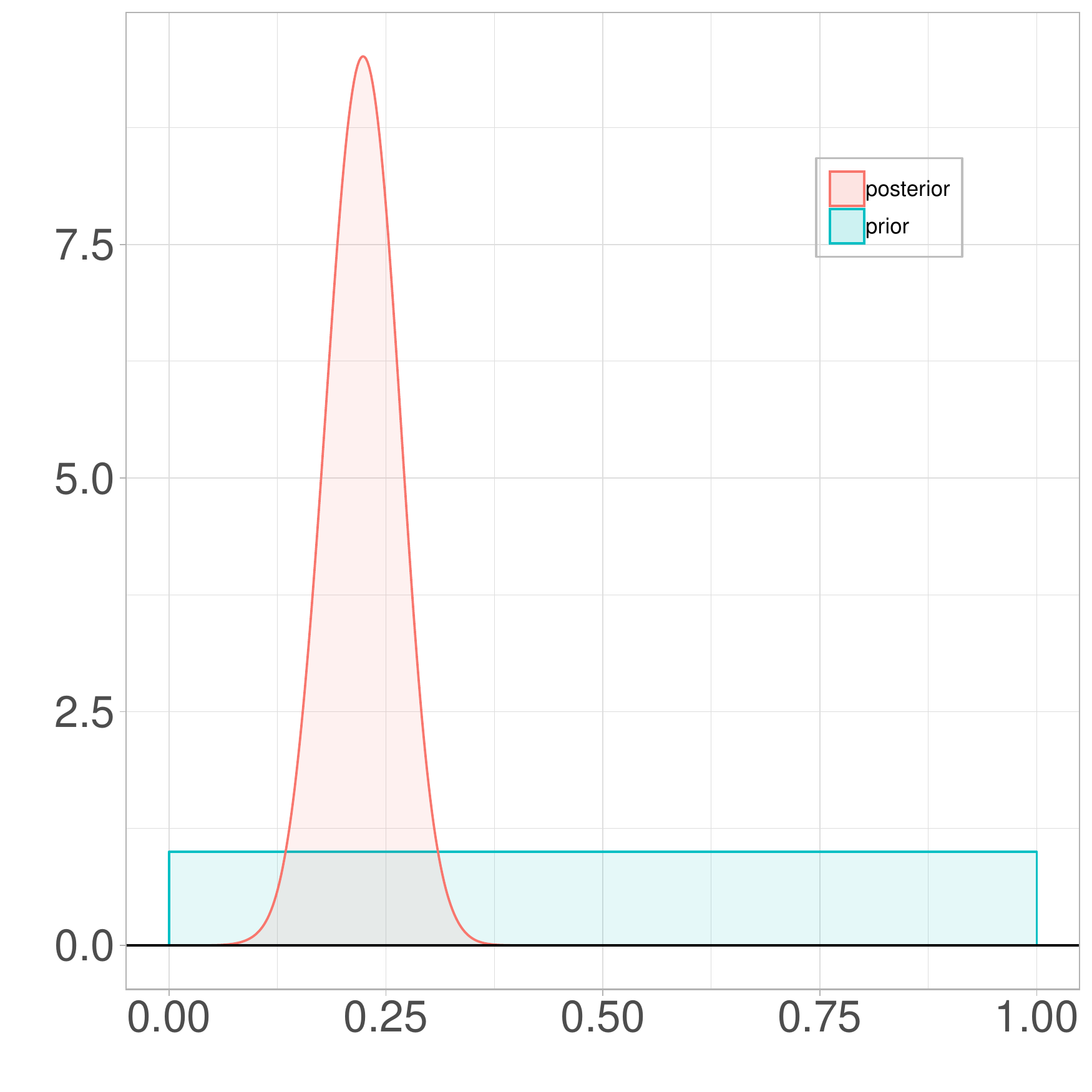}
	&  \includegraphics[width=.2\textwidth]{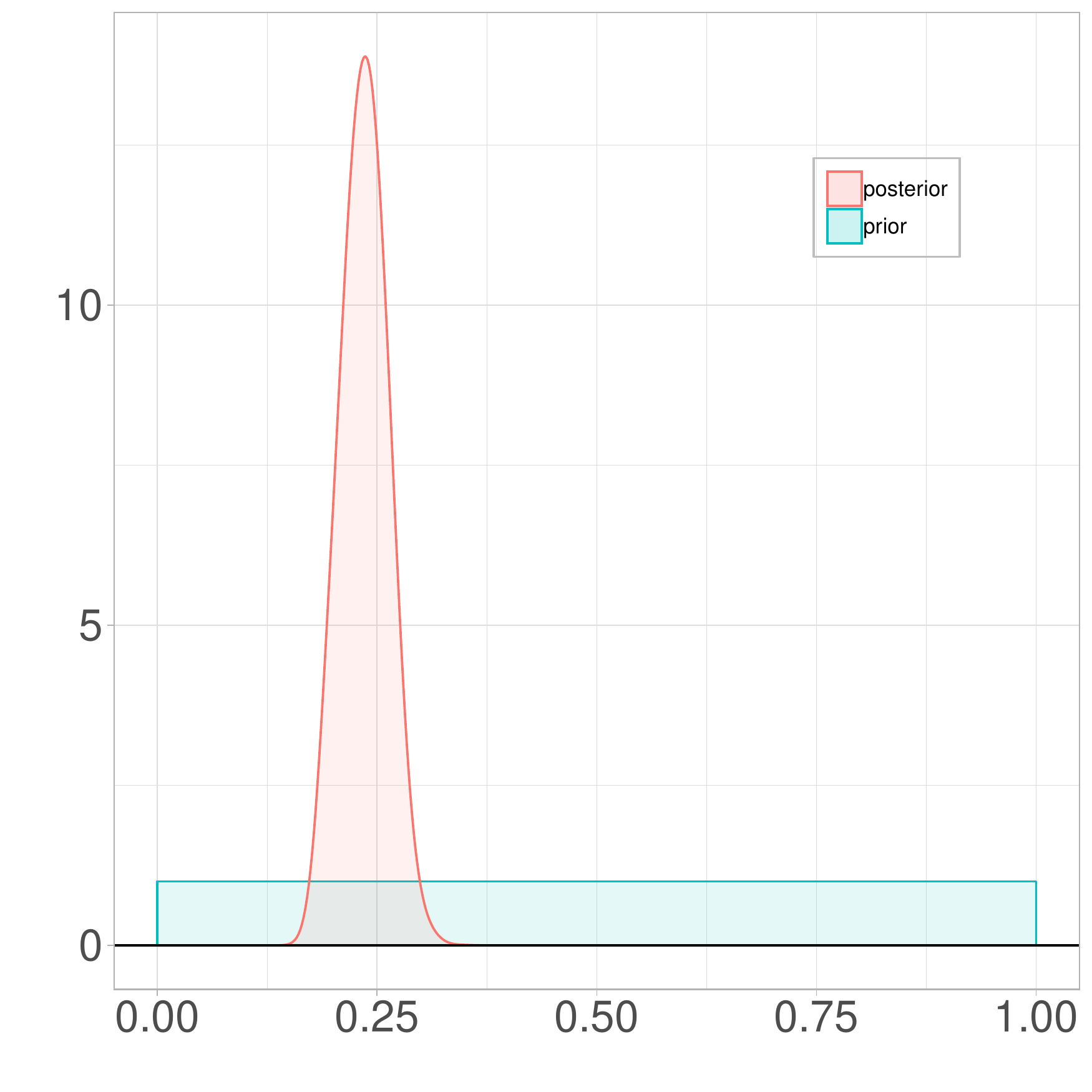}\\
	&\multicolumn{3}{c}{$\psi_{\delta}$}\\
  \end{tabular}   
\caption{\textit{Prior} (in blue) and \textit{posterior} (in red) densities of $\sigma_{\delta}^2$ and $\psi_{\delta}$ for $\mathcal{M}_3$, $\mathcal{M}_4$ and $\mathcal{M}_4'$.}
\label{fig:comparisionDensities2}
\end{center}
\end{figure}

Figure \ref{fig:comparisionDensities2} illustrates the estimation of the parameters from the discrepancy term.
As expected, learning from data has improved our \textit{prior} belief by decreasing the \textit{prior} uncertainty of the parameters. 
It shows that in both cases (with and without emulator) that convergence seems to be reached at some point. \newline

\begin{figure}[htbp!]
\begin{center}
  \begin{tabular}{cccccc}
    \rotatebox{90}{ \hspace{4em} \footnotesize $\mu_t$}
    & \includegraphics[width=.2\textwidth]{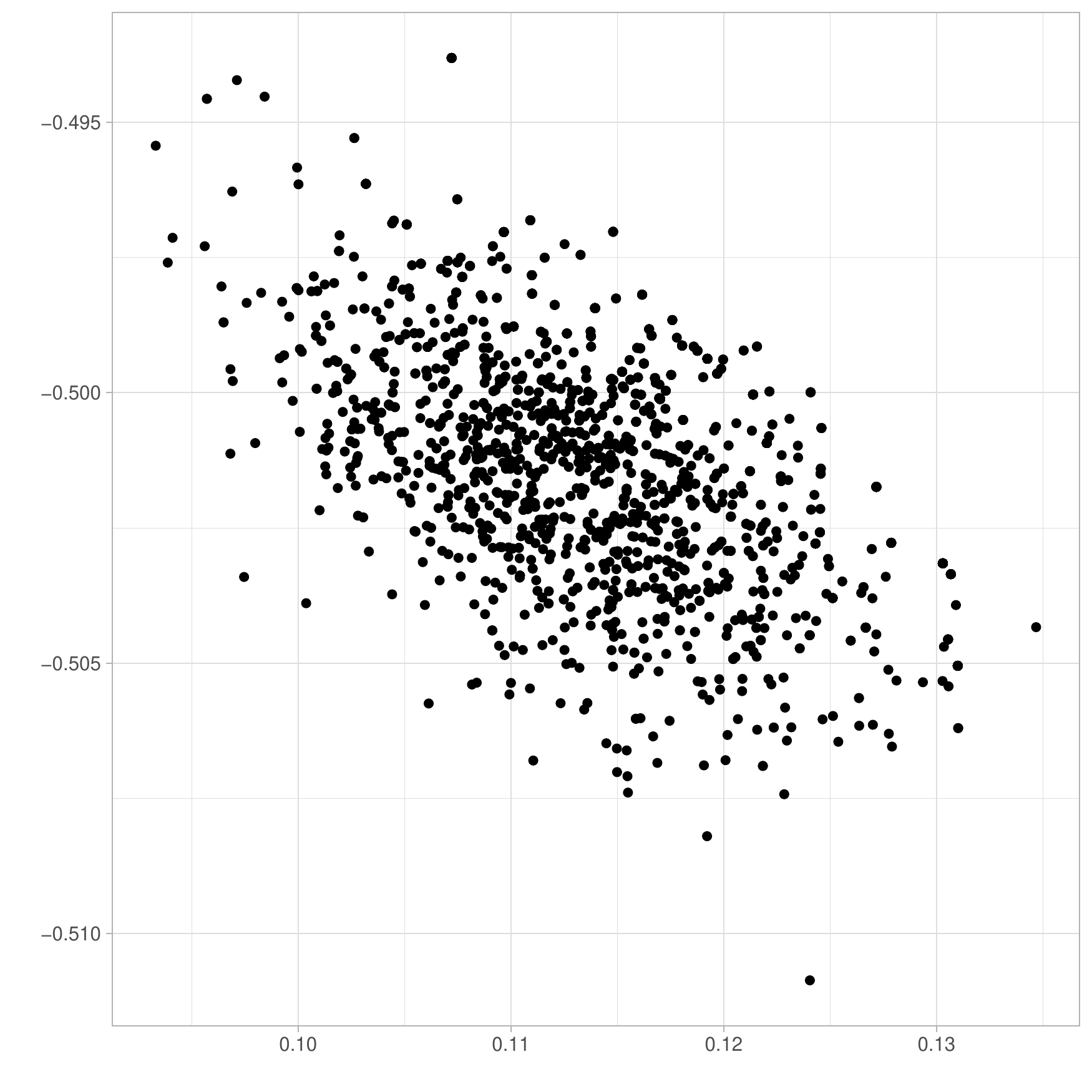} 
    &\rotatebox{90}{ \hspace{4em} \footnotesize $a_r$}
    &  \includegraphics[width=.2\textwidth]{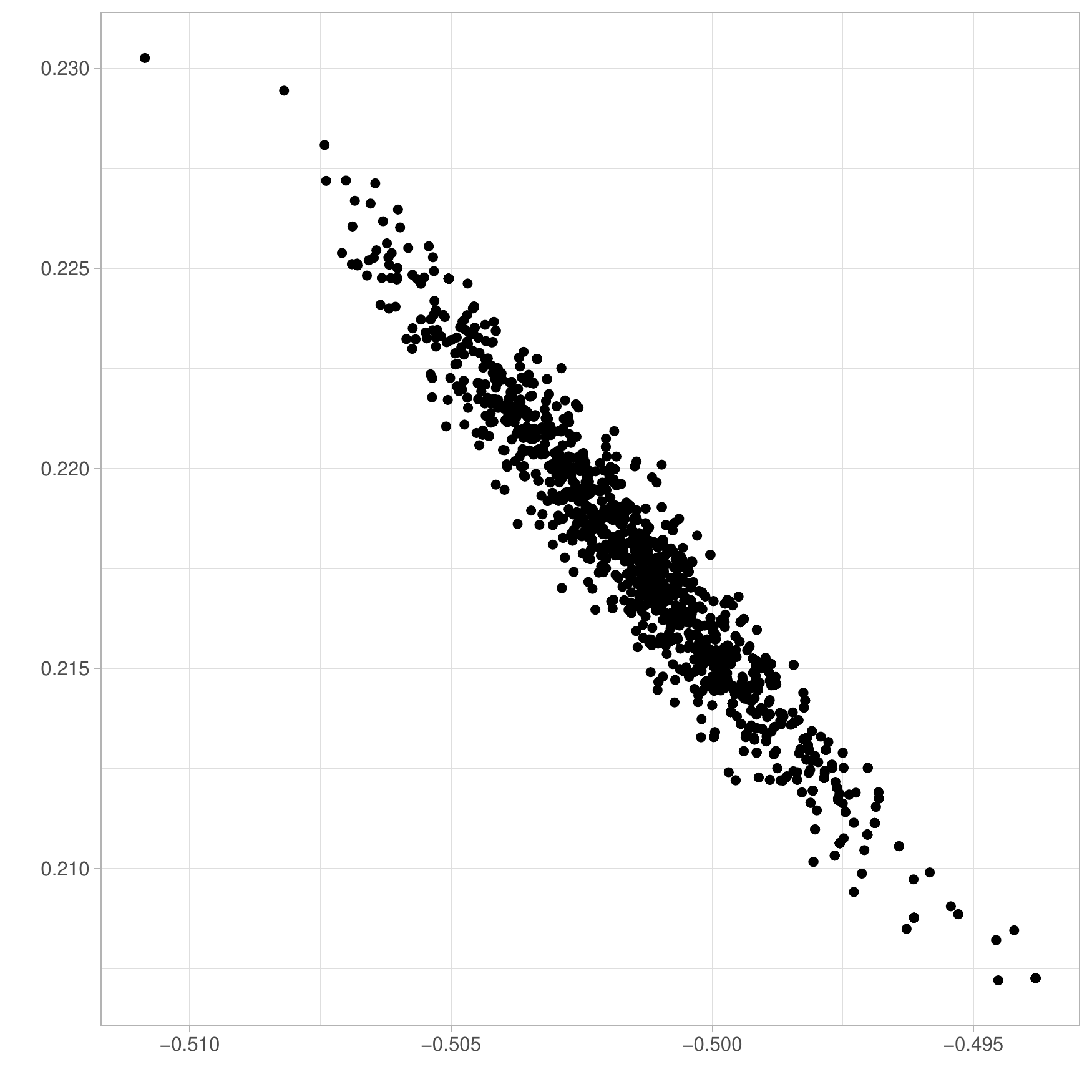}
    &\rotatebox{90}{ \hspace{4em} \footnotesize $\eta$}
	&  \includegraphics[width=.2\textwidth]{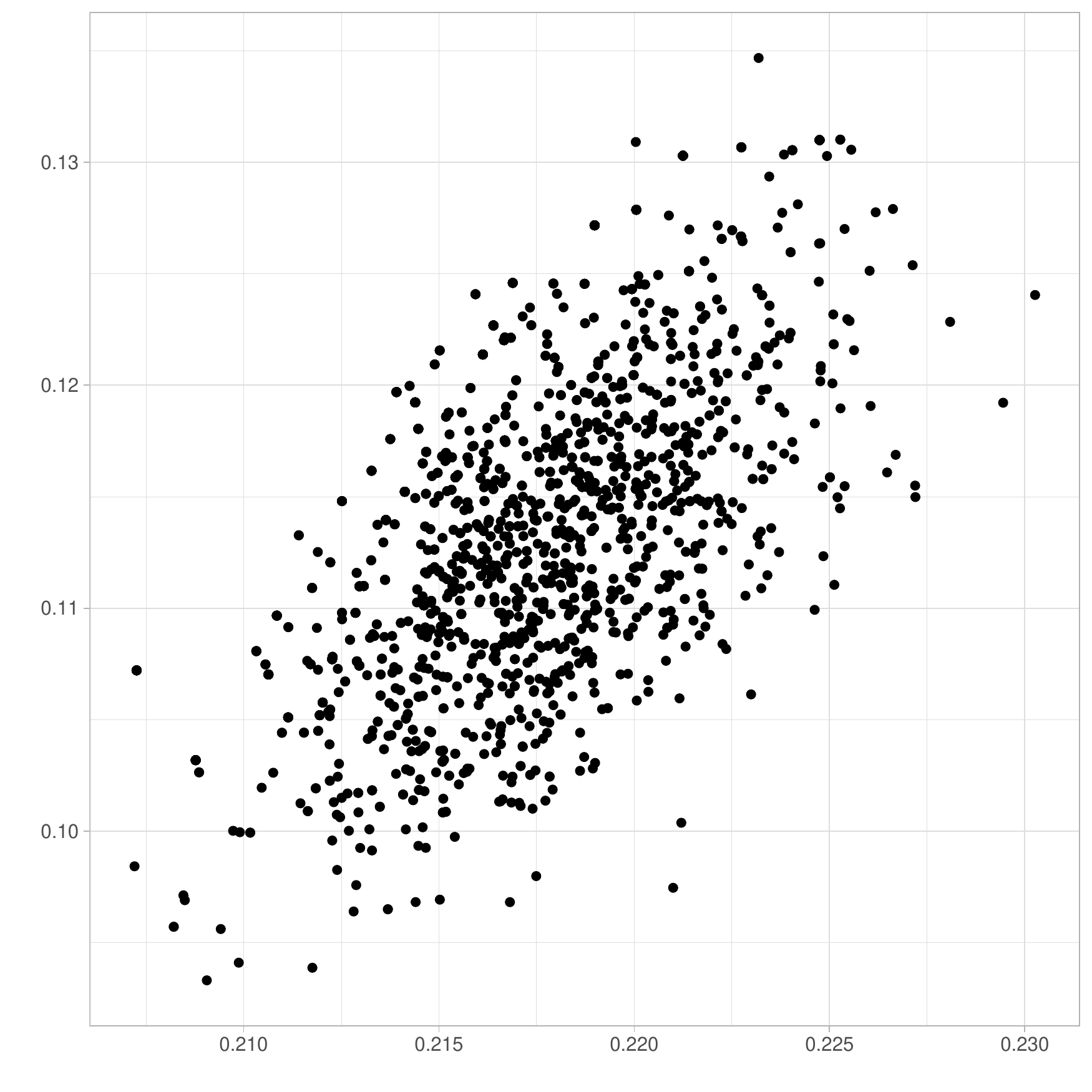}
\\
	&$\eta$ & &$\mu_t$ & & $a_r$\\
  \end{tabular}
\caption{Correlation between parameters of Model $\mathcal{M}_1$}
\label{fig:corrPlot}
\end{center}
\end{figure}

In the \textit{posterior} densities generated, we also depict correlation between the parameters. As a matter of fact, a strong positive and linear correlation links all the parameters ($\eta$, $\mu_t$ and $a_r$) with one another as illustrated in Figure \ref{fig:corrPlot}. A strong correlation can be seen between $\mu_t$ and $a_r$. A lower, but still meaningful, correlation is also visible between $\eta$ and $\mu_r$, as well as between $a_r$ and $\eta$.\newline

\subsection{Comparison}

To compare the prediction ability of the four models, a cross validation (CV) was performed. Three days of data (randomly chosen) were taken from the calibration dataset for each of the $100$ repetitions of the CV. 
The densities, generated from the MCMC samples, allow us to compute, for each model, the $90\%$ predictive credibility intervals for power. 
The coverage rate at $90\%$ represents the quantity of validation experiments contained in these credibility intervals.
The Root Mean Square Error (RMSE) is also computed for the instantaneous power. The results are displayed in Table 
\ref{tab:comparison}.\newline

\begin{table}[htbp!]
\centering
\caption{Comparison of the RMSEs and coverage rates in prediction of 100 test-sets on three randomly selected days where $\mathcal{M}_2'$ and $\mathcal{M}_4'$ are the models based on the Gaussian process established after the sequential design}
\label{tab:comparison}
\begin{tabular}{c|c|c|c|c|c|c}

& $\mathcal{M}_1$ & $\mathcal{M}_2$ & $\mathcal{M}_3$ & $\mathcal{M}_4$ & $\mathcal{M}_2'$&  $\mathcal{M}_4'$ \\
\hline
\hline
coverage rate at 90\% (in \%) & 91 & 44 & 85 & 42 & 71 & 68 \\
\hline
RMSE of the instantaneous power ($W$) & 5.103 & 21.79 & 4.56 & 18.78 & 10.94 & 9.29 \\
\end{tabular}
\end{table}

\new{The coverage rates for $\mathcal{M}_1$ and $\mathcal{M}_3$ correspond to the chosen credibility level.
However for $\mathcal{M}_2$ and $\mathcal{M}_4$ the coverage rates are below this level. As expected, the coverage rates of $\mathcal{M}_2'$ and $\mathcal{M}_4'$ increase since the quality of the emulator has been improved. The coverage levels remain below $90\%$ which may 
result from the negligence of certain sources of uncertainty
such as the estimation of the nuisance parameters. As shown in Figure \ref{fig:calibrationSeq}, it has also led to spikier \textit{posterior} distributions for some of the parameters than the ones obtained with $\mathcal{M}_1$ or $\mathcal{M}_3$.}
\newline

\new{Overall, the model $\mathcal{M}_3$ gives better results than the others in two respects. First, the code realizes a better prediction than the emulator. Second, a correlation structure remains in the error. Adding the discrepancy in the model makes it possible to reproduce the real results. In the case when an emulator is used instead of the code, the CV produces worse results which was expected since the number of points chosen in the DOE is not sufficient to reproduce the exact behavior of the code. The use of the sequential design (even adding only $10$ points) has allowed to drastically improves the models for both calibration and prediction.\newline
}

\section{Conclusion and discussion}

\new{This article focuses on code calibration which is a part of uncertainty quantification in numerical experiments. Although the code used in this paper is a quick code which predicts power generated from a small PV plant, it was also treated as time consuming in order to investigate the consequences of emulation on calibration. In particular, it has been shown that sequential designs \citep{damblin2018} could help to perform a better calibration by improving the emulator.
This work can then be extended to bigger computational codes in application at larger PV plants where emulation is required. As we are working with a physical code, it is important to keep in mind the meaning of the physical boundaries. Indeed, this has led
to confirm and to interpret the presence of the discrepancy term.}
\newline

%

\new{When using models with the discrepancy term ($\mathcal{M}_3$ and $\mathcal{M}_4$), the mean of the GP was set to $0$ since we consider the parameter $\boldsymbol{\theta}$ as a best fitting parameter. However, in spite of this hypothesis, a confounding can still occur between the posterior distributions of $\boldsymbol{\theta}$ and $\delta$. That is why more recent works \citep{plumlee2017,gu} advocate  adding constraints on the GP which models the discrepancy. These methods result in additional computational burden for the estimation procedure but seem  promising ways to deal with the confounding effect.
Other works \citep{jenny} make the case for setting strong hypotheses on the discrepancy term but this needs a deep elicitation which is not always possible.} 
\newline


\new{One may wonder which model to use in a  particular case study. 
If the code is time consuming, only $\mathcal{M}_2$ or $\mathcal{M}_4$ are practicable. Then,
the relevance of the discrepancy term is questionable. A first attempt to answer this question was developed in  \citet{damblin2016}. 
Models with or without the discrepancy term are compared by computing a Bayes factor. 
This is done in a simplified context where the code is assumed to be linear with respect to the parameters to be calibrated. The extension to
the general case is challenging since Bayes factors are burdensome to compute and extremely sensitive to the \textit{prior} distributions of the parameters.
}

\section{Acknowledgement}
This work was supported by the research contract CIFRE n$^{\circ}$2015/0974 between Électricité de France and AgroParisTech.

\newpage

\begin{appendices}
	\addtocontents{toc}{\protect\renewcommand{\protect\cftchappresnum}{Appendix }}
	\renewcommand{\chaptername}{Appendix}
	\section{Gaussian processes \label{ap:GaussianProcesses}}
	Let us consider a probability space $(\Omega,\mathcal{F},\mathbb{P})$ where $\Omega$ stands for a sample space, $\mathcal{F}$ a $\sigma$-algebra on $\Omega$ and $\mathbb{P}$ a probability on $\mathcal{F}$. A stochastic process $X$ is a family such as $\{ X_t\ ;\ t\in\mathcal{T}\}$ where $\mathcal{T}\subset\mathbb{R}^d$. It is said that the random process is indexed by the indexes of $\mathcal{T}$. At $t$ fixed, the application $X_t \ : \ \Omega \rightarrow \mathbb{R}$ is a random variable. However at $\omega \in \Omega$ fixed, the application $ t \rightarrow X_t(\omega)$ is a trajectory of the stochastic process.\newline

For $t_1 \in \mathcal{T},\dots, t_n\in\mathcal{T}$, the probability distribution of the random vector $(X_{t_1},\dots,X_{t_n})$ is called finite-dimensional distribution of the stochastic process $\{X_t\}_{t\in\mathcal{T}}$. Hence, the probability distribution of an aleatory process is determined by its finite-dimensional distributions. Kolmogorov's theorem guaranties the existence of such a stochastic process if a suitable collection of coherent finite-dimensional distributions is provided.\newline

A random vector $\boldsymbol{Z}$ such as $\boldsymbol{Z}=(Z_1,\dots,Z_n)$ is Gaussian if $\forall \lambda_1,\dots,\lambda_n \in \mathbb{R}$ the random variable $\sum_{i=1}^n\lambda_iZ_i$ is Gaussian. The distribution of $Z$ is straightforwardly determined by its first two moments \hc the mean $\boldsymbol{\mu}=(\mathbb{E}[Z_1],\dots,\mathbb{E}[Z_n])$ and the variance covariance matrix $\Sigma = cov(Z_i,Z_j)_{1\leq i,\ j\leq n}$. When $\Sigma$ is positive definite, $Z$ has a probability density defined by equation (\ref{eq:densityGaussian}). \newline

\begin{equation}
f(\bm{z})=\frac{|\Sigma|^{-1/2}}{(2\pi)^{n/2}}\exp \Big\{-\frac{1}{2}(\boldsymbol{z}-\boldsymbol{\mu})^T\Sigma^{-1}(\boldsymbol{z}-\boldsymbol{\mu}) \Big\}
\label{eq:densityGaussian}
\end{equation}

Let us consider two Gaussian vectors called $\bm{U_1}$ and $\boldsymbol{U_2}$ such that\hc \newline

\begin{equation*}
\begin{pmatrix}
\bm{U_1}\\
\bm{U_2}
\end{pmatrix} \sim \mathcal{N} \Big( \begin{pmatrix}
\bm{\mu_1}\\
\bm{\mu_2}
\end{pmatrix}, \begin{pmatrix}
\Sigma_{1,1} & \Sigma_{1,2}\\
\Sigma_{2,1} & \Sigma_{2,2}
\end{pmatrix} \Big)
\end{equation*}

The conditional distribution $\bm{U_2}|\bm{U_1}$ is also Gaussian (Equation (\ref{eq:conditionalGaussian})). This property is especially useful when an emulator is created from a code. \newline

\begin{equation}
\bm{U_2}|\bm{U_1} \sim \mathcal{N}(\bm{\mu_2}+\Sigma_{2,1}\Sigma_{1,1}^{-1}(\bm{U_1}-\bm{\mu_1}), \Sigma_{2,2}-\Sigma_{2,1}\Sigma_{1,1}^{-1}\Sigma_{1,2})
\label{eq:conditionalGaussian}
\end{equation}

A stochastic process $\{X_t\}_{t\in\mathcal{T}}$ is a Gaussian process if each of its finite-dimensional distributions is Gaussian. Let us introduce the mean function such that $m : t\in\mathcal{T} \rightarrow m(t)=\mathbb{E}[X_t]$ and the correlation function such that $ K : (t,t')\in\mathcal{T}\times \mathcal{T}\rightarrow K(t,t')=corr(X_t,X_{t'})$. A Gaussian process with a scale parameter noted $\sigma^2$ will be defined as equation (\ref{eq:GaussianDef}).

\begin{equation}
X(.) \sim \mathcal{PG}(m(.),\sigma^2K(.,.))
\label{eq:GaussianDef}
\end{equation}

Gaussian processes are used in this article in two cases. In the first one, $f$ is a code function with a long runtime and the Gaussian process emulates its behavior. The Gaussian process is called the emulator of the code. The second case is when we want to estimate the error made by the code (called code error or discrepancy in this article). For the former, we want to create an emulator $\tilde{f}$ of a deterministic function $f$. In a Bayesian framework, the Gaussian process is a "functional" \textit{a priori} on $f$ \citep{currin1991}.\newline

Let us note\hc
\begin{equation}
f(.) \sim \mathcal{PG}(h(.)^T\boldsymbol{\beta}_f,\sigma_f^2K_{\boldsymbol{\psi}_f}(.,.))
\end{equation}
where $\boldsymbol{\beta}_f$, $\sigma_f^2$, $\boldsymbol{\psi}_f$ are the parameters specifying the mean and the variance-covariance structure of the process and $h(t)=(h_1(t),\dots,h_n(t))$ is a vector of regressors. For $(t,t')\in \mathcal{T}\times\mathcal{T}$\hc
\begin{equation}
cov(f(t),f(t'))=\sigma_f^2K_{\boldsymbol{\psi}_f}(t,t')
\end{equation}

Let us consider that the code has been tested on $N$ points \textit{i.e.} on $N$ different vectors $\bm{t}$. The design of experiments (DOE) is noted $D=(t_1,\dots,t_N)^T$ and the outputs of $D$ by $f$ will be defined as $y=(f(t_1),\dots,f(t_N))^T$. The correlation matrix  induced by $y$ can be defined by the correlation function $K_{\boldsymbol{\psi}_f}(.,.)$ and can be written as $\Sigma_{\boldsymbol{\psi}_f}(D)=\Sigma_{\boldsymbol{\psi}_f}(D,D)$ such that $\forall (i,j) \in [1,\dots,n] \  \Sigma_{\boldsymbol{\psi}_f}(D)(i,j)=K_{\boldsymbol{\psi}_f}(t_i,t_j)$.

\begin{equation}
\begin{pmatrix}
f(t)\\
f(D)
\end{pmatrix} \sim \mathcal{N}\Big( \begin{pmatrix}
h(t)^T\boldsymbol{\beta}_f\\
h(D)^T\boldsymbol{\beta}_f
\end{pmatrix}, \sigma_f^2\begin{pmatrix}
\Sigma_{\boldsymbol{\psi}_f}(t) &  \Sigma_{\boldsymbol{\psi}_f}(t,D) \\
\Sigma_{\boldsymbol{\psi}_f}(t,D)^T & \Sigma_{\boldsymbol{\psi}_f}(D)
\end{pmatrix}\Big)
\label{eq:Conditionnal}
\end{equation}

From Equation (\ref{eq:conditionalGaussian}), it follows that $f(t)|f(D)\sim\mathcal{PG}(\mu_p(t),\Sigma_p(t))$. This conditional is called \textit{posterior} distribution with \hc
\begin{equation*}
\mu_p(t)=h(t)^T\boldsymbol{\beta}_f+\Sigma_{\boldsymbol{\psi}_f}(t,D)\Sigma_{\boldsymbol{\psi}_f}(D)^{-1}(f(D)-h(D)^T\boldsymbol{\beta}_f)
\end{equation*} 
\begin{equation*}
\Sigma_p(t,t')=\sigma_f^2\Big(\Sigma_{\boldsymbol{\psi}_f}(t,t')-\Sigma_{\boldsymbol{\psi}_f}(t,D)^T\Sigma_{\boldsymbol{\psi}_f}(D)^{-1}\Sigma_{\boldsymbol{\psi}_f}(t',D)\Big)
\end{equation*}

The mean obtained \textit{a posteriori} is called the Best Linear Unbiased Predictor (BLUP) which is the linear predictor without bias $\tilde{f}$ of $f$ which minimizes the Mean Square Error (MSE) \hc
\begin{equation}
MSE(\tilde{f})=\mathbb{E}[(f-\tilde{f})^2]
\end{equation}

In this appendix, we will not discuss the choice of $K_{\boldsymbol{\psi}_f}$, the parameter estimation, nor the validation of the Gaussian process.
\newpage

\end{appendices}

\bibliography{biblio}

\end{document}